\documentclass{aa}
\usepackage{graphicx}
\usepackage{amsmath}
\usepackage{txfonts}
\usepackage{lscape}
\usepackage{longtable}
\usepackage{rotating}
\usepackage{appendix}
\usepackage{url}
\usepackage{float}  
\begin{document}
\def\HII{H\,{\sc{ii}}}
\def\HI{H\,{\sc{i}}}
\def\farcs{\hbox{$.\!\!^{\prime\prime}$}}
\def\Ks{$K_{\rm{s}}$}
\def\sun{\hbox{$_\odot$}}
\def\degr{\hbox{$^\circ$}}

\def\h{\hbox{$^{\reset@font\r@mn{h}}$}}
\def\m{\hbox{$^{\reset@font\r@mn{m}}$}}
\def\s{\hbox{$^{\reset@font\r@mn{s}}$}}

\def\msol{\hbox{\kern 0.20em $M_\odot$}}
\def\lsol{\hbox{\kern 0.20em $L_\odot$}}

\def\kms{\hbox{\kern 0.20em km\kern 0.20em s$^{-1}$}}
\def\pc{\hbox{\kern 0.20em pc$^{2}$}}

   \title{Interstellar matter and star formation in W5-E}   
   \subtitle{A {\bf {\it Herschel}}\thanks{\emph{Herschel} is an ESA space observatory with science instruments provided by European-led Principal Investigator consortia and with important participation from NASA.} view}
\author{L. Deharveng\inst{1}
	\and A. Zavagno\inst{1}
    \and L.D. Anderson\inst{2}
    \and F. Motte\inst{3}
    \and A. Abergel\inst{4}
    \and Ph. Andr\'e\inst{3}
    \and S. Bontemps\inst{5}
    \and G. Leleu\inst{1}
    \and H. Roussel\inst{6}
    \and D. Russeil\inst{1}	}
    	
\offprints{lise.deharveng@oamp.fr }
\institute{Laboratoire d'Astrophysique de Marseille (UMR 7326 CNRS \& 
Universit\'e d'Aix-Marseille), 38 rue F. Joliot-Curie, 13388 Marseille 
CEDEX 13, France 
	\and Physics Department, West Virginia University, Morgantown, WV 26506, USA
    \and Laboratoire AIM Paris-Saclay, CEA/IRFU - CNRS/INSU - Universit\'e Paris Diderot, 
    Service d'Astrophysique, B\^at. 709, CEA-Saclay, 91191 Gif-sur-Yvette CEDEX, France
    \and Institut d'Astrophysique Spatiale, UMR 8617, CNRS, Universit\'e Paris-Sud 11, 
    91405, Orsay, France
    \and CNRS/INSU, Laboratoire d'Astrophysique de Bordeaux, UMR 5804, BP 89, 
    33271, Floirac CEDEX, France
    \and Institut d'Astrophysique de Paris, UMR 7095 CNRS, Universit\'e Pierre \& Marie Curie, 
    98 bis Boulevard Arago, 75014 Paris, France	}
   \date{Received ; Accepted }
  \abstract
   {}
   {We identify the young stellar objects (YSOs) present in the vicinity of the W5-E \HII\ region, and study the influence of this \HII\ region 
   on the star formation process in its surrounding molecular material.}
   {W5-E has been observed with the {\it Herschel}-PACS and -SPIRE photometers, as part of the HOBYS key program; maps have been obtained 
   at 100~$\mu$m, 160~$\mu$m, 250~$\mu$m, 350~$\mu$m, and 500~$\mu$m. The dust temperature and column density have been obtained by fitting spectral 
   energy distributions (SEDs). Point sources have been detected and measured using PSF photometry with DAOPHOT.}
   {The dust temperature map shows a rather uniform temperature, in the range 17.5~K -- 20~K in the dense condensations or filaments, 
   in the range 21~K -- 22~K in the photodissociation regions (PDRs), and in the range 24~K -- 31~K in the direction of the ionized regions. The values 
   in the column density map are rather low, everywhere lower than 10$^{23}$~cm$^{-2}$, and of the order of a few 10$^{21}$~cm$^{-2}$ in the PDRs. 
   About 8000~$\msol$ of neutral material surrounds the ionized region, which is low with respect to the volume 
   of this \HII\ region; we suggest that the exciting stars of the W5-E, W5-W, Sh~201, A and B \HII\ regions formed along a dense filament or sheet  
   rather than inside a more spherical cloud. Fifty point sources have been detected at 100~$\mu$m. Most of them are Class 0/I YSOs. 
   The SEDs of their envelopes have been fitted using a modified blackbody model. These envelopes are cold, with a mean temperature of 15.7$\pm$1.8~K. 
   Their masses are in the range 1.3~$\msol$ -- 47~$\msol$. Eleven of these point sources are candidate Class~0 YSOs. Twelve of these 
   point sources are possibly at the origin of bipolar outflows detected in this region. None of the YSOs contain a massive central object, but a few may 
   form a massive star as they have both a massive envelope and also a high envelope accretion rate. Most of the Class~0/I YSOs are observed 
   in the direction of high column density material, for example in the direction of the massive condensations present at the waist of the bipolar Sh~201 \HII\ region  
   or enclosed by the bright-rimmed cloud BRC14. The overdensity of Class~0/I YSOs on the borders of the \HII\ regions present in the field strongly suggests 
   that triggered star formation is at work in this region but, due to insufficient resolution, the exact processes at the 
   origin of the triggering are difficult to determine.}
   {}
\titlerunning{Star Formation in W5-E}
\authorrunning{L.~Deharveng et al.}
\maketitle
%

\section{Introduction}

W5 is a Galactic \HII\ region located in the Perseus arm, at a 
distance of about 2~kpc (Sect.~2). It is composed of two adjacent \HII\ regions, W5-E and W5-W, each surrounding 
an exciting stellar cluster.  Due to its simple morphology and its proximity to the Sun, W5 is an excellent laboratory to study 
the feedback of massive stars on the surrounding material in the context of ongoing star formation.

W5 is a well-studied \HII\ region -- what more can we learn from {\it Herschel} observations? {\it Herschel} provides the 
complete, deep image of all the gas available for a new generation of stars.  First, the 
high sensitivity and high resolution far-infrared (FIR) {\it Herschel} maps trace the emission from cold dust and thus we can have
a better view of the distribution of the dense molecular material surrounding the ionized region.  It is precisely this dense
material from which new generations of stars may form. Second, the {\it Herschel} fluxes from 100~$\mu$m  to 500~$\mu$m of the young stellar objects (YSOs) 
allow us to create well-sampled spectral energy distributions (SEDs).  {\it Herschel} allows us to
determine the peak wavelength of the SED, which is especially important for the determination of dust temperature.
With a well-sampled SED we may estimate the parameters of YSOs such as the mass of their envelopes,
and better constrain their evolutionary stages. Third, with {\it Herschel} we can detect rare young Class~0 sources and candidate prestellar cores.
Such sources are unresolved with {\it Herschel} and have faint or no {\it Spitzer} 24~$\mu$m counterparts. 
Thus, with {\it Herschel} we may create a complete census of the massive YSOs in W5. We wish to learn the prevalence of triggering in 
the formation of the young sources. We pay special attention  to the formation of massive stars.

This paper is organised as follows: we present the W5 region in Sect.~2; we discuss the {\it Herschel} observations and data reduction in Sect.~3; 
we discuss the distribution and characteristics of the neutral material in Sect.~4; we give the method of detection and the estimation of the physical parameters of YSOs
in Sect.~5 (we only discuss one zone here and we describe the other star forming zones in the Appendix). We discuss star formation in the W5-E complex 
in Sect.~6, and conclude in Sect.~7.
 
\section{Presentation of the region}

Based on their proximity to one another on the sky and their similar radial velocities, it is often assumed that W5 and the nearby W3 and W4 \HII\ regions are 
part of the same large star forming complex.  
A complete description of the W3/W4/W5 complexes is given in Megeath et al. (\cite{meg08}).  
The distance to W3, 2.00$\pm$0.05~kpc, is known accurately
from maser parallax measurements (Xu et al. \cite{xu06}; Hachisuka et al. \cite{hac06}). W4, which is adjacent to W3, probably lies at the same distance 
(Megeath et al. \cite{meg08}).   
The situation for W5 is more uncertain, as it is separated in angle (by more than one degree) from the two other \HII\ regions. 
Spectrophotometric observations, however, give very similar distances  
for the OB clusters IC~1805 and IC~1848 located respectively in W4 and W5  
(in the range 2.1~kpc--2.4~kpc; Becker \& Fenkart \cite{bec71}, Moffat \cite{mof72}, Massey et al. \cite{mas95}, Chauhan et al. \cite{cha09} \& \cite{cha11a}). In the following 
we adopt a distance of 2.0~kpc for the W5 region; this allows us to compare directly with the results of Koenig et al. (\cite{koe08}; hereafter KOE08), based on {\it Spitzer} observations. 

\begin{figure*}[tb]
\centering
\includegraphics[width=170mm]{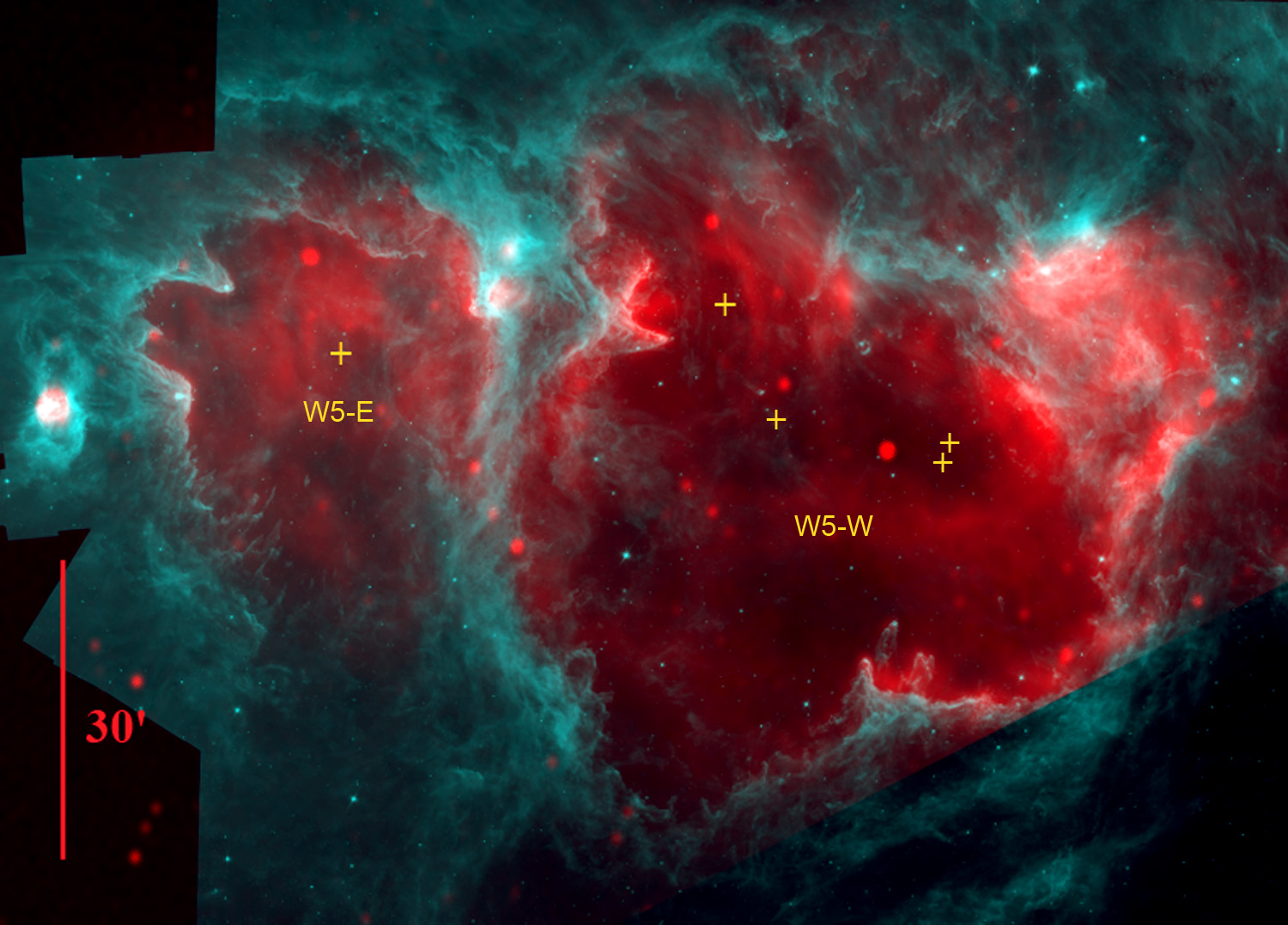}
  \caption{Composite colour image of the W5 region. Red is the radio-continuum emission of the ionized gas at 1.42~GHz (CGPS map); turquoise is 
  the {\it Spitzer}-IRAC 8.0~$\mu$m PAH emission. The two 8.0~$\mu$m bubbles surround the \HII\ regions W5-E and W5-W. The yellow crosses
  lie at the position of the exciting stars of W5-E and W5-W. Most of the compact radio sources in the field are extragalactic sources. North is up and east is left.}  
\label{radio}
\end{figure*}

W5 is an optically visible \HII\ region. Megeath et al. (\cite{meg08}; their fig.~3) show an optical view of the entire region. 
In their figure the [SII] emission, emitted by the low excitation zones close to the ionization fronts (IFs), clearly shows 
the presence of two distinct adjacent \HII\ regions; they have been named W5-E and W5-W by Karr and Martin (\cite{kar03}).  
In the optical W5-E appears almost circular in projection with a diameter of $\sim24$~pc, centered near its exciting star HD~18326. 
At radio wavelengths W5-E has the same aspect and appears almost circular around
 the exciting star. In Fig.~\ref{radio} we show the radio-continuum emission at 1.42~GHz (Canadian Galactic Plane Survey, Taylor et al. \cite{tay03}); the radio-continuum intensity is slightly enhanced 
in the central regions and on the borders of bright-rimmed clouds (identified on Fig.~\ref{presentation}). As shown by Fig.~\ref{radio} the W5-E and W5-W \HII\ regions are 
enclosed in two adjacent bubbles, observed by {\it Spitzer} at 8.0~$\mu$m (KOE08, their figure 4; the 8.0~$\mu$m band is dominated by the emission from polycyclic aromatic hydrocarbons - PAHs - from the photo-dissociation regions or PDRs). The W5-E \HII\ region is  slightly extended in the north-south direction (diameter $\sim$30~pc). 
This is likely due to a density gradient (the density being higher in the north than in the south).   
This density gradient would also explain why the 8~$\mu$m shell surrounding the ionized region opens towards the south. This is discussed in Sect.~4.

The main exciting star of W5-E, HD~18326 (also BD+59$\degr$0578), has a spectral type of O7($\pm$0.5)V (Chauhan et al. \cite{cha11a} and references therein). 
Its coordinates are $\alpha$(2000)$=02^{\rm h} 59^{\rm m} 23.171^{\rm s}$, $\delta$(2000)$=60\degr$33$\arcmin$59.50$\arcsec$ 
(or $l=$138\fdg025970, $b=+$1\fdg500243; Ma\'{i}z-Apell\'aniz et al. \cite{mai04}). 
Its colour B$-$V=0.362 and its spectral type indicate a colour excess E(B$-$V)=0.632 or a visual extinction $A_V=1.96$~mag. 
HD~18326 is part of a rich cluster studied by Chauhan et al. (\cite{cha11a}). They estimate a mean age of 1.3~Myr for the cluster, and a varying extinction in the 
range 1.92~mag--2.50~mag.  This gives an estimate of the interstellar extinction in the direction of W5-E. 

Heyer and Terebey (\cite{hey98}) show large-scale maps of the $^{12}$CO\,(1-0) emission over the entire W3-W4-W5 complex, obtained with a resolution of 45\arcsec.
Their figure 2 shows the molecular material associated with W5-E in the channels 
at $-37.5$~km~s$^{-1}$ and  $-42.5$~km~s$^{-1}$; this material is mainly located in the northern areas of the bubble (thus the suggested density gradient). (See also figure 2 
in Karr and Martin \cite{kar03}, based on the same observational data.) Detailed maps of CO emission can be found in Carpenter et al. (\cite{car00}; $^{13}$CO\,(1-0) maps) 
and in Karr and Martin (\cite{kar03}; $^{12}$CO\,(1-0) maps). These maps show an 
accumulation of molecular material in thin layers following the ionization fronts, and in some well defined regions like inside the bright-rimmed 
cloud BRC14, near the \HII\ region Sh~201, and in the region between W5-E and W5-W; these regions are identified in Fig.~\ref{presentation}. 
High resolution observations of W5-E (HPBW=15.6\arcsec) have been obtained in the $^{13}$CO\,(1-0) and C$^{18}$O\,(1-0) lines 
by Niwa et al. (\cite{niw09}). Eight clouds are identified (three of them are associated with the bright-rimmed clouds BRC12, BRC13, and BRC14), with masses in the range 
460$\msol$ to 36,\,000$\msol$\footnote{These high masses are somewhat puzzling: Karr and Martin (\cite{kar03}) estimate the mass of the molecular 
material in the whole W5 region (including both W5-E and W5-W) to be 44,\,000\msol, a large fraction of this material being associated with 
W5-W. In addition, the most massive clouds in Carpenter et al. (\cite{car00}) are about ten times less massive than those in Niwa et al. (\cite{niw09}).}. 
 
Three bright-rimmed clouds (BRCs) are located on the border of W5-E, BRC12, BRC13, and BRC14 (Fig.~\ref{presentation}). Their morphology and young stellar content 
have been the subject of numerous studies whose aim was to establish if the star formation 
observed in these clouds was triggered by the nearby W5-E \HII\ region. Using a combination of radio, IR, submillimeter, and CO 
observations, Morgan et al. (\cite{mor09}, and references therein) conclude that these clouds are excellent candidates for triggered star formation. 
A similar conclusion is reached by Chauhan et al. (\cite{cha11b}), based on age determination and on the observed 
small scale sequential star formation. 
These regions will be discussed individually in Sect.~5 or in the Appendix.

The  W5 region has been observed by {\it Spitzer} with the IRAC and MIPS instruments. KOE08  
used these data to discuss star formation in W5. They produced a catalogue containing 18,\,518 stellar objects, among which are 2,\,064 Class~I and Class~II YSOs. 
These YSOs are identified and classified using their {\it Spitzer} colours, after an extinction correction based on 2MASS data. 
These authors showed that Class~I and Class~II YSOs are not distributed uniformly, but are primarily in clustered or filamentary structures. 
They also showed that within the cavity 
carved by the \HII\ region, the ratio of Class~II (older) YSOs to Class~I (younger) YSOs is $\sim$7 times higher than within the molecular material at the 
periphery of the ionized gas. KOE08 attributed this result to an age difference between sources in these two locations, the formation 
of the younger objects possibly being triggered by the W5-E \HII\ region. We discuss these results in Sect.~6.

\section{Herschel: observations and reduction}

\subsection{{\it Observations with Herschel}}

W5-E was observed with the PACS photometer (Poglitsch et al.~\cite{pog10}) as part of the HOBYS key program 
(Motte et al. \cite{mot10}) on 23 February 2010. The PACS photometer was used to 
make simultaneous observations in two photometric bands (100~$\mu$m and 160~$\mu$m).  
 Two cross-scan maps were performed at angles of 45$\degr$ and 135$\degr$ with 
a scanning speed of 20$\arcsec$/second. The maps cover an area of 
$70\arcmin\times70\arcmin$ and are centered at $\alpha$(2000)$=02^{\rm h} 59^{\rm m} 26^{\rm s}$, $\delta$(2000)$=60\degr$31$\arcmin$05$\arcsec$.  
The total observing time was 3.6 hours.

W5-E was observed with the SPIRE photometer (Griffin et al. \cite{gri10}, Swinyard et al.~\cite{swi10}) as part of the key program 
``Evolution of Interstellar Dust'' (Abergel et al. \cite{abe10}) on 11 March 2010. The SPIRE photometer was used to make 
simultaneous photometric observations in the three photometric bands (250~$\mu$m, 350~$\mu$m and 500~$\mu$m). The scan speed was
30$\arcsec$/second. Cross-linked scanning is achieved by scanning at angles of 42$\degr$ and $-42\degr$ (one map is obtained for each scanning angle). 
The maps cover an area of $65\arcmin$ $\times$ $60\arcmin$ and are centered at 
$\alpha$(2000)$=02^{\rm h} 59^{\rm m} 25^{\rm s}71$, $\delta$(2000)$= 60\degr$ 32$\arcmin$ 32$\farcs$1. The total observing time is 1.2 hours. 

These data are described and used in Anderson et al. (\cite{and12}; herafter AND12).  AND12 (their Sect.~3) describes more fully these observations 
and their subsequent reduction using the HIPE (version 7.1; Ott \cite{ott10}) and {\it Scanamorphos} (version 9; Roussel \cite{rou12}) softwares. We use the AND12 maps.\\

The {\it Herschel} maps contain a number of point sources that have counterparts in {\it Spitzer} maps. 
The comparison of the positions of the 100~$\mu$m 
point sources with that of their 24~$\mu$m counterparts has shown a shift of a few arcsec 
($\sim$ 9.4$\arcsec$ in right ascension and 0.8$\arcsec$ in declination) between the two sets of coordinates. A smaller shift also affects the SPIRE maps. 
These shifts have been corrected, so that our coordinates are now compatible with the {\it Spitzer} data (see Sect.~3.1 in AND12).

\subsection{Photometry of the YSOs}

In what follows we measure the flux of YSO candidates detected by {\it Herschel}.
We consider all sources that are not resolved in the PACS 100~$\mu$m band as YSO candidates, and analyse these sources at all  
{\it Herschel} wavelengths. We pay special attention to the YSOs classified as Class~I by KOE08, to see if they have {\it Herschel} counterparts. 

We measure {\it Herschel} fluxes using the DAOPHOT stellar photometry package with PSF fitting (Stetson \cite{ste87}), in addition to
aperture photometry on bright isolated sources. 
DAOPHOT, which was designed for crowded fields, is over-performing 
for the {\it Herschel} images containing a few tens of sources. These fields, however, are very difficult to reduce
because we often find sources located in the photodissociation regions surrounding the ionized ones, in regions of
bright and highly spatially variable background emission. 
We use DAOPHOT in an iterative process. In the first step we use bright isolated sources to construct 
a PSF at each photometric band.  We use this PSF to obtain a first estimate of the source positions and magnitudes. Second, we subtract the identified sources 
from the original frame (\# 1).  We smooth the resulting image with a median filter (with a filter window of 15~pixels$\times$15~pixels, larger than the defects resulting from 
the first imperfect reduction; this gives an image, at lower resolution, of the background emission alone.  We subtract this image
from the original one. The resulting image (\# 2) shows the point sources superimposed on a more uniform background 
(the only remaining low-brightness features are smaller than the filter window). In our third step we run DAOPHOT 
on the improved image \# 2. We subtract the identified sources from the original \#1 image to determine if the  
reduction has been improved.  We check the results by eye: 
if the reduction is good no brightness holes or peaks are seen at the position of the sources on the subtracted image. 
We could in principle iterate this process, but we find that it is not necessary.

The frames on which we have performed the DAOPHOT reduction on have a pixel size of 1$\farcs$7 (PACS 100~$\mu$m), 2$\farcs$85 (PACS 160~$\mu$m), 4$\farcs$5 
(SPIRE 250~$\mu$m), 6$\farcs$25 (SPIRE 350~$\mu$m), and 9$\farcs$0 (SPIRE 500~$\mu$m). The HPBW (of the PSF) given by 
DAOPHOT is always slightly larger that the resolution (FWHM) given
 in the Observers' Manuals: $\sim$ 7\farcs 5 at 100~$\mu$m, $\sim$ 12\farcs 5 at 160~$\mu$m, 
$\sim$ 25\arcsec\ at 250~$\mu$m, $\sim$ 34\arcsec\ at 350~$\mu$m, and $\sim$ 48\arcsec\ at 500~$\mu$m. This is probably the signature 
of non-Gaussian PSFs and of our choice for the value of the parameter ``FITTING RADIUS'' in DAOPHOT.
 
We also measure the {\it Herschel} fluxes of the isolated stars used to construct the PSF using aperture photometry, employing
a circular aperture of radius 8 pixels (at all wavelengths 
the PSF radius is slightly larger than 2 pixels). We estimate the
underlying background using an annular aperture centered on the source of inner radius 8 pixels and outer radius 10 pixels.  We apply an aperture correction 
to take into account the fact that: i) all the flux is not enclosed in an aperture of radius 8 pixels; 
ii) a few percent of the source flux is present in the annular zone used to estimate the background. These corrections 
are of the order of 18\% at 100~$\mu$m, 15\% at 160~$\mu$m (based on the PACS instrumental manual), 10\% at 250~$\mu$m, 
8\% at 350~$\mu$m, and 7\% at 500~$\mu$m (our own estimates). The comparison between the DAOPHOT and the aperture magnitudes 
allow us to estimate the accuracy of our measurements. The dispersion is $\sim$0.1~mag (or 10\% of the flux) at 100~$\mu$m and 250~$\mu$m, 
0.15~mag (15\% of the flux) at 160~$\mu$m, 0.2~mag (20\%) at 350~$\mu$m, and 0.3~mag (30\%) at 500~$\mu$m. These numbers are for rather bright 
sources or for fainter ones superimposed on a flat background. The accuracy is the photometry measurements is probably worse for faint sources superimposed on a bright and variable background, such as those located in the vicinity of PDRs. This shows the difficulty of such measurements.

\section{The dense neutral material associated with W5-E}

\begin{figure*}[tb]
\centering
\includegraphics[width=170mm]{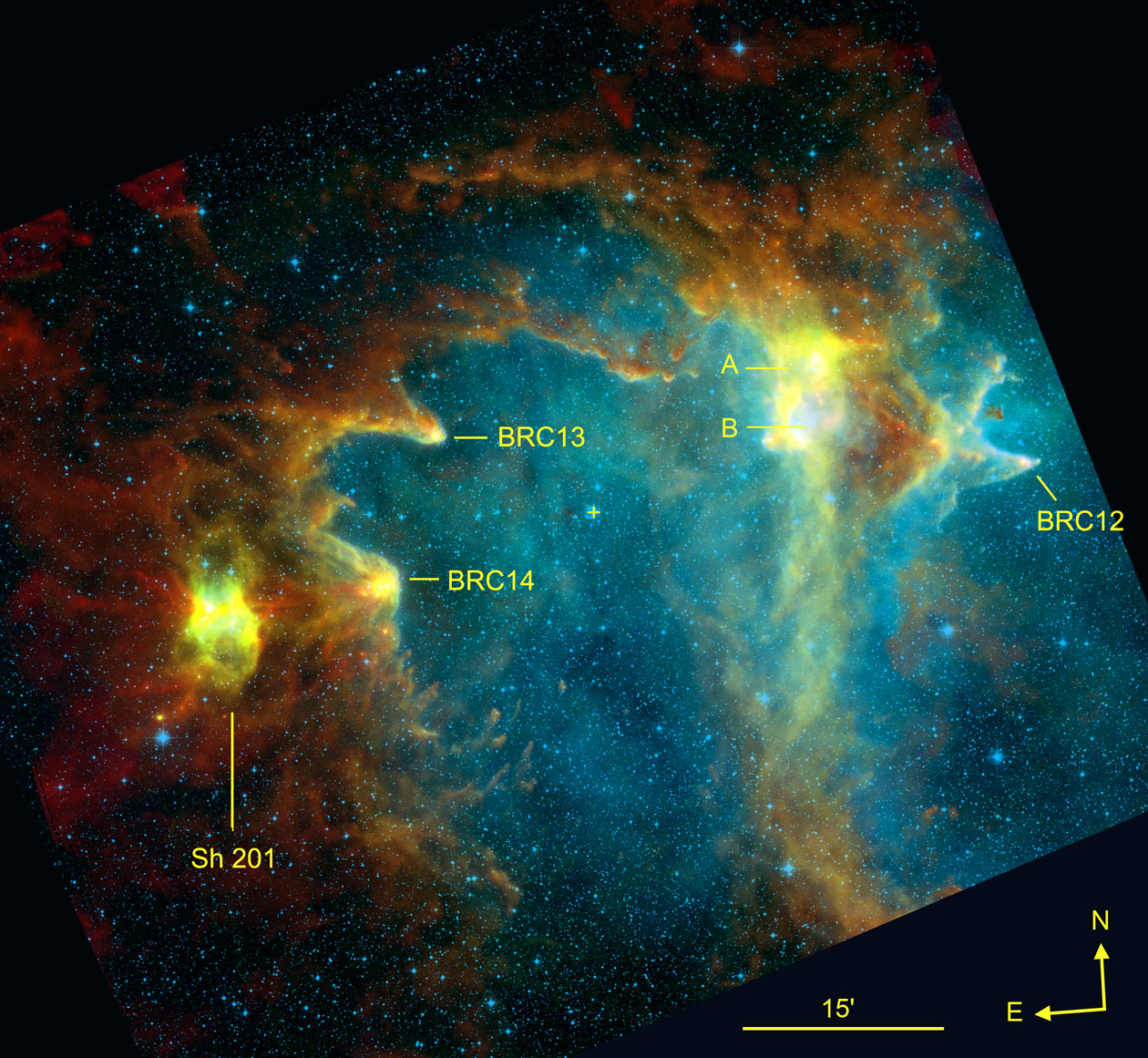}
  \caption{Composite colour image of the W5-E field observed by {\it Herschel}. Red is the 
  250~$\mu$m {\it Herschel}-SPIRE data, which traces cold dust emission, green is the 100~$\mu$m {\it Herschel}-PACS data, which
traces emission mainly from dust at higher temperature located in the PDRs, and blue is 
  the DSS2-red survey, which traces the H$\alpha$ emission of the ionized gas. The yellow cross lies at the position 
  of HD~18326, the exciting star of W5-E.}  
\label{presentation}
\end{figure*}

The overall morphology of W5 is more complicated than a simple spherical morphology around the central exciting star.
Fig.~\ref{presentation} is a composite colour image of the observed field that shows the overall morphology of 
this region and allows us to identify its different components. In red we 
show the cold dust emission at 250~$\mu$m from the {\it Herschel}-SPIRE data; these data trace molecular material associated with the region. 
In green we show the 100~$\mu$m dust emission from the {\it Herschel}-PACS data; these data trace warmer dust located in the PDRs 
surrounding the ionized regions or associated with massive YSOs. In blue we show the H$\alpha$
emission of the ionized gas from the DSS2-red frame. Fig.~\ref{presentation} clearly shows that the dust bubble is opened towards the south.  
This configuration was already apparent in the CO maps and {\it Spitzer} images. In the north there are two parallel ionization
fronts and associated PDRs, running 
from east to west, with ionized gas between the two.  At the periphery of the bubble we find bright-rimmed structures enclosing bright sources
such as YSOs, \HII\ regions, and small clusters.

\subsection{Determination of the mass and column density}

It is generally assumed that the emission from cold dust at {\it Herschel}-SPIRE wavelengths 
is optically thin. This property allows us 
to estimate the distribution of the dense molecular material hosting this cold dust, and, 
via various assumptions, to determine the H$_2$ column density, $N(\mathrm H_2)$, and 
the mass of various structures (condensations, filaments, etc...; see AND12). The derivation
and the limitations of such determinations can be found in Hildebrand (\cite{hil83}). 
To summarize, the total (gas+dust) mass of a feature is related to its flux density
$S_{\nu}$ by:
\begin{equation}
M_{\mathrm{(gas+dust)}} = 
100\,\,\frac{S_\nu\,\,
D^2}{\kappa_\nu\,\,
B_\nu(T_{\mathrm{dust}})},
\end{equation}
where $D$ is the distance to the source, $\kappa_{\nu}$ is 
the dust opacity per unit mass of dust at the frequency $\nu$, 
and $B_{\nu} (T_{\mathrm{dust}})$ 
is the Planck function for a temperature $T_{\mathrm{dust}}$. 
For all estimates in this paper, we assume a gas-to-dust 
ratio of 100. This value is uncertain and the subject of discussion. The dust opacities however are even more uncertain. 
As discussed by Henning et al. (\cite{hen95}) the dust opacity depends 
on the size, shape, chemical composition, physical structure, and temperature of the grains, thus of the dust environment. This has been discussed 
for example by Martin et al. (\cite{mar11}). Table \ref{opacity} 
gives the opacities at the {\it Herschel}-SPIRE wavelengths estimated from various models. The values obtained 
by Ossenkopf and Henning (\cite{oss94}; OH1 and OH2) are in columns 2 and 3; they correspond respectively to their models 
``thick ice mantle'' and ``thin ice mantle'' (density 10$^6$~cm$^{-3}$ and age $\geq$10$^5$~yr). 
We calculated the values in column 4 using the formula: 
\begin{equation}
\kappa_{\nu}=10\times\left(\frac{\nu}{1000~GHz}\right)^2\,,
\end{equation}
which assumes a spectral index of 2 for the dust emissivity, and gives at 1.3~mm the dust opacity 
recommended by Preibisch et al. (\cite{pre93}) for cloud envelopes.
The thick ice mantle model (OH1) is possibly more appropriate for cold and dense cores (see also Henning et al. \cite{hen95})\footnote{A dust opacity of 
4.3~cm$^2$~g$^{-1}$ at 250~$\mu$m has been estimated for dust in the Galactic ISM at high latitude; two to four times larger values are estimated for dust 
in the diffuse ISM of the Galactic plane (Martin et al. \cite{mar11}), showing that the dust properties probably vary with the environment. 
Little is known about the opacity of dust in dense cores or envelopes.}. 
In the following we use the opacities given in the last column of Table~\ref{opacity} to estimate the mass and column density, but we 
note that, due to the unknown properties of the dust grains, the dust opacities and thus the derived clouds' masses are uncertain 
by at least a factor $\sim$2.

\begin{table}
\caption{Opacity of the dust at {\it Herschel}-SPIRE wavelengths}
\begin{tabular}{rrrr}
\hline\hline
$\lambda$ & OH1 & OH2 & Equation (2) \\
($\mu$m) & (cm$^2$~g$^{-1}$) & (cm$^2$~g$^{-1}$) & (cm$^2$~g$^{-1}$) \\
\hline
250  & 21.1  & 17.5 & 14.4 \\ 
350  & 11.2  & 10.1 & 7.3 \\
500  & 5.5   & 5.0  & 3.6 \\
850  & 1.95  & 1.8  & 1.25 \\
1300 & 1.0   & 0.9  & 0.5 \\ 
\hline
\label{opacity}
\end{tabular}\\
\end{table}

\begin{figure*}[tb]
\sidecaption
\includegraphics[width=12cm]{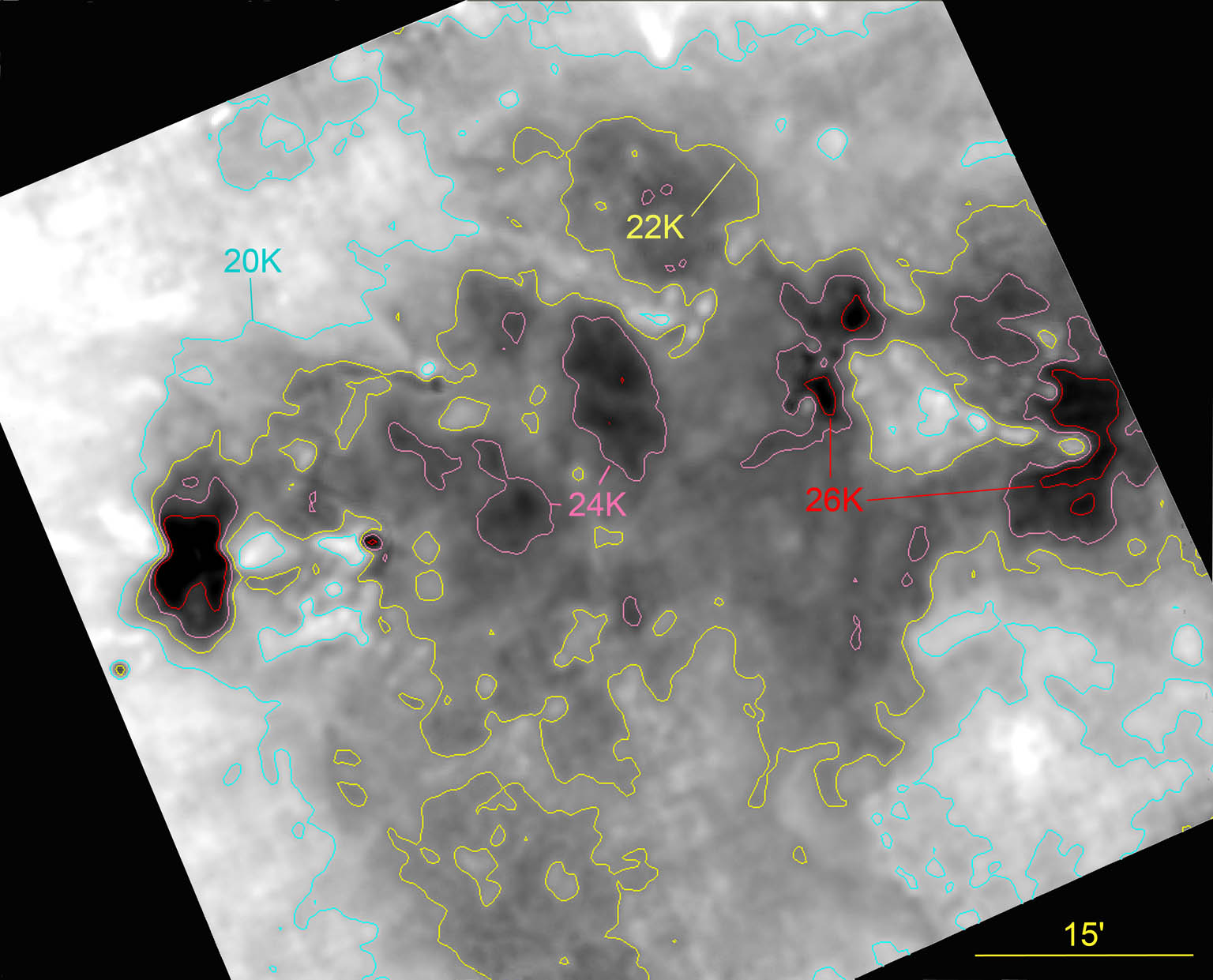}
  \caption{The temperature map of W5.  Iso-temperature contours (20~K in blue, 22~K in yellow, 24~K in pink, and 26~K in red)
   are superimposed on the grey-scale temperature map. The resolution is that of the 500~$\mu$m map, 37~$\arcsec$. North is up and east is left.}  
\label{temperature}
\end{figure*}

Here we use the map of dust temperature over the W5-E region, similar to the map shown in AND12,
using PACS and SPIRE data and a simple model of modified blackbody emission.  While AND12 fitted simultanously
for the dust temperature and the column density, we here use the {\it Herschel} fluxes to derive a temperature
at each location, leaving the column density as a free parameter.  We smoothed the {\it Herschel} images 
to the same resolution using a two-dimentional Gaussian representing the SPIRE beam at 500~$\mu$m and reprojected
the data to a uniform pixel spacing of the 500~$\mu$m data.  Fig.~\ref{temperature} shows the temperature map and some iso-temperature contours; 
cold dust appears in white, hot dust in black. The contours correspond to 18~K, 20~K, 22~K, 24~K and 
26~K. The temperature does not vary strongly over the field: $T_{\mathrm{dust}}$ lies in the range 17~K  to 31~K. 
Temperatures higher than 26~K are observed in the central cavities of W5-E and W5-W, 
in the direction of Sh~201, in the direction of the small A and B \HII\ regions lying between W5-E and W5-W, 
and in the direction of the IR sources inside BRC14.  Due to the relatively low fluxes, the temperatures
in the direction of the interior of W5-E are especially uncertain.  Temperatures in the range 17~K--20~K are observed in zones 
of high column density such as the filaments in the vicinity of Sh~201 or inside BRC12, BRC13, and BRC14. The PDR regions, 
adjacent to the IFs, are at an intermediate temperature of 21~K--22~K.  

The knowledge of the dust temperature allows us to estimate the column density of the gas. With the same assumptions, 
we can estimate the H$_2$ column density from the surface 
brightness $F_{\nu}$, using the formula: 
\begin{equation}
N(\mathrm H_2) =
\frac{100\,\, F_{\nu}}
{\kappa_{\nu}\,\,
B_{\nu}(T_{\mathrm{dust}})\,\,
2.8\,\,m_{\mathrm H}\,\,\Omega_\mathrm{beam}}\mathrm,
\end{equation}
where $F_{\nu}$ is expressed in Jy~beam$^{-1}$, $B_{\nu}$ in Jy, the factor of 2.8 is the mean molecular weight,
$m_{\mathrm H}$ is the hydrogen atom mass, and $\Omega_\mathrm{beam}$ is the beam solid angle. The visual extinction can be estimated from the column density. 
From the classical relations,
$N(\mathrm H+H_2)/E(B-V)=5.8 \times 10^{21}$~ particles~cm$^{-2}$~mag$^{-1}$
(Bohlin et al.~\cite{boh78}) and $A_V=3.1\,\,E(B-V)$, we obtain  $A_V=5.34 \times 10^{-22}~N(\mathrm H+H_2)$ or $A_V=1.07 \times 10^{-21}~N(\mathrm H_2)$ in a molecular 
medium.
  
\begin{figure*}[tb]
\sidecaption
\includegraphics[width=12cm]{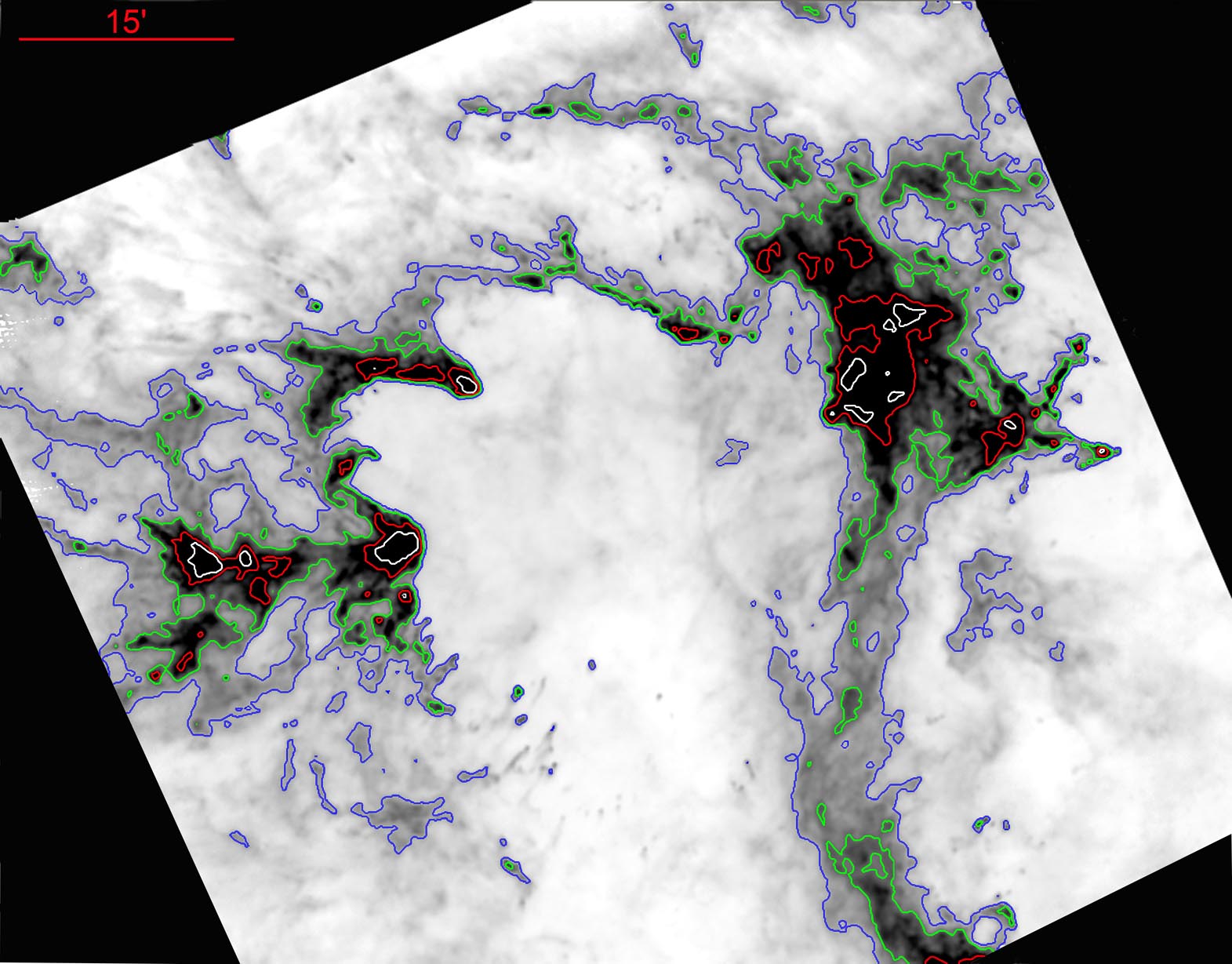}
  \caption{The column density map of W5.  Contours of equal column density (0.5$\times$10$^{21}$~cm$^{-2}$ in blue, 10$^{21}$~cm$^{-2}$ in green, 2.5$\times$10$^{21}$~cm$^{-2}$ in red, 
  and 5$\times$10$^{21}$~cm$^{-2}$ in white) are superimposed on a grey-scale
  column density map. The resolution is that of the {\it Herschel} 250~$\mu$m data, $18\arcsec$. North is up and east is left.}  
\label{column}
\end{figure*}

We used the SPIRE 250~$\mu$m image, which has the highest angular resolution at SPIRE wavelengths, to create a map of the gas column density\footnote{The column density 
map in AND12 was created from the 500~$\mu$m (lower resolution) map using a slightly different method.}. 
We reprojected the temperature map previously obtained to the 4.5\arcsec\ pixel grid of the 250~$\mu$m image and used a dust opacity of 
14.4~cm$^2$~g$^{-1}$ (Table~\ref{opacity}).  
In Figure~\ref{column} we show the column density map. The column density does not reach very high values; the highest values are found 
in the direction of the cloud enclosed by BRC14 (9.0$\times$10$^{22}$~cm$^{-2}$, or $A_V=96$~mag), in the cloud east of Sh~201 
($\sim$6.2$\times$10$^{22}$~cm$^{-2}$, or $A_V=66$~mag), inside BRC13 (1.7$\times$10$^{22}$~cm$^{-2}$, or  $A_V$=18.2~mag), or in the vicinity of the B \HII\ region 
($\sim$9.6$\times$10$^{21}$~cm$^{-2}$). Everywhere else in the field the column density is lower.  

Assuming the same kind of dust previously described, we can use the column density
to estimate the optical depth of the dust at 100~$\mu$m. Our maximum column density N(H$_2$)$\sim 10^{23}$~cm$^{-2}$ corresponds to an optical
depth $\tau_{100\mu{\rm m}}\sim$0.42. The column density is everywhere much lower; thus we consider in the following that the dust is optically thin
at 100~$\mu$m. (This justify using the 100~$\mu$m fluxes to estimate the dust temperature.)

\section{Star formation: YSOs and dense condensations}

The protostars were originaly grouped into three "Classes" according to the spectral index $\alpha$ of their SED, measured in the range 2.2~$\mu$m--10~$\mu$m 
(Lada \cite{lad87}; Greene et al. \cite{gre94}):
\begin{equation}
\alpha=\frac{dlog(\lambda F_\lambda)}{dlog(\lambda)},
\end{equation}
Class~I sources have $\alpha>0.3$; they are highly embedded protostars those luminosity is dominated by a spherical infalling envelope. Class~II YSOs, with $-2<\alpha<-0.3$, are protostars surrounded by a substantial accreting disk.  Class~III sources having  $\alpha<-2$ are young stars having dissipated most of their disk. Sources with $-0.3\leq \alpha \leq +0.3$ are called 
flat spectrum sources.

Yet, Robitaille et al. (\cite{rob06}) showed that the same object can be classified differently depending on the viewing angle; for example,
a Class~II source with  an edge-on disk can have a positive spectral index that would make it resemble a Class~I object. Thus these authors prefer to
discuss the evolutionary stages of their model by adopting a ``Stage'' classification analogous to the Class scheme, but based on the physical
properties of the YSO (e.g., based on disk mass or envelope accretion rate) rather than properties of its SED (e.g., slope). Stage~0 and I objects
(which they do not differentiate)  have significant infalling envelopes and possibly disks, Stage~II objects have optically thick disks (and the
possible remains of a tenuous infalling envelope), and Stage~III objects have optically thin disks. The boundaries between the different stages are
somewhat arbitrary, in the same way as the Class scheme. They define Stage~0/I objects as those that have $\dot{M}_{\rm env}$/M$_*>10^{-6}$~yr$^{-1}$, Stage~II objects as those that have 
$\dot{M}_{\rm env}$/M$_*<10^{-6}$~yr$^{-1}$ and $M_{\rm disk}$/M$_*>10^{-6}$, and Stage~III objects as those that have 
$\dot{M}_{\rm env}$/M$_*<10^{-6}$~yr$^{-1}$ and $M_{\rm disk}$/M$_*<10^{-6}$. 

Class~0 sources were added later to the Class scheme (Andr\'e et al. \cite{and93}, \cite{and00}). They are younger than Class~I sources 
(younger than 10$^5$ yr). They differ from the Class~I sources by the shape of their SEDs, which resembles a cold single-temperature
blackbody. Class~0 protostars have M$_*$/M$_{\rm env}<1$ (or L$_{\rm sub-mm}$/L$_{\rm bol}>$~10$^{-2}$; see Sect.~6.3.3).\\

In what follows we comment on the point-like sources detected by {\it Herschel}, and on the Class~I sources identified by KOE08. 
We discuss separately several well-known regions, some associated with small \HII\ regions and some others associated with bright-rimmed clouds (BRCs).  
Three well known BRCs are present in the field observed by {\it Herschel}; they are BRC12, BRC13, and BRC14 (containing the IR source AFGL~4029). 
The \HII\ regions are Sh~201, region A and region B (that we discuss in ``The region between W5-E and W5-W``, Appendix D). All these regions are 
identified on Fig.~\ref{presentation}. In this section we present the BRC12 region to illustrate how we conduct the study of the individual regions.  
We give the same details about all the other zones in Appendixes A to E.

The following section, and the corresponding sections in the Appendix share the same format:
\begin{enumerate}
\item  We describe each region. 
\item We describe and comment on the {\it Herschel} data. 
We identify in a figure the {\it Herschel} point sources discussed in the text. Often the detected point sources have a {\it Spitzer} 
counterpart (at least at 24~$\mu$m), and are listed
in the KOE08 catalogue. For each region a first table (``\#1'') gives, according to KOE08, the identification number (column 2), 
the coordinates (columns 3 and 4), the 2MASS magnitudes (columns 5 to 7), the {\it Spitzer}-IRAC and MIPS magnitudes (columns 8 to 12), 
and the classification (column 13) of the {\it Spitzer} counterpart. Columns 14 to 18 of the same table contain the measured {\it Herschel} fluxes.
\item Whereas the classifications of KOE08 are based on IR colours and magnitudes up to 24~$\mu$m, 
we try to take advantage of the now more complete SEDs and use the SED fitting-tool of Robitaille et al. (\cite{rob07}; hereafter ROB07) to estimate 
the evolutionary stage of the sources. We list the parameters necessary to estimate the evolutionary stage (the mass of the central source, the mass and accretion rate of the disk, and the accretion rate of the envelope) in a second table (``\#2'');   
we consider only the models with $\chi ^2-\chi^2$(best) per data point $\leq$3; the first numbers in table \#2 correspond 
to the best model, and the range of values in brackets corresponds to the other selected models. Some limitations of the ROB07 SED fitting tool are discussed in Robitaille (\cite{rob08}). Offner et al. (\cite{off12}) also discuss the accuracy of protostellar properties inferred from the SEDs; they caution us against the use of fits to the Robitaille grid of models for constraining the stellar and disk properties of unresolved protostellar sources. Thus, one must be very cautious when considering the parameters given in tables \#2.  A large fraction of the {\it Herschel} point-sources have
a dominant envelope. We show in Sect.~6.3.1 that these envelopes are not 
well treated in ROB07 models. We prefer to use a modified blackbody model to obtain the temperatures and masses of the envelopes; these are given in Sect.~6.3.1. This situation 
probably has some repercussions on the disk's parameters obtained with the SED fitting tool; we often find that the parameters of the disk are not well constrained. 
In turn, this has some repercussions on the luminosity\footnote{ The luminosity of the central object is the difference between the total luminosity of the source and the 
luminosity of the accreting disk, based on the disk accretion rate; the luminosity of the accreting envelope is not taken into account in ROB07 models.} of the central object, and 
thus on its mass\footnote{ The mass of the central object 
is derived from its luminosity and temperature, placing these on evolutionary tracks in an Hertzprung-Russel diagram. The mass may be wrong if the evolutionary tracks are wrong. 
The pre-main sequence tracks used, for example those of Siess et al. (\cite{sie00}) for stars in the mass range 0.1 to 7.0~$\msol$, are for non accreting stars. Since then, Hosokawa et al. (\cite{hos09}, \cite{hos10}) have shown that protostars with high accretion rates may occupy different positions in the HR diagram, depending of the accretion rate and of the geometry of the accretion.}.
\item Keeping this in mind, we show the SEDs of some interesting sources, and comment on individual objects. 
\end{enumerate}

Some point sources are only observed at {\it Herschel} wavelengths. These are potentially the most interesting sources as they may be Class~0 YSOs or starless cores
(candidate prestellar condensations). Their coordinates and {\it Herschel} fluxes are also given in our tables (\#1). We discuss the candidates Class~0 and 
starless cores in Sect.~6.3.3 and Sect.~6.6 respectively.

\subsection{The bright-rim cloud BRC12 and vicinity}

\begin{figure*}[tb]
\sidecaption
\includegraphics[width=12cm]{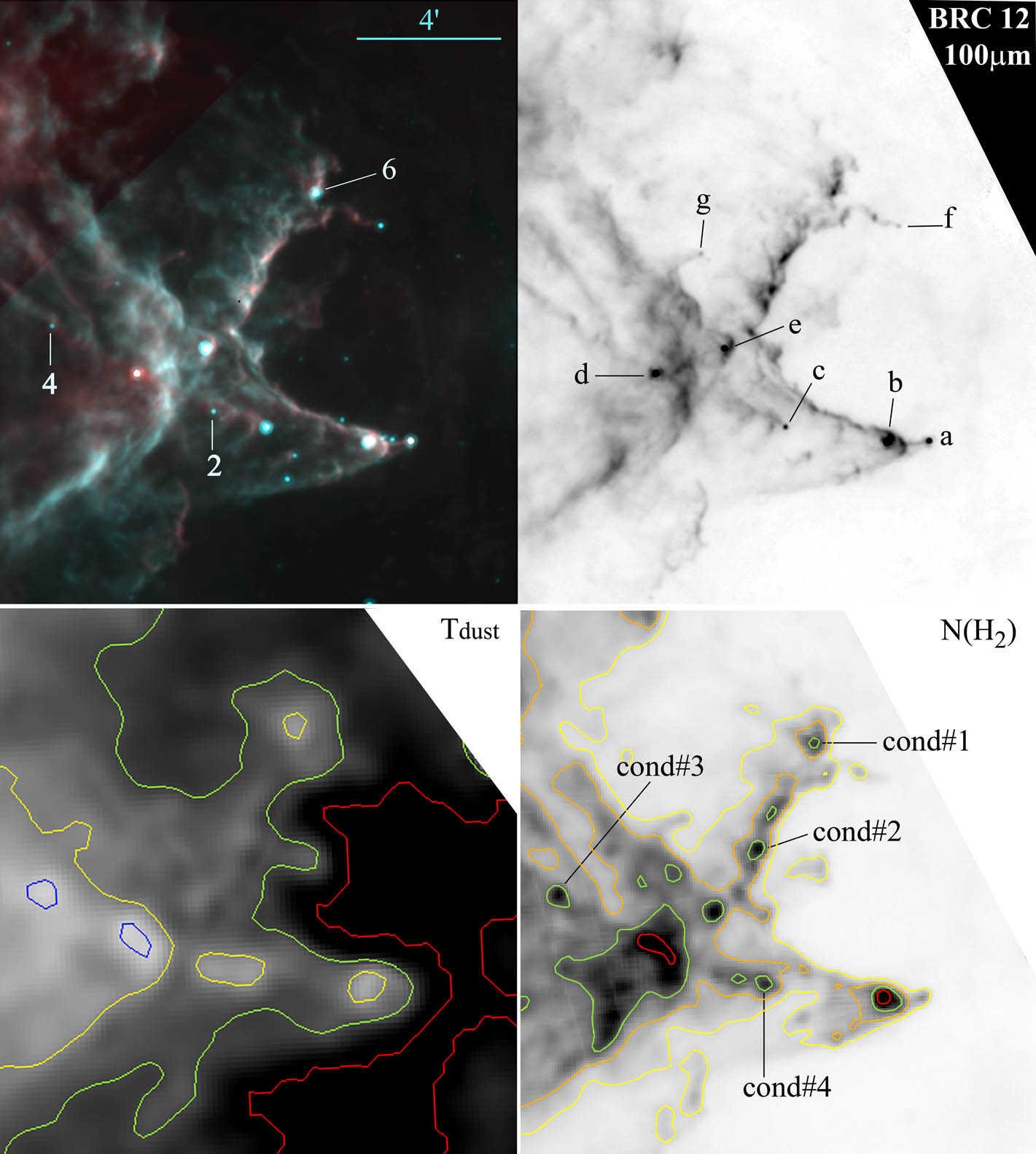}
  \caption{Field containing the bright-rimmed cloud BRC12. {\it Top left:} Composite colour image: red is the 
  {\it Herschel}-PACS emission at 100~$\mu$m and turquoise is the {\it Spitzer}-MIPS emission at 24~$\mu$m. {\it Top right:} The YSOs 
  discussed in the text and present in Table~\ref{BRC12tablea} are identified. The underlying grey-scale image is the {\it Herschel}-PACS image at 100~$\mu$m.
  {\it Bottom left:} Temperature map of the dust: the red contour is 26~K, the green is 24~K, the yellow is 22~K, and the dark blue is 20~K. {\it Bottom right:} 
  Column density map: the red contour is 4$\times$10$^{21}$~cm$^{-2}$, the green is 2$\times$10$^{21}$~cm$^{-2}$, the orange is 1$\times$10$^{21}$~cm$^{-2}$, and the yellow 
  is 0.5$\times$10$^{21}$~cm$^{-2}$. The column density is high mostly in the direction of the point sources (it is the column density of their envelopes, smoothed by the beam).  Some other condensations appear, which may be prestellar (see Sect.~6.6). North is up and east is left.}
  \label{BRC12a}
\end{figure*}

BRC12 (or SFO12) is part of the catalogue of bright-rimmed clouds observed in the direction of northern \HII\ regions  
by Sugitani et al. (\cite{sug91}). It is associated with IRAS 02511+6023, an IR point source of 640~\lsol\ according to Sugitani et al. (\cite{sug91}; 
from IRAS fluxes and a distance d=2~kpc) or 
220~\lsol\ according to Morgan et al. (\cite{mor08}; from SCUBA fluxes and d=2~kpc). BRC12 is associated with W5-W and not W5-E. 
It is part of the bubble surrounding W5-W, and is facing BD+60$\degr$0586, an O8III star that contributes to the ionization of    
W5-W. Morgan et al. (\cite{mor04}) hypothesize that this star is the main source 
of ionizing photons reaching the rim (it is the closest O star in projection). 
The NVSS radio-continuum map at 20~cm (Condon et al. \cite{con98}) shows strong emission following the bright rim BRC12 and the  
nearby ionization front (IF). The emission at 20~cm traces the dense ionized boundary layer (IBL) bordering the enclosed molecular material 
(see fig.~1 in Morgan et al. \cite{mor04}; online material). From the integrated radio flux of this feature Morgan et al. (\cite{mor04}) calculate
an electron density in the IBL of 340~cm$^{-3}$. BRC12 has been observed at 450~$\mu$m and 850~$\mu$m with SCUBA by Morgan et al. (\cite{mor08}). 
A dense core is detected at the tip of the bright rim (their figure~A.9). This core corresponds to the source BRC12-b discussed below (Table~\ref{BRC12tablea} and Fig.~\ref{BRC12a}).  \\

Fig.~\ref{BRC12a} shows a colour image of this region with {\it Herschel} 100~$\mu$m emission in red, 
and {\it Spitzer} 24~$\mu$m emission in turquoise ({\it Top left}). We have detected seven 100~$\mu$m point sources in this region; they are identified by  
letters in Fig.~\ref{BRC12a} ({\it Top right}). Six of them have {\it Spitzer} counterparts. 
Table~\ref{BRC12tablea} gives their parameters, according to KOE08, and the measured {\it Herschel} fluxes. 
Five more Class~I and one bright Class~II YSO, identified by KOE08 but without 100~$\mu$m counterparts, are present in the field covered by Fig.~\ref{BRC12a}. 

\begin{figure*}[tb]
\centering
 \includegraphics[angle=0,width=150mm]{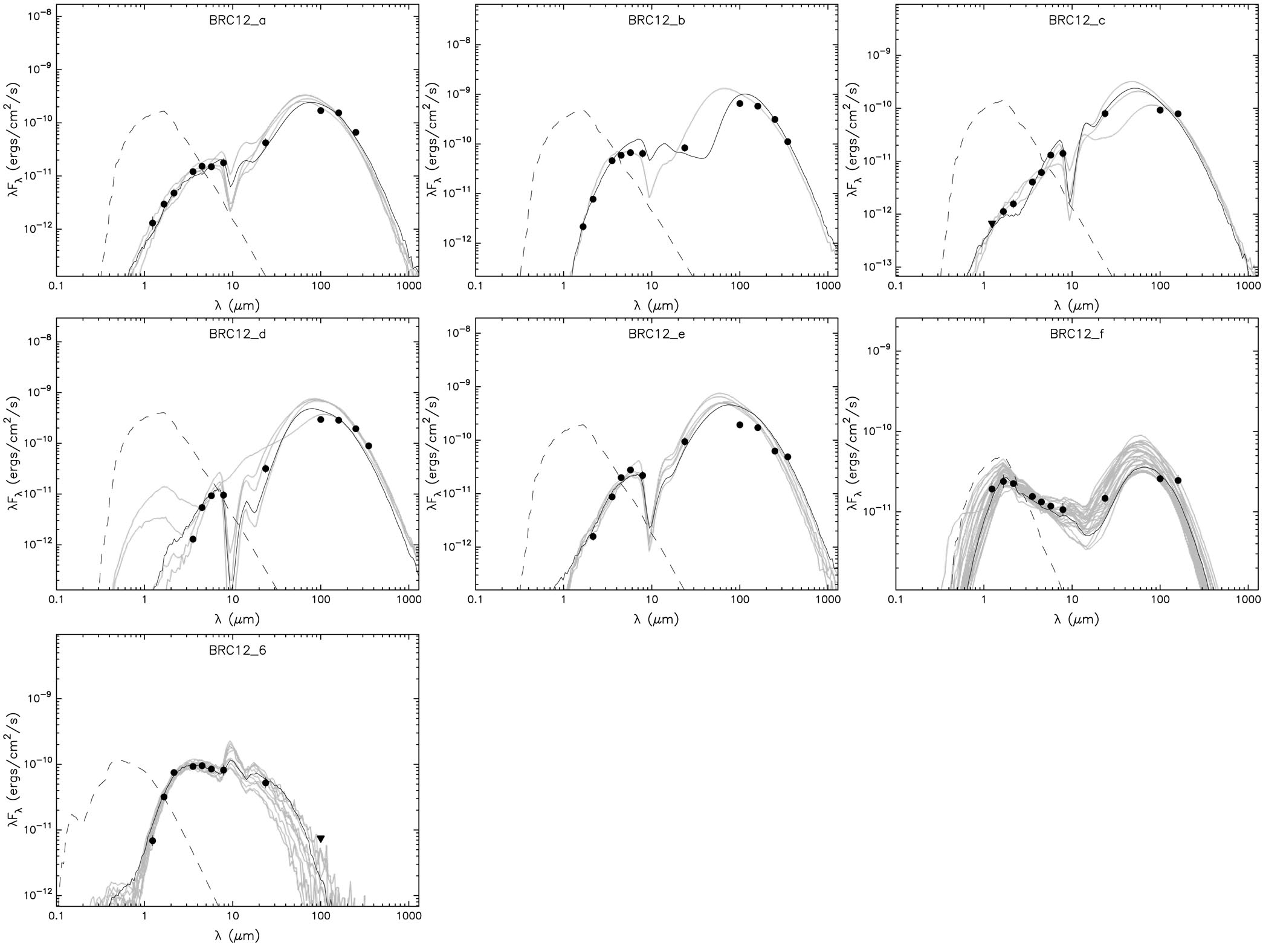}
  \caption{Spectral energy distributions of point sources in the BRC12 field. We used the SED fitting-tool 
  of ROB07.  
  The filled circles represent the input fluxes and the black line shows the best fit. The grey lines show subsequent fits,  all corresponding 
   to $\chi ^2-\chi^2$(best) per data point $\leq$3. The dashed line shows the stellar photosphere corresponding to the central source of the best fitting model, 
   as it would look in the absence of circumstellar dust, but including interstellar extinction. }
  \label{BRC12b}
\end{figure*}

We fit the SEDs of the YSO candidates in Table~\ref{BRC12tablea} using the SED fitting-tool of ROB07. We have assumed a distance of 2~kpc 
and an interstellar extinction of 1.96~mag (the extinction of HD~18326, that we consider of interstellar origin). We show several SEDs in Fig.~\ref{BRC12b}. 
In the following we comment on individual sources:

$\bullet$ BRC12-a: this YSO lies at the extreme tip of BRC12, linked to the main BRC by a very thin filament. It is a Stage~I YSO with, possibly, a central 
object of intermediate mass. The disk's mass and accretion rate are not well constrained.
 
$\bullet$ BRC12-b: this is the brightest mid-IR source in the field, located at the tip of BRC12. It is probably the counterpart of IRAS 02511+6023; it is however less bright 
than estimated for the  IRAS source  (due to the low resolution of IRAS, the IRAS fluxes probably include some of the PDR emission). 
The two possible models reported in Table~\ref{BRC12tableb} 
are somewhat different, but both point to a rather massive Stage~I YSO. The dense core observed by Morgan et al. (\cite{mor08}) at 
450~$\mu$m et 850~$\mu$m with SCUBA is co-spatial with this YSO. Morgan et al. (\cite{mor08}) obtain for the temperature of the cold component of the source 21~K, 
which differs for the temperature of 17.9~K that we estimate from the four Herschel fluxes (Table \ref{temperatureYSOs}). Using their integrated flux at 850~$\mu$m, an opacity of 2.0~cm$^2$~g$^{-1}$ at this wavelength, and a distance of 1.9~kpc they estimate a mass of 18.9~$\msol$ for this source. Their flux, assuming our opacity law (Table \ref{opacity}), 
 and a distance of 2.0~kpc gives a mass of 32~$\msol$, 
very similar to our estimate (Table~\ref{temperatureYSOs}). Thus, BRC12 harbors a rather massive Stage~I YSO, embedded in a high mass envelope
with, possibly, a high envelope accretion rate. This suggests that a high mass star is forming there. 

$\bullet$ BRC12-c: the three models with $\chi ^2-\chi^2$(best) per data point $\leq$3 that fit this source's SED are rather similar. They all point to a Stage~I YSO of intermediate mass.

$\bullet$ BRC12-d: this is also a bright mid-IR source. Five models, somewhat different, fit the SED. All correspond 
to a Stage~I YSO, possibly of intermediate mass; the parameters of the disk are not well constrained.

$\bullet$ BRC12-e: this YSO seems to be at the tip of another bright rim cloud. It resembles BRC12-b, and is a Stage~I YSO. But its SED is not well fitted. 

$\bullet$ BRC12-f: as for BRC12-a, this source lies at the extremity of a very narrow pillar.  It is, however, less bright and less massive than BRC12-a. 
The SED is not well-constrained, as shown by Fig.~\ref{BRC12b}, especially the disk parameters. The best model corresponds to an uncertain Stage~I/II YSO.

$\bullet$ BRC12-g: this source is situated at the extreme tip of a narrow pillar. It has a very faint 24~$\mu$m counterpart that possibly is extended (its flux was not measured by KOE08).

Five other sources in the field have been classified as Class~I YSOs by KOE08 (Table~\ref{BRC12tablea}). None of these have counterparts in the 
{\it Herschel}-PACS image at 100~$\mu$m. Their SEDs are not well-constrained. 
YSO~\#1 is observed only in the {\it Spitzer} bands and lies very close to the bright YSO \#6; thus its classification is very uncertain. 
YSO~\#5 lies $\sim$ 8\arcsec\ away from the very bright BRC12-b, and the two sources cannot be  
separated at 24~$\mu$m and at longer wavelengths;  therefore, its classification is also uncertain. 
YSOs \#2 and \#4 lie in the vicinity of condensations \#4 and \#3 respectively (Fig.~\ref{BRC12a}). Their SEDs are not well constrained; however $\dot{M}_{\rm env}$ is never equal to zero; thus they could be Stage I or Stage I/II YSOs.   
YSO~\#6 lies very close to the IF, in the direction of condensation~\#1; its SED is not well constrained, especially the disk's parameters.  
But it is clearly a Stage~II YSO as its envelope accretion rate is null in all models. 

\begin{table*}[tb]
\caption{Photometry of point sources: the bright-rim cloud BRC12 and its vicinity}
\resizebox{18cm}{!}{
\begin{tabular}{llllrrrrrrrrrrrrrr}
\hline\hline \\[0.5ex]
Name & \# & RA(2000) & Dec(2000) & $J$    & $H$   & $K$   & [3.6] & [4.5]  & [5.8]  & [8.0] & [24]     & Class & S(100)  & S(160)   & S(250)  & S(350) & S(500) \\
     &         & ($\degr$) & ($\degr$) & mag & mag & mag & mag & mag & mag & mag & mag & & Jy & Jy & Jy & Jy & Jy \\          
\hline \\[0.5ex]
a & 10337 & 43.718074 & 60.595320    & 16.18  & 14.49 & 13.22 & 10.73 &   9.74 &   9.01 &   7.85 &   3.33  &  I     &  5.68 &   8.24  &   5.55 &         & \\
b &	10566 & 43.758246 & 60.594901    &        & 14.83 & 12.70 &  9.28 &   8.27 &   7.39 &   6.45 &   2.59  &  I     & 21.86 &  31.00  & 26.14  & 13.03  & \\
c & 11055 & 43.854618 & 60.602491    &        & 15.54 & 14.43 & 11.92 &  10.73 &   9.16 &   8.10 &   2.64  &  I     &  3.09 &   4.22  &         &         & \\
d & 11604 & 43.976307 & 60.628341    &        &       &       & 13.16 &  10.86 &   9.53 &   8.52 &   3.64  &  I     &  9.87 &  15.29  & 16.19  & 10.42  & \\
e &	11311 & 43.911143 & 60.639847    &        &       & 14.42 & 11.09 &   9.45 &   8.34 &   7.62 &   2.46  &  I     &  6.47 &   9.17  &  5.25  &  5.71  & \\
f &	10484 & 43.743111 & 60.695488    & 13.26  & 12.22 & 11.54 & 10.46 &   9.89 &   9.27 &   8.40 &   4.47  &  II    &  0.87 &   1.33  &         &         & \\
g &       &           &              &        &       &       &       &        &        &        &         &        &  0.98 &   2.50  &  2.28  &         & \\
1 &	10810 & 43.806725 & 60.707375    &        &       &       & 13.87 &  13.50 &  11.58 &   9.93 &         &  I     &        &          &         &         & \\
2 &	11275 & 43.903961 & 60.610319    & 15.36  & 14.43 & 13.76 & 12.67 &  12.04 &  10.85 &   9.14 &   5.01  &  I     &        &          &         &         & \\ 
3 &	11886 & 44.053573 & 60.603121    &        &       &       & 15.31 &  14.10 &  12.87 &  11.54 &   7.26  &  I     &        &          &         &         & \\  
4 &	11895 & 44.056178 & 60.650891    & 16.55  & 14.51 & 13.43 & 11.98 &  11.06 &  10.35 &   9.42 &   5.24  &  I     &        &          &         &         & \\   
5 &	10592 & 43.763089 & 60.594435    & 16.52  & 14.82 & 13.85 & 12.26 &  11.49 &  10.56 &   9.70 &         &  I     &        &          &         &         & \\   
6 & 10797 & 43.803910 & 60.711229    & 14.38  & 11.91 & 10.23 &  8.52 &   7.75 &   7.13 &   6.19 &   3.10  &  II    &        &          &         &         & \\  [0.5ex]  
\hline
\label{BRC12tablea}
\end{tabular}\\
}
\end{table*}

\begin{table*}[tb]
\caption{Characteristics of the YSOs: the bright-rim cloud BRC12 and its vicinity}
\resizebox{18cm}{!}{
\begin{tabular}{llllllllll}
\hline\hline \\ [0.5ex]
Name &  $\chi ^2$ & $M_*$   & $T_*$ & $M_{\rm disk}$  & $\dot{M}_{\rm disk}$  & $\dot{M}_{\rm env}$  & $L$     & $i$     & Stage \\
     &            & (\msol )& (K)   &(\msol)          &(\msol~yr$^{-1}$)      &(\msol~yr$^{-1}$)     & (\lsol) & (\degr) &       \\          
\hline \\ [0.5ex]
a &  39 (--72)  & 2.3 (1.9--3.0)  & 4334  & 3.3e-2 (4.0e-4--1.3e-1)  & 3.3e-8 (2.4e-9--1.4e-6)  & 2.5e-4 (--1.1e-4)  & 51 (44--77)  & 41 (31--56)  &  I \\
b &	 58 (--91)  & 2.9 (6.9)    & 4175  & 8.4e-3 (6.4e-3)  & 3.8e-7 (2.3e-7)   & 1.3e-3 (4.0e-4) &  139 (364)  & 18 (56)  & I \\
c &  49 (--76)  & 1.6 (--2.0)   & 4130  & 3.2e-2 (--7.8e-3) & 7.3e-7 (1.2e-8--1.0e-6)  & 4.8e-5 (3.6e-5--9.1e-5) &  43 (36--66) & 31 (--63)  & I \\
d &  125 (--152) & 4.1 (1.6--5.1)  & 4420  & 4.5e-2 (1.6e-3--1.5e-1)  & 5.3e-7 (1.1e-8--4.5e-6)   & 3.1e-4 (--9.7e-4)  & 133 (52--183)  & 69 (--18)  & I \\
e &	 126 (--159) & 2.5 (--5.3) & 4329  & 5.4e-2 (2.5e-3--2.1e-1) & 4.3e-6 (--1.8e-10)  & 4.1e-4 (--1.4e-4)  & 80 (--184)  & 31 (--63)  & I \\
f &	 37 (--70)   & 1.2 (0.7--4.7) & 4228 & 4.0e-4 (2.0e-4--7.9e-2)  & 2.5e-9 (--7e-7)  & 1.0e-5 (3.3e-7--9.1e-5) & 14 (10--25)  & 56 (18--87)  & I/II ? \\
6 &  13 (--37)   & 4.8 (--9.1)  & 16000 (--24000) & 4.0e-4 (2.9e-5--1.7e-2) & 1.7e-8 (2.3e-11--6.0e-7) & 0  & 400 (--4400)  & 81 (--87)  &  II \\ [0.5ex]
\hline
\label{BRC12tableb}
\end{tabular}\\
}
\end{table*}

\section{Discussion}

\subsection{The overall morphology of W5-E and the long-distance influence of the ionizing radiation}

\begin{figure*}[tb]
\includegraphics[width=170mm]{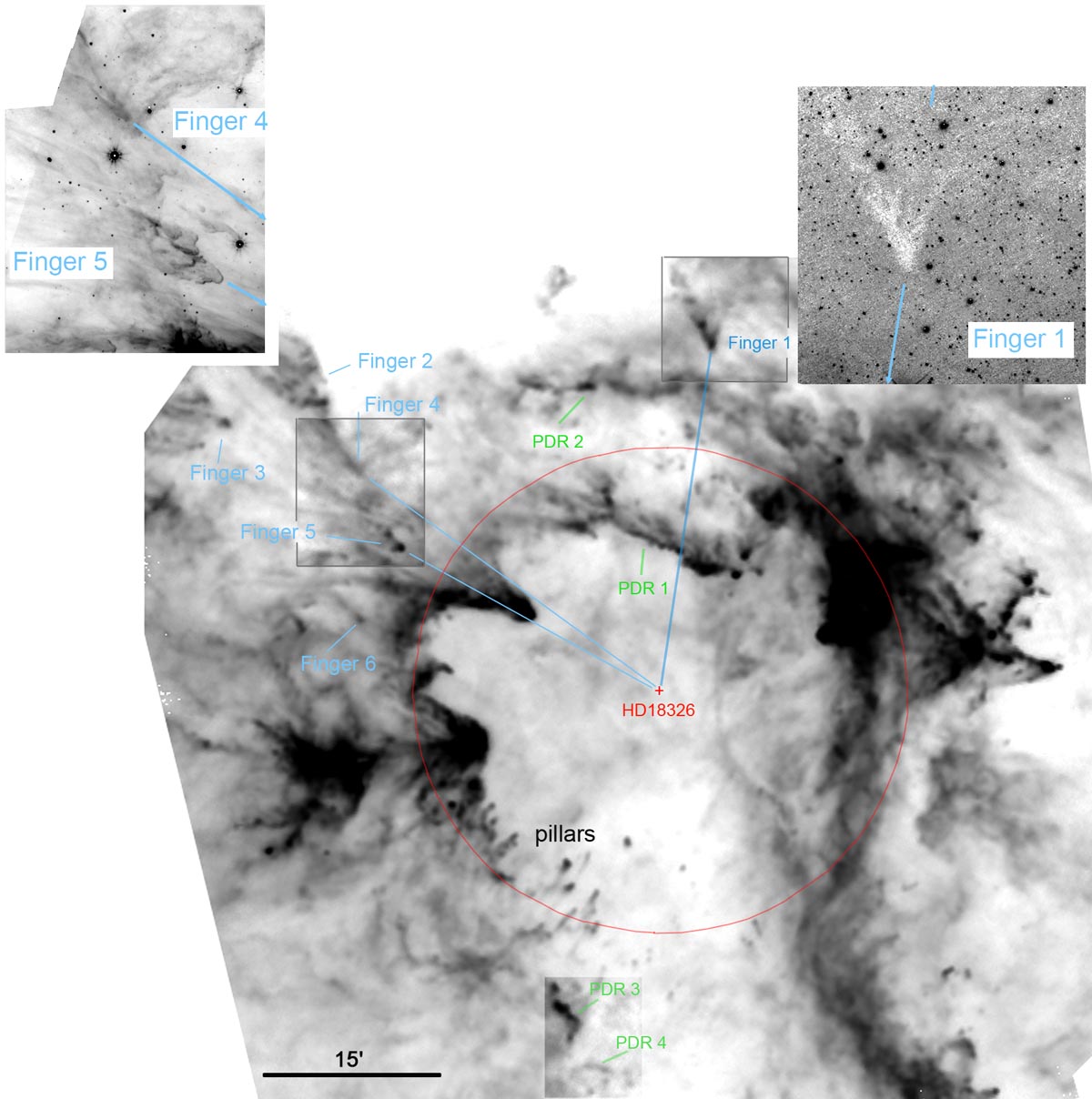}
  \caption{The morphology of W5-E. Peculiar structures discussed in the text (PDRs, fingers, pillars
  ) are identified on the 350~$\mu$m image (logarithmic scale). 
  The red circle shows the approximate extent of the optical \HII\ region (radius 12~pc). The right insert shows finger~\#1 in absorption, on a DSS-red image. 
  The left insert shows fingers \#4 and \#5 pointing towards the exciting star, as they appear on the {\it Spitzer} 8.0~$\mu$m image 
  (fingers \#2 and \#3 lie outside the {\it Spitzer} field). North is up and east is left.}
  \label{global}
\end{figure*}

W5-E appears to have the simple morphology of a large bubble \HII\ region of radius $\sim$12~pc (the radius of the circular H$\alpha$ \HII\ region).  Its morphology is more complicated though, as shown by Fig.~\ref{global} and discussed below:

$\bullet$ In the north two PDRs are present, separated by some 7.5~pc. PDR~1 is $\sim$7~pc from the HD~18326 exciting star and PDR~2 is $\sim$14.5~pc from it  
(these are projected distances, measured in the tangential plane). H$\alpha$ and radio-continuum emissions are observed between these two PDRs. This
morphology probably results from a projection effect. PDR~1 is seen in absorption in the optical (on the DSS-red or H$\alpha$ images), and thus lies in front of the northern part of the ionized region. 

$\bullet$ In the south the PDRs lie far away from the exciting star: PDR~3 is $\sim$16~pc from HD~18326,  
PDR~4 is $\sim$18~pc from it. That the region has expanded more in the south than in the north
shows that during the evolution of the \HII\ region its IF has reached regions of lower density in the south than in the north.
The evolution of an \HII\ region in 
a medium with a density gradient has been studied by Franco et al. (\cite{fra90}). If the density gradient is steep (steeper than in $r^{-1.5}$
where $r$ is the distance to the exciting star), the ionization occurs very quickly and the \HII\ region opens to the outside.
In this case, no neutral material is collected. This is probably what happened in the southern parts of W5-E. W5-E is mostly
ionization-bounded in the east, north, and west, and the ionization radiation escapes
freely in the south. The fact that the ionized gas escapes freely in the south probably slows down the expansion of the \HII\ region in the other directions. 

$\bullet$ The tips of the bright-rimmed clouds BRC13 and BRC14 are located close to the exciting star (respectively at 7.3~pc and 8.9~pc from it in projection).
In these directions the ionizing radiation has reached, somewhere along its path, a relatively high density medium. Also, in these directions,
the column density is higher (by a factor 2--3) than in the directions of PDR~1 or PDR~2: more neutral material has been collected in the PDR
associated with BRC13 and BRC14. Like PDR~1, BRC~13 and BRC~14 are seen in absorption in the optical; thus, the 
enclosed condensations lie in front of the ionized region.

$\bullet$ Some structures resembling fingers pointing toward the exciting star are observed at large distances from this star
(especially in the north and east). They are observed at 250~$\mu$m, 350~$\mu$m, and 500~$\mu$m. The tips of fingers \#1, \#2, and \#3
(Fig.~\ref{global}) are respectively at 17~pc, 21.5~pc, and 25~pc from the exciting star (in projection). 
These fingers point very clearly to this star. 
This suggests that the main IF has holes through which the ionization radiation escapes, shaping the neutral material far away from the star.  
Finger~\#1 is especially interesting, as it is seen in absorption on the DSS-red image (right insert in Fig.~\ref{global}); it lies in front of the stellar field, and is related to W5-E as it clearly points to its exciting star. Molecular condensations are always present at the head of these fingers; the masses of these condensations are in the range 1.4\msol\ -- 5.5\msol\ (Sect.~6.6, Table~\ref{divers}). They are susceptible to form, by gravitational collapse later-on, low- or intermediate-mass stars, but not massive stars.

$\bullet$ There is a large quantity of material between W5-E and W5-W.  This material is ionized in the A and B \HII\ regions, and  is
molecular in the long ridge extending north-south, south of the A and B \HII\ regions.
This material is possibly compressed by the two expanding W5-E and W5-W \HII\ regions; however its temperature and column density 
are similar to those of PDR~1 and PDR~2.

\subsection{The pillars at the south-east border of W5-E}

\begin{figure*}[tb]
\sidecaption
\includegraphics[width=12cm]{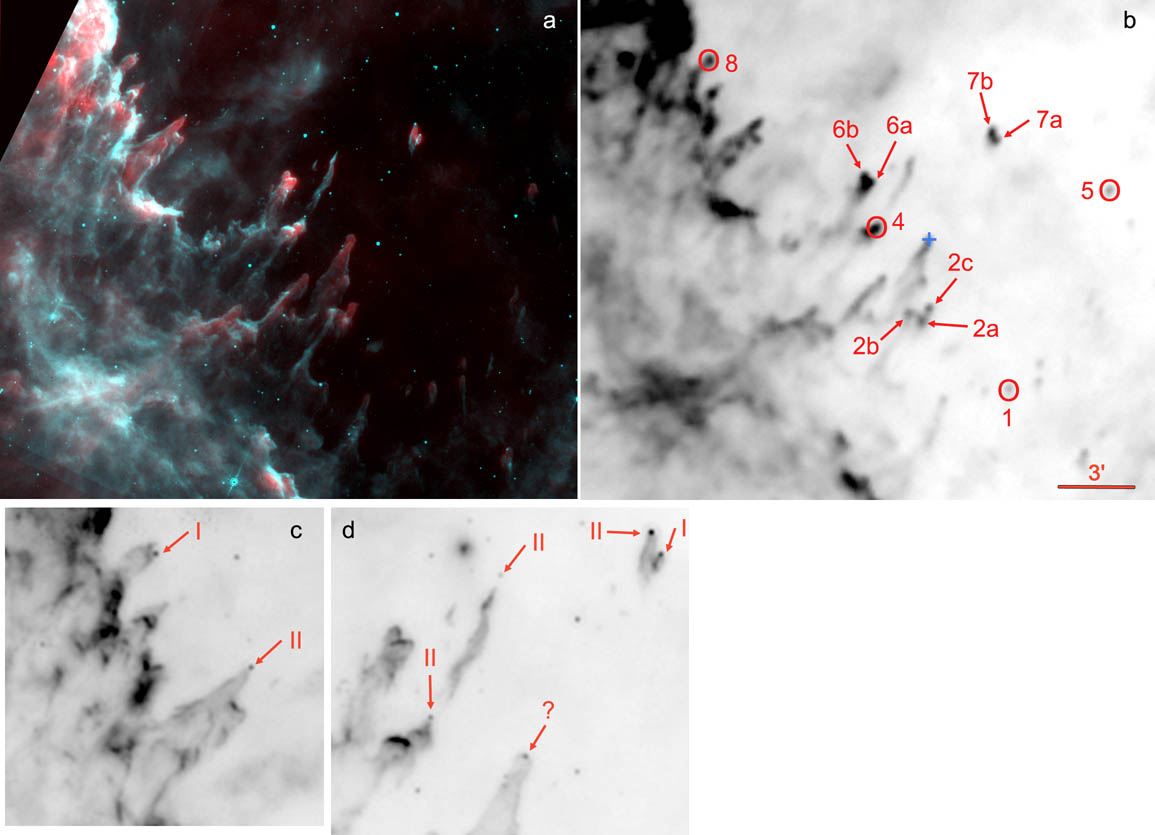}
  \caption{The IR emission from the ``pillars'' located south-east of W5-E.  
 {\it a:} Colour composite image showing in red the 250~$\mu$m cold dust emission and in turquoise the 8.0~$\mu$m PAH emission. 
  {\it b:} Identification of a few pillars on the grey-scale {\it Herschel} 250~$\mu$m image. {\it c, d:} Details of the pillars: the grey-scale underlying {\it Spitzer} 24~$\mu$m image shows Class~I/II sources lying at the tip of the pillars (from KOE08). North is up and east is left.}
  \label{pillars}
\end{figure*}

``Pillars,'' also called ``elephant trunks'', are small structures located at the interface between an \HII\ region and the adjacent molecular cloud
that look like ``fingers''.  They have a ``head'', that is nearest the ionizing star, and a filamentary ``tail'' that leads from the head away from the star.

The W5-E \HII\ bubble is open in the south, a direction of low density, and many pillars are observed there, 
especially on the south-east border. These pillars are especially conspicuous at 8.0~$\mu$m as shown in Fig.~\ref{pillars}. 
They have counterparts in the {\it Herschel}-SPIRE images. What we see at 8.0~$\mu$m is mainly the emission of PAHs from the ``skin'' of these 
dense molecular structures. 
All such pillars have the same basic morphology, with a rounded head at the tip of a thin filament. Most of these structures point toward the exciting star HD~18326. 
The column density is the highest in the head (Fig.~\ref{pillars}; Table~\ref{pillarsbis}) and it 
decreases from the head to the tail.  Some heads are separated from the parental molecular cloud, but most of them are not (this is especially 
evident in the 8.0~$\mu$m map). These structures are not fully resolved at 250~$\mu$m since the diameters of the heads and the widths of the filaments 
are of the order of the PSF. 

We list some parameters associated with these structures in Table~\ref{pillarsbis}. To estimate their mass and column density we need their temperature. 
We could use the dust temperatures from the temperature map (Sect.~4.1 and Fig.~\ref{temperature}),
but the size of the pilars is smaller than the resolution of the temperature map (which is the resolution of the 500~$\mu$m map), 
and thus the temperature map values are not necessarily accurate for these structures.  We also measured the fluxes of isolated
structures at all {\it Herschel} wavelengths, using an aperture of radius
45\arcsec\ (thus much larger than the beam at 500~$\mu$m), and fitted their SEDs with a modified blackbody model.   
(This is not necessarily better than the temperature map, as what we get is an average temperature within the aperture.) 
The first three columns of Table~\ref{pillarsbis} give the identification of these structures 
according to Fig.~\ref{pillars} and the coordinates of the condensations at the heads of the pillars.  
Column four gives the dust temperature obtained from the temperature map. These temperatures are very similar: for the 15 
structures in Table~\ref{pillars} we obtain T$_{\rm dust}$=22.7$\pm$0.5~K. This temperature is characteristic of the PDRs in W5
(Fig.~\ref{temperature}). The temperatures obtained by fitting the SEDs (aperture photometry) are also given 
in column four (with an asterisk). Again these temperatures (T$_{\rm dust}$=21.5$\pm$1.9K) are compatible with the temperature of a PDR. 
In the following, we use a mean temperature of 22~K to estimate the mass and column density of all these structures.   
Column five gives the flux density at 250~$\mu$m of the heads of the pillars, calculated by integrating over an aperture of radius 22.5\arcsec (much 
larger than the beam at 250~$\mu$m). Columns six and seven give the mass of the heads and their column density at the peak emission.  
The masses are all small, in the range 0.3\msol--1.4\msol\ (except for pillar \#10). These masses 
are uncertain mostly because of the uncertainty on the opacity.  We find, however, that no massive stars can form at the head of these pillars.
These structures are not well-resolved at 250~$\mu$m but we may estimate their sizes using the 8.0~$\mu$m emission, which traces the 
ionization fronts bordering these structures. For example, the column density in the direction $\alpha$(2000)$=03^{\rm h}00^{\rm m}08.46^{\rm s}$ $\delta$(2000)$=60\degr$17$\arcmin$02$\arcsec$ (the head of the pillar marked  
with a blue cross in Fig.~\ref{pillars}{\bf b}) is 7.15~10$^{20}$~cm$^{-3}$.  The diameter of the head given by the 8.0~$\mu$m image is $\sim$0.145~pc, 
and thus we find a mean density $\sim$1600~molecules~cm$^{-3}$ in the head of this standard pillar (estimated assuming a uniform density and temperature).\\ 

\begin{figure*}[tb]
\sidecaption
\includegraphics[width=12cm]{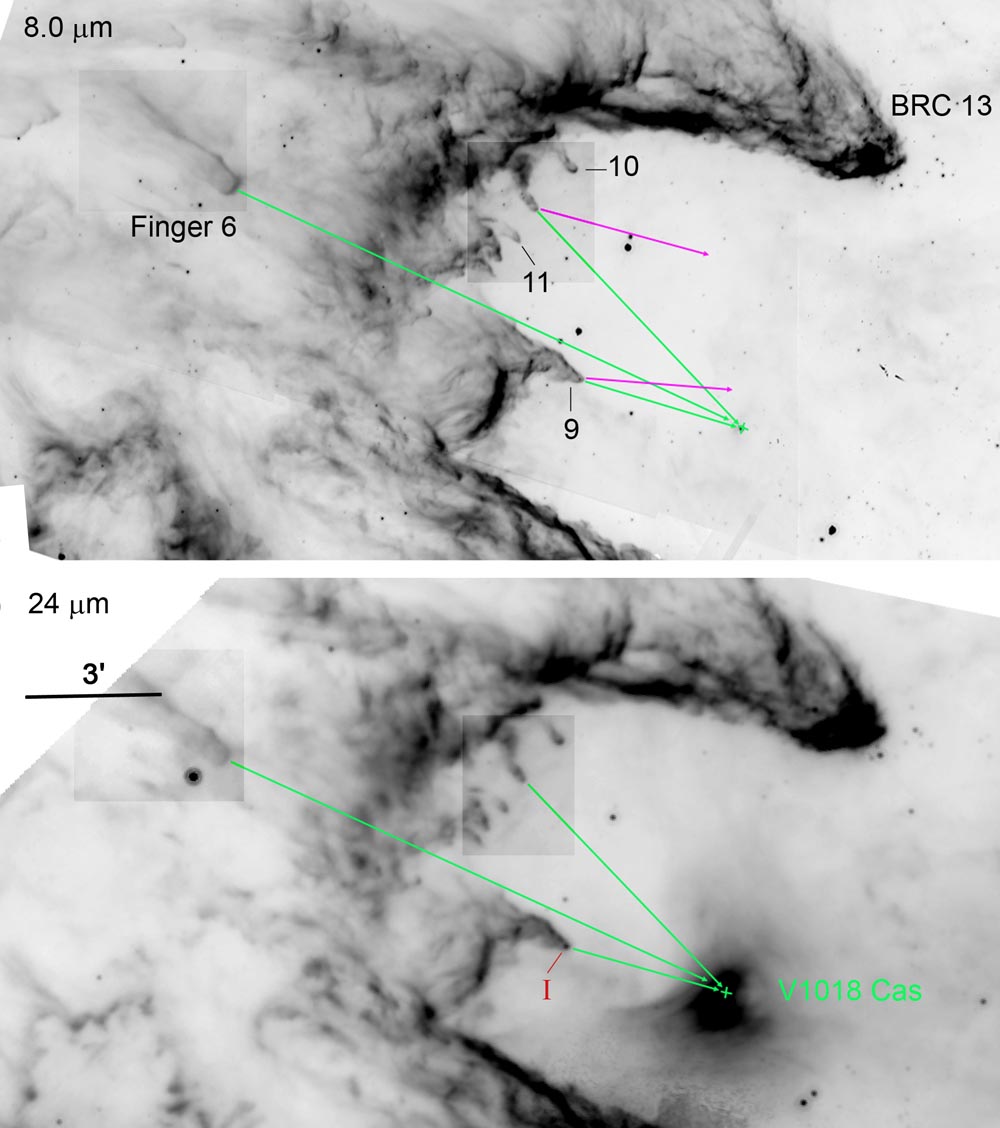}
  \caption{ Details of the pillars situated between BRC13 and BRC14. These pillars seem to point towards the massive star V1018 Cas. {\it Top:} The pillars are 
  identified on the {\it Spitzer} 8.0~$\mu$m image,  
  according to their number in Tables~\ref{pillarsbis} and Table~\ref{divers}. The green arrows indicate the direction of V1018 Cas (green cross), 
  the pink arrows that of HD~18326. {\it Bottom:} The {\it Spitzer} 24~$\mu$m image shows the bright extended emission possibly associated with V1018 Cas. 
  A Class~I YSO has been detected at the head of pillar~\#9 by KOE08. North is up and east is left.}
  \label{pillarsbisbis}
\end{figure*}

All the pillars at the south-east border of W5-E point to the exciting star HD~18326. A group of pillars on the east side of W5-E between BRC13 and BRC14 (including pillars \#9, \#10, and \#11 in Table~\ref{pillarsbis} and Finger \#6 in Table~\ref{divers}), however, seem to point towards a different direction. They point to a zone 
where extended 24~$\mu$m emission is observed (see Fig.~\ref{pillarsbisbis}). There is another massive star V1018 Cas in this direction
(green cross in Fig.~\ref{pillarsbisbis}). This star is an eclipsing binary    
(Bulut \& Demircan \cite{bul07}). Its spectral type is O7V (Otero et al. \cite{ote05}) or B1
(Reed \cite{ree03}). This star is possibly responsible for the 
surrounding extended 24~$\mu$m emission but its radial velocity ($\sim$10~km~s$^{-1}$, Duflot et al. \cite{duf95}) 
indicates that it likely does not belong to the W5 complex (the ionized gas has a velocity in the range $-34$ to $-39$~km~s$^{-1}$). 
This situation is difficult to understand, and the velocity of V1018~Cas needs to be confirmed.\\

\begin{table}[tb]
\caption{Physical parameters of the pillars}
\resizebox{85mm}{!}{
\begin{tabular}{lllllll}
\hline\hline
Name  & RA(2000)    & Dec(2000)     & T$_{\rm dust} $ & S$_{\nu}$(250$\mu$m)   & M        & N(H$_2$) \\
      &             &               & (K)             & (Jy)        & (\msol)  & (cm$^{-2}$) \\          
\hline
1   & 02:59:42.18 & +60:11:09 & 23.7  & 0.50 & 0.33 & 3.7e20 \\
1$^*$ &           &           & 20.8$^*$  &      &      &     \\
2a  & 03:00:10.34 & +60:13:53 & 22.8  & 1.75 & 1.16 & 6.9e20 \\
2b  & 03:00:14.58 & +60:14:10 & 22.95 & 0.89 & 0.59 & 6.1e20 \\
2c  & 03:00:07.93 & +60:14:21 & 22.75 & 1.25 & 0.83 & 6.7e20 \\
3   & 02:58:21.41 & +60:14:32 & 22.3  & 1.46 & 0.97 & 7.0e20 \\
4   & 03:00:24.93 & +60:17:35 & 22.6  & 2.25 & 1.49 & 1.2e20 \\
4$^*$ &           &           & 24.3$^*$  &      &      &  \\
5   & 02:59:09.89 & +60:19:07 & 23.5  & 0.71 & 0.47 & 4.5e20 \\
5$^*$ &           &           & 19.1$^*$  &      &      &  \\
6a  & 03:00:26.96 & +60:19:24 & 22.0  & 2.08 & 1.38 & 1.5e20 \\
6b  & 03:00:28.86 & +60:19:36 & 22.0  & 2.08 & 1.38 & 1.5e20 \\
6a$^*$+6b$^*$ &   &           & 21.2$^*$  &      &      &  \\ 
7a  & 02:59:46.24 & +60:21:08 & 22.75 & 1.45 & 0.96 & 8.8e20 \\
7b  & 02:59:47.53 & +60:21:27 & 22.75 & 1.64 & 1.09 & 9.6e20 \\
7a$^*$+7b$^*$ &   &           & 22.3$^*$  &      &      &  \\
8   & 03:01:19.28 & +60:24:13 & 22.5  & 1.72 & 1.14 & 1.1e20 \\
\hline
9  & 03:01:48.07  & +60:35:31 & 22.6  & 3.71 & 2.46 & 1.5e21 \\
10 & 03:01:49.13  & +60:40:01 & 23.0  & 1.45 & 0.96 & 7.6e20 \\
11 & 03:01:58.76  & +60:38:29 & 23.2  & 0.66 & 0.44 & 4.9e20 \\
\hline
\label{pillarsbis}
\end{tabular}\\
}
\end{table}

What is the origin of the pillars? There are several models that attempt to explain these structures. All these models are based 
on the influence of the UV radiation field of the central exciting star upon an inhomogeneous surrounding 
medium. They differ in the origin of the inhomogeneities. In the well-known radiation-driven implosion models (RDI), pre-existing 
dense condensations are present.
The RDI models describe the compression of a spherical isothermal isolated globule (Bertoldi \cite{ber89}; Lefloch \& Lazareff \cite{lef94}; 
Kessel-Deynet \& Burkert \cite{kes03}; Miao et al. \cite{mia09}; Bisbas et al. \cite{bis09}, \cite{bis11}; Haworth \& Harries \cite{haw11}; Tremblin et al. \cite{tre11}). 
In another model, 
the inhomogeneities are due to turbulence (Gritschneder et al. \cite{gri09a}, \cite{gri10}). Both models are able to produce 
structures similar to the observed pillars with a dense core (the head) at the tip of a filament. As discussed in Gritschneder at al. (\cite{gri10}) 
the velocity field in principle allows us to distinguish one process from the other.  Unfortunately, we do not have the velocity field to  
discriminate easily between the two processes.

Both models predict star formation in the pillars, under specific conditions. 
In Gritschneder et al.'s simulations, gravitational collapse occurs at the tip of the pillars, leading to the formation of cores and then stars. This occurs after 
a few 10$^{5}$~yr, depending of the initial conditions of the simulations 
(mean density and Mach number). The formed cores are not very massive, with masses $\leq$1\msol. In the simulations of Bisbas et al. (\cite{bis11}),
the critical parameter is the incident ionizing flux reaching the globule $\Phi_{\rm LyC}$. Star formation occurs for $\Phi_{\rm LyC}$ over a particular range. 
We estimate that $\Phi$$_{\rm LyC}$$\sim$8$\times$10$^{8}$~cm$^{-2}$~s$^{-1}$ at the head of pillar \#4 or at the position of the 
blue cross in Fig.~\ref{pillars}; this value allows star formation according to Bisbas et al. (\cite{bis11}); in these conditions the  first star forms  
rather quickly (in less than 0.1~Myr) once the
pre-existing globule is reached by the ionizing photons, and the new stars have low masses.   
YSOs are indeed present at the tip of some pillars, as shown by Fig.~\ref{pillars} (see also Fig.~1 in Chauhan et al. \cite{cha11a}). 
According to KOE08 they are Class~I or Class~II YSOs. Chauhan et al. (\cite{cha11b}) have used $V$ versus $V-I$ diagrams 
and YSO evolutionary tracks to estimate the age of a few sources present at the tip of the pillars. They estimate that they are pre-main sequence 
low-mass stars having ages of one Myr or less. All this is consistant  with all the simulations.

The RDI simulations of  Haworth \& Harries (\cite{haw11}) suggest a somewhat different origin for the pillars: they could be due 
to small-scale 
instabilities of the IF at the periphery of a dense globule, and not to the compression of the globule. The ionizing photon flux is also a critical parameter in their simulations. The ionizing photon flux is low at the position of the pillars (position of the blue cross, Fig.~\ref{pillars}). As shown by 
the simulations this configuration exhibits the largest susceptibility to develop instabilities (but the simulations also show that the resulting pillars are smoothed out by the diffuse field radiation). Chauhan et al. (\cite{cha11b}) favour this explanation for the origin of the pillars. 

That there are numerous pillars observed at the south-west border of W5-E, a region of low ionizing flux, favours the explanation put forth by Gritschneder et al. (\cite{gri10}) involving turbulence, or the explanation of Haworth \& Harries (\cite{haw11}) involving instabilities; however, following 
 the model of  Haworth \& Harries, we would expect the formation of pillars at the periphery of dense condensations; these are not observed with {\it Herschel}. 
 Thus, the formation of the pillars, the cores at their heads, and the subsequent low-mass YSOs are likely triggered by the W5-E \HII\ region, resulting from the 
interaction of the UV radiation of its exciting star with a low density inhomogeneous medium, probably of turbulent origin.

\subsection{The properties of the young stellar objects}

We detect 50 100~$\mu$m point-like sources in the field covered by {\it Herschel}.  The number of detected sources decreases at longer wavelengths 
(respectively 43, 37, 23, and 11 at 160~$\mu$m, 250~$\mu$m, 350~$\mu$m, and 500~$\mu$m). Due to the temperature of their envelopes (Sect.~6.3.1) their fluxes are maximum in the 100~$\mu$m--160~$\mu$m range, and decrease at longer wavelengths; also, due to the decreasing angular resolution, it becomes more and more difficult to separate point sources from their bright background.

\subsubsection{Temperatures and masses of the envelopes}

\begin{figure*}[tb]
\sidecaption
\includegraphics[width=12cm]{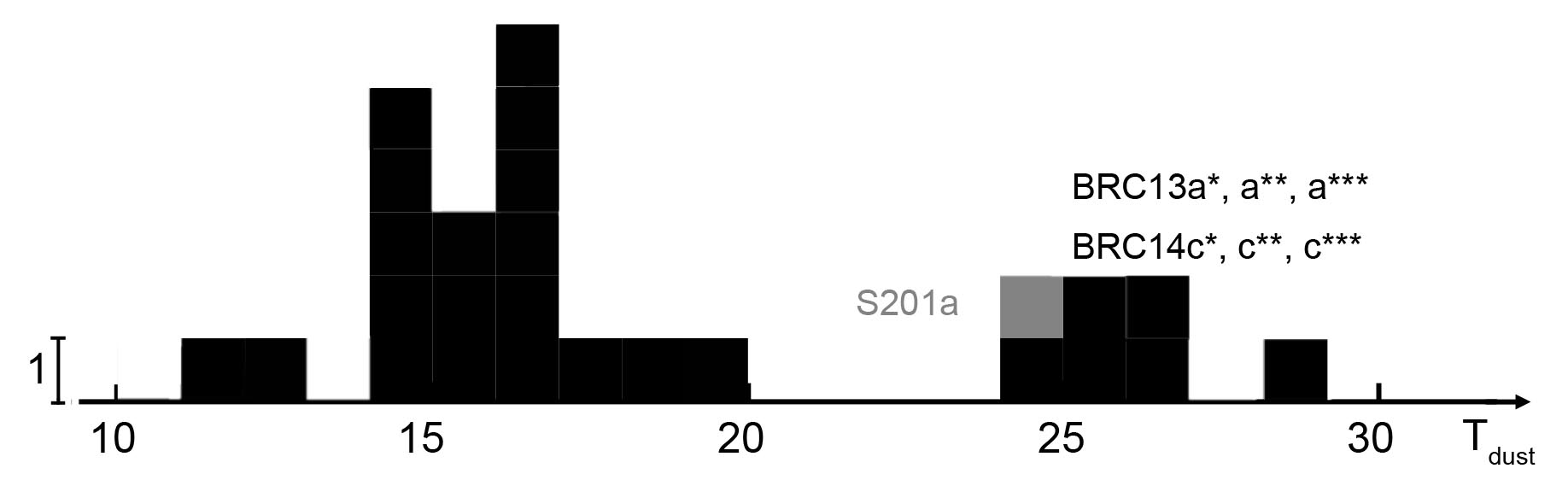}
  \caption{The dust temperature of YSO envelopes or of compact condensations, estimated by fitting the {\it Herschel} fluxes using a modified blackbody model.}
  \label{YSOstemperature}
\end{figure*}

The {\it Herschel} emission from YSOs comes mostly from their envelopes. We can use their fluxes and a modified blackbody model (see Sect.~4) to estimate 
the envelope temperatures and masses. We use the fluxes given in Tables \ref{BRC12tablea}, \ref{BRC13}, \ref{BRC14tablea}, \ref{S201}, \ref{between}, and \ref{isolated}. 
We apply a colour correction, which takes into account the varying spectral response with dust temperature (see Sect.~4.1 in AND12).  
We compute the masses from the 350~$\mu$m fluxes, using a spectral index $\beta$=2 and the temperatures previously estimated. 
We list the temperatures and masses in Table~\ref{temperatureYSOs}, in columns 3 and 4 respectively. This Table contains only the sources that are detected in at least four {\it Herschel} bands. The first column gives the identification of the sources, the second one the number of flux measurements ($n$ equal to 4 or 5). 

Fig.~\ref{YSOstemperature} displays the temperatures of the detected sources. The histogram shows two maxima: one around 15.7~K, containing the majority of the point sources, and one around 25.8~K. Thus we find that the envelopes of YSOs are generally cold, with a characteristic temperature of 15.7$\pm$1.8~K. 
The second peak contains extended sources associated with clusters: the 
extended BRC13-a source around the luminous BRC13-a Class~II YSO (three apertures, the smallest one is the hottest), the extended BRC14-c1+c2 sources 
associated with IRS1 and IRS2 in AFGL~4029 (again three apertures and the smallest is the hottest), and the YSO S201-a.  
The high temperature of S201-a is also apparent in the temperature map in Fig.~\ref{temperature}
and therefore this source differs from the other YSOs that have cold envelopes. S201-a seems to be extended at 8.0~$\mu$m, and unresolved at the other {\it Spitzer} wavelengths; it is possibly surrounded by PAH emission. Its SED points to a luminous Class~I YSO, observed with the disk face-on (through the cavity, at least for the best model). 
As mentioned in Sect.~6.4, this YSO is probably not associated with W5-E, and instead lies in the far background; it displays a massive envelope.\\

\begin{figure}[tb]
\centering
\includegraphics[width=85mm]{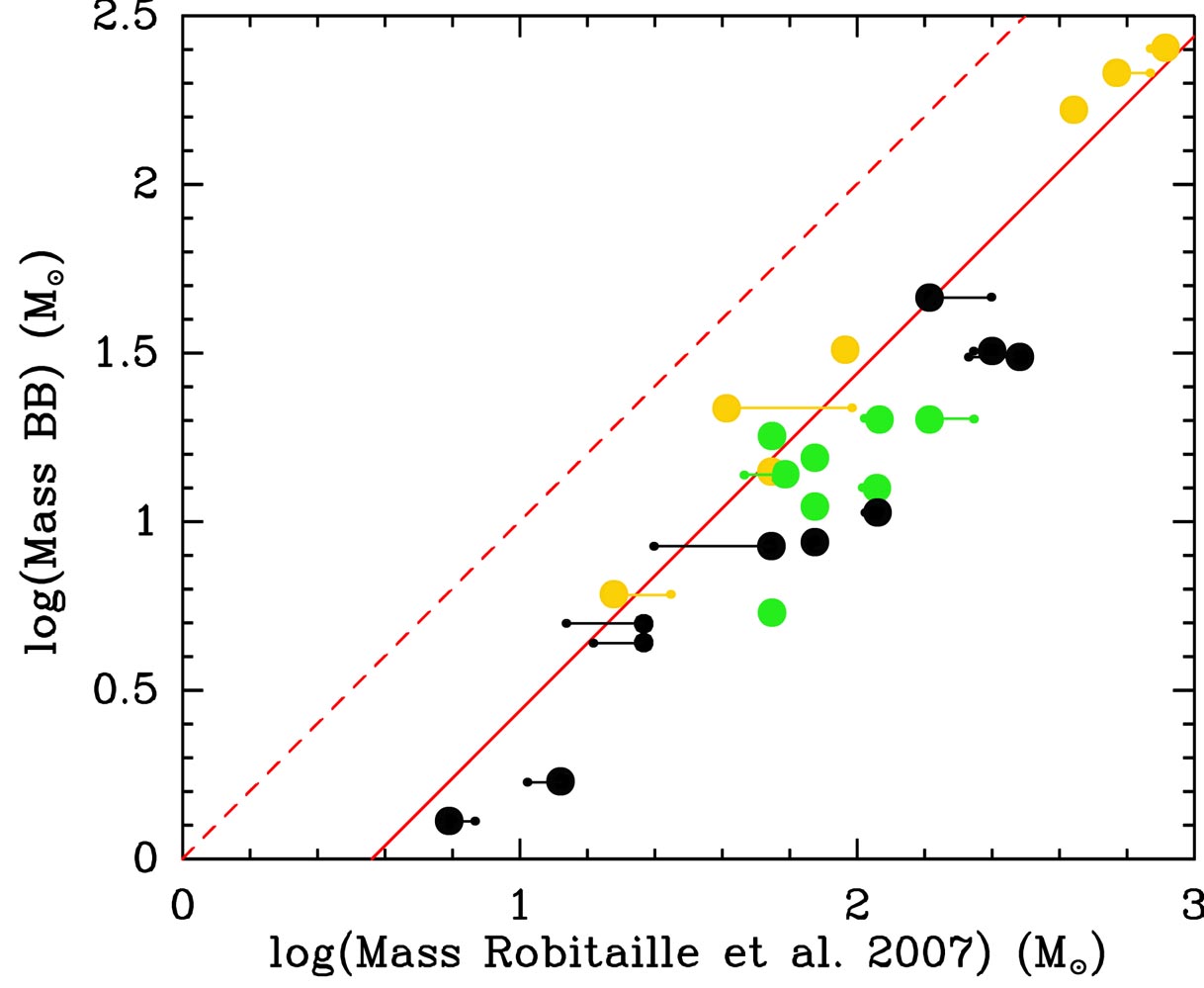}
  \caption{The masses of the YSO envelopes or of compact condensations estimated from the {\it Herschel} fluxes.  Here we use only sources that
   have four or five measured {\it Herschel} fluxes.  
   We estimate the masses M(BB) using a modified blackbody model. We estimate the masses M(Robitaille et al.) using ROB07 SED fitting-tool. 
   The red dashed line shows equality between the two masses and we find no sources located along it.  The red solid line shows where the points should lie, taking into 
   account the different dust opacities used in the mass estimations. The black and coloured dots are respectively for sources with 4 and 5 measured fluxes. 
   Yellow and green dots are respectively for hot (T$\sim$25.8~K in Fig.~\ref{YSOstemperature}) and cold (T$\sim$15.7~K in Fig.~\ref{YSOstemperature}) sources. 
   The horizontal bars indicate the range of values obtained using ROB07's SED fitting-tool.}
  \label{YSOsmasse}
\end{figure}

\begin{figure*}[tb]
\centering
\includegraphics[width=180mm]{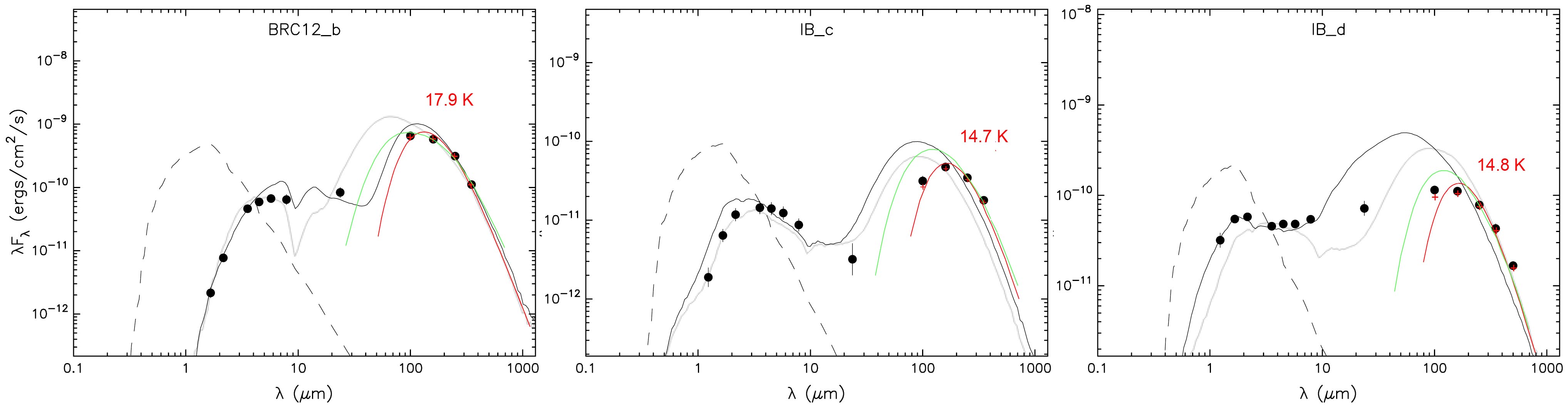}
  \caption{Fitting the SEDs of the envelopes of three YSOs. The black curves show the fits of the whole SEDs, obtained using the SED fitting tool of ROB07; the green curves show the fits obtained for the envelopes, using the same models, but considering only the {\it Herschel} fluxes. The red curves are the fits obtained using the blackbody modified model.}
  \label{fits}
\end{figure*}

Table~\ref{temperatureYSOs} allows us to compare the masses of the envelopes estimated with two different methods. Column 4 gives the masses obtained using the modified blackbody model and the flux at 350~$\mu$m.  
Column 5 gives the masses estimated using the SED fitting-tool of ROB07, fit using the Herschel fluxes alone.
We also give in column 5 the range of masses
corresponding to the ten best models. Fig.~\ref{YSOsmasse} shows that the masses obtained using the SED fitting-tool of ROB07  
are always much larger than the masses estimated with the modified blackbody model. This results mainly from the different opacities used in the two models. 
At 350~$\mu$m, ROB07 use 
an opacity of 2~cm$^2$~g$^{-1}$ whereas we use an opacity of 7.3~cm$^2$~g$^{-1}$. The masses derived with the two methods therefore  
should be in the ratio 7.3/2=3.65 (the red line in Fig.~\ref{YSOsmasse}).  The discrepancy however is even greater: for the 25 regions 
in Table~\ref{YSOstemperature} we obtain a mass ratio of 5.58$\pm$2.54. In Fig.~\ref{YSOsmasse}, we show the masses estimated using fluxes in 
five {\it Herschel} bands in colour (yellow and green),
whereas the masses estimated using only four {\it Herschel} fluxes are in black. The discrepancy is worse for the sources with only 4 flux measurements 
(ratio equal to 6.60$\pm$2.26 and 4.78$\pm$2.53 for respectively 11 sources with $n=4$ and 14 sources with $n=5$); but the main difference is between 
sources surrounded by hot dust (second maxima in the histogram of Fig.~\ref{YSOstemperature}; yellow dots) or cold dust 
(first peak in the histogram of Fig.~\ref{YSOstemperature}; green and black dots): ratio equal to 2.96$\pm$0.79 and 6.59$\pm$2.23 for respectively 7 sources 
surrounded by hot dust or 18 sources with cold envelopes. The largest discrepancies are for the cold envelopes. 

The transfer models presently used in the SED-fitting tool of ROB07 do not include cold dust (the external radii
 of the envelopes correspond to a dust temperature of 30~K; T. Robitaille, private communication). To account for the {\it Herschel} measured 
fluxes - due to cold dust, the SED-fitting tool overestimates the mass of the envelope.  Additionally, the emission peak of the envelope is shifted
toward shorter wavelengths (as the SED-fitting tool overestimates the temperature of the envelope). These two effects (overestimation of the 
envelope mass and temperature) are conspicuous in Fig.~\ref{fits} which shows the fits obtained by the two methods.  Obviously, the modified blackbody model (red curves) fits 
the {\it Herschel} data points better than ROB07's models (green curves).  New dust radiative transfer models are in 
preparation, including PAH emission, heating by external radiation, and larger envelope radii (Sewilo et al. \cite{sew10}). The first tests 
of these models are discussed by these authors. Presently it is probably better to use the modified blackbody model to estimate the characteristics 
of the YSOs' envelopes.

The masses of the envelopes detected by {\it Herschel} around 100~$\mu$m point sources, estimated using the modified blackbody model, 
lie in the range 1.3~$\msol$ -- 47~$\msol$.\\


\begin{table}[tb]
\caption{Temperatures and masses of the envelopes of YSOs and compact molecular 
condensations}
\begin{tabular}{rrrrr}
\hline\hline
Identification & $n$ & T$_{\rm d}$(BB) & M$_{\rm env}$(BB) & M$_{\rm env}$(Rob.) \\
 &  & (K) & ($\msol$ ) & ($\msol$ ) \\
\hline
BRC12-b & 4 & 17.9 & 30.9 & 306.0 (--215)\\
BRC12-d & 4 & 16.3 & 32.1 &251.0 (--223) \\
BRC12-e & 4 & 18.7 & 12.3 &115.0 (--106) \\
BRC13-a* & 5 & 28.3 & 6.1 & 19.2(--28) \\
BRC13-a** & 5 & 26.0 & 13.8 & 61.5 (--46) \\
BRC13-a*** & 5 & 24.8 & 21.8 & 41.2 (--97)\\
BRC14-a3 & 5 & 14.4 & 15.6 & 75.4 (--79)\\
BRC14-b & 5 & 16.7 & 20.2 & 165.0 (223)\\
BRC14-c & 5 & 11.8 & 22.5 &  \\
BRC14-c* & 5 & 26.4 & 166.6 & 440.0 (--460) \\
BRC14-c** & 5 & 25.5 & 214.7 & 591 (--741)\\
BRC14-c*** & 5 & 25.0 & 254.9 & 823 (--741) \\
S201-a\tablefootmark{a} & 5 & 24.9 & 32.6 & 92.8 \\
S201-b & 4 & 19.4 & 1.3 & 6.2 (--7.4) \\
S201-c & 5 & 12.9 & 18.0 & 56.0\\
S201-d & 5 & 15.9 & 5.4 & 56.0 \\
S201-g & 4 & 14.5 & 46.5 & 165 (--251) \\
S201-h & 4 & 16.1 & 4.4 & 23.4 (--16.6)\\
IB-a & 4 & 14.1 & 14.1 & 56.0 (--54.7) \\
IB-b & 4 & 16.4 & 8.7 & 75.4 (--54.7) \\
IB-c & 4 & 14.7 & 8.5 & 56.0 (--25.1) \\
IB-d & 5 & 14.8 & 20.3 & 115.2 (--106.0)\\
IB-e & 5 & 16.0 & 10.7 & 115.2 (--106.0)\\
Isolated-5 & 4 & 16.6 & 1.7 & 13.6 (--10.6) \\
Isolated-11 & 5 & 15.5 & 11.1 & 75.4 (--78.7)\\
Isolated-13 & 4 & 15.6 & 5.0 & 23.4 (--13.8) \\
\hline
\label{temperatureYSOs}
\end{tabular}
\tablefoot{
\tablefoottext{a}{For a distance of 5.45~kpc (Sect.~6.5).}
}
\end{table}

\subsubsection{Evolutionary stages - Comparison with Koenig et al.'s (\cite{koe08}) classification}

\begin{figure}[tb]
\centering
\includegraphics[width=85mm]{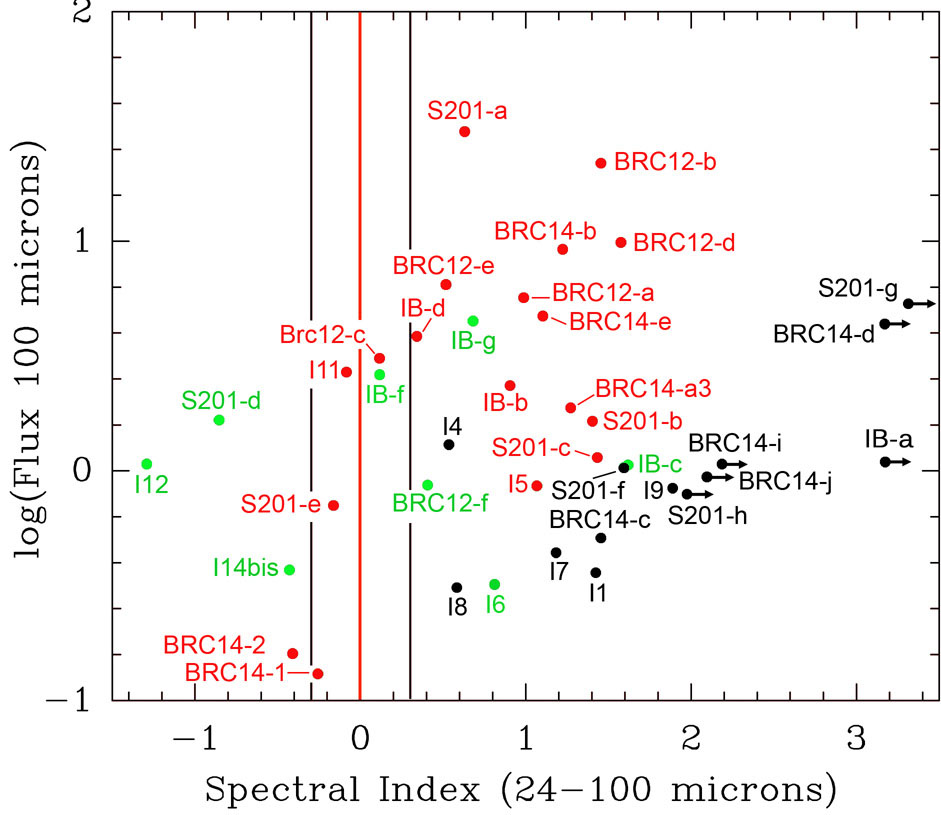}
  \caption{The integrated flux at 100~$\mu$m of YSOs versus their spectral index between 24~$\mu$m and 100~$\mu$m. 
  The sources classified as Class~I and Class~II YSOs by KOE08 are respectively represented by red and green dots. The black dots show sources with no counterparts 
  in KOE08. }
  \label{flux}
\end{figure}

\begin{figure*}[tb]
\includegraphics[width=162mm]{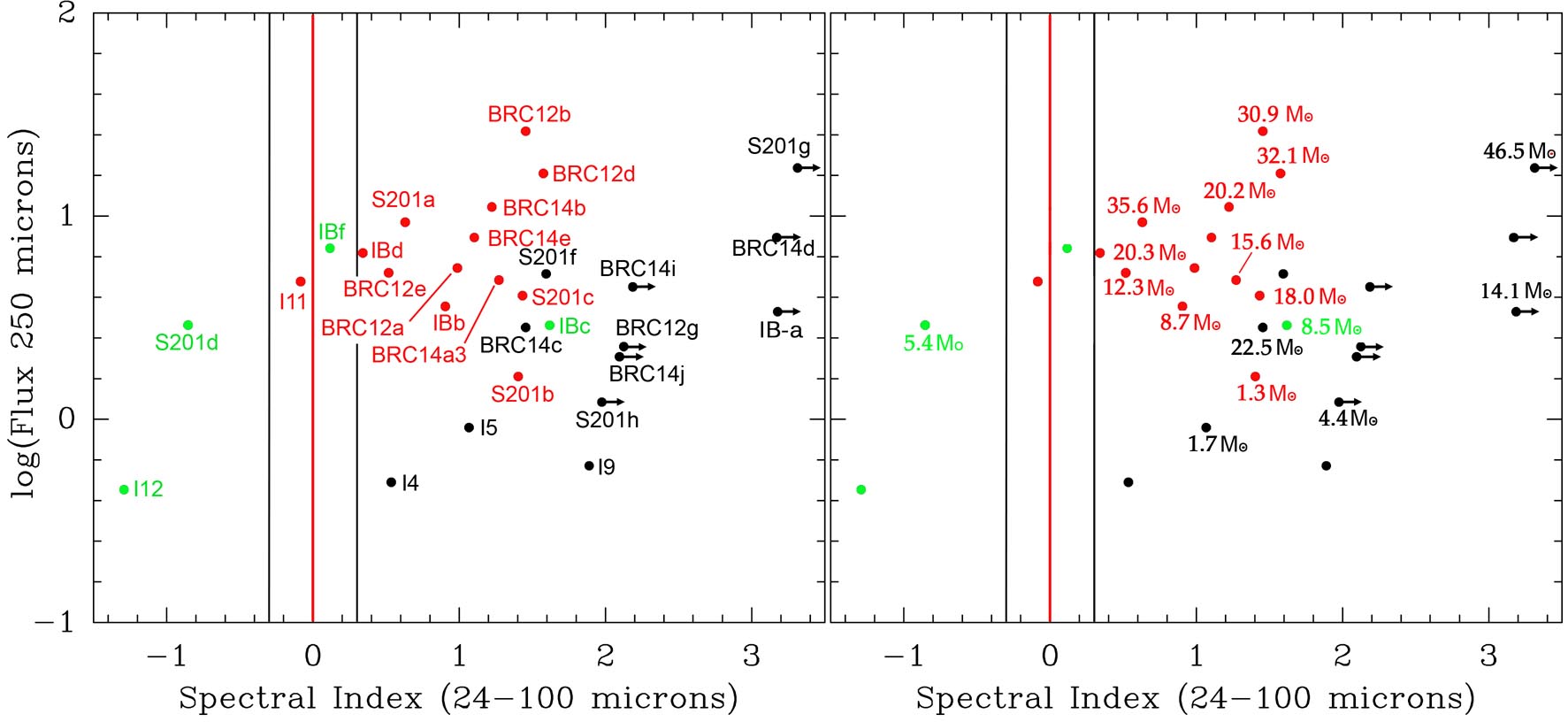}
\centering
  \caption{The integrated flux at 250~$\mu$m of YSOs versus their spectral index between 24~$\mu$m and 100~$\mu$m.
  The symbols are the same symbols as in Fig.~\ref{flux}. {\it Left:} identification of 
  the YSOs. {\it Right:} the masses of the envelopes are indicated, estimated using the modified blackbody model (we require 
  at least four {\it Herschel} fluxes for a mass estimate).}
  \label{fluxbis}
\end{figure*}

The field covered by {\it Herschel} contains 50 point sources detected at 100~$\mu$m  
(we are considering here the point sources and not the condensations). Twenty eight of these 100~$\mu$m  point sources (56\%) have a counterpart in the KOE08 catalogue: 
19 are Class~I sources, and 9 are Class~II; the remaining eleven sources are candidate Class~0 sources. We fit the SEDs of these sources using the 
SED fitting-tool of ROB07 and 2MASS, {\it Spitzer}, and {\it Herschel} fluxes. Among the 19 Class~I YSOs of KOE08, we confirm that 16 are Stage~I sources (84\% of the cases); among the 9 Class~II YSOs we confirm with certainty that 3 are in Stage~II (30\% of the cases). Thus the agreement with KOE08 is rather good, especially for the Class~I sources that, due to their cold envelopes, we expect {\it Herschel} to detect.  We comment on these results below.

Fig~\ref{flux} and Fig.~\ref{fluxbis} show the flux versus spectral index diagrams for these sources (the equivalent of the more classical magnitude -- colour diagrams) using {\it Herschel} measurements. We estimate the spectral index between 24~$\mu$m and 100~$\mu$m.  The spectral index indicates whether the flux is increasing between these wavelengths (a possible signature of the presence of an envelope).  We have used colours to indicate the 
classification of KOE08, red for their Class~I and green for their Class~II sources. The black symbols correspond to sources that are not in KOE08 catalogue. 

Fig.~\ref{flux} shows that most Class~I YSOs in KOE08 have a positive spectral index between 24~$\mu$m and 100~$\mu$m; 
however a few have a negative one. KOE08 Class~II YSOs have positive as well as negative spectral indices.  
One explanation for this large range of spectral indices may be, as Robitaille et al. (\cite{rob06}) demonstrate, the influence of the inclination angle 
(between the line of sight and the plane of the disk) 
upon the shape of the SED.  Their fig.~7 shows that: i) Stage~0/I YSOs, dominated by their envelopes, have a positive spectral
index if the central source is of intermediate or low mass and a negative index if the central source is massive; ii) Stage~II YSOs, 
dominated by their disks, have a spectral index that depends of the inclination angle.  For example, Stage~II YSOs with a central object of intermediate
 mass, seen edge-on, have a double-peaked SED and a positive spectral index; and again if the central object is massive, the 
spectral index is negative. Several of the Class~II YSOs of KOE08 that have a positive spectral index present a double-peaked
 SED (BRC12-f, IB-c, IB-f, IB-g). It is dificult to classify these sources, they are possibly intermediate mass Class~II YSOs
viewed edge-on. We do not confirm the classification of BRC14-1 and -2 as Class~I YSOs, as proposed by KOE08.
 The classification of S201-e, I-11, and I-6, which have flat SEDs, is also uncertain. We conclude that the spectral index
is not a great indicator of the stage of the YSO. Sources with a highly positive spectral index, however, are Class~0/I sources (or Stage~0/I); among them, 
as is discussed in Sect.~6.3.3, BRC14-c, BRC14-d, BRC14-i, BRC14-j, S201-g,  S201-h, and IB-a are candidate Class~0 YSOs; they have no 24~$\mu$m counterparts
or very faint ones.

Fig.~\ref{fluxbis} uses the flux at 250~$\mu$m in ordinate and the same spectral index from Fig.~\ref{flux}
in abcissa; the flux at 250~$\mu$m can be considered as an indicator of the mass of the envelope. Indeed Fig.~\ref{fluxbis} 
({\it Right}) shows a decrease in envelope masses from top ($\sim$47~\msol) to bottom ($\sim$1~\msol). The mass depends on
the dust temperature of the envelope, which is sometimes difficult to estimate accurately (see the dispersion of the temperatures 
in Fig.~\ref{YSOstemperature}). This explains why a few nearby sources in this diagram can have rather different masses.  (This is the case
for IB-c and BRC14-c.  BRC14-c has the coldest envelope temperature of the whole sample; its temperature is possibly
underestimated and thus its mass overestimated). Fig.~\ref{fluxbis} shows that several YSOs have an accreting 
envelope mass higher than 20~\msol; these YSOs will possibly form massive stars (for example, BRC12-b, with an envelope accretion rate 
of $\sim$1~$\times$10$^{-3}$~\msol~yr$^{-1}$ or BRC12-d, with an accretion rate of $\sim$3~$\times$10$^{-4}$~\msol~yr$^{-1}$, 
or BRC14-a3, BRC14-b, BRC14-c, BRC14-e, or S201-c, which have accretion rates $\geq$4~$\times$10$^{-4}$~\msol~yr$^{-1}$; we again caution that the envelope accretion rate, 
which is estimated using the ROB07 SED fitting-tool, is very uncertain). 
Among the candidate Class~0 sources, S201-g has a high mass envelope; it is probably also the case of BRC14-d (15.6~\msol\, assuming a 
temperature of 15.7~K for the envelope).

The field covered by {\it Herschel} contains 55 Class~I sources according to KOE08. Only 19 of these have a {\it Herschel} counterpart at 100~$\mu$m, 
16 of which are confirmed as Stage~I sources. Thus, most KOE08 Class~I sources (65\%) are not detected by {\it Herschel}. Most of these sources are also faint at 24~$\mu$m; we cannot say anything new about them. A few of the KOE08 Class~I sources, however, are bright at 24~$\mu$m; their non-detection at 100~$\mu$m makes their identification as Class~I sources doubtful (see for example the case of the 3 bright sources in the Sh~201 region, Appendix~C). Also, a few KOE08 Class~I sources belong to small stellar groups (for example in the BRC14 field, Appendix~B), or lie close to another bright YSO (for example in the BRC12 field, Sect.~5.1); the resolution of {\it Herschel} does not allow us to separate these sources.

Several {\it Spitzer} sources (KOE08 Class~I and Class~II) are sometimes observed in the direction of {\it Herschel} point sources. Sometimes the resolution at 100~$\mu$m is sufficient to associate one {\it Spitzer} source with the {\it Herschel} source (for example BRC14-a3, Appendix~B), but not always. In these cases {\it Herschel} may be detecting one compact condensation inside which a small stellar group is forming (for example BRC14-g+h, Appendix~B, or IB-e, Appendix~D). Higher resolution observations are needed to better understand these situations.

\subsubsection{Candidate Class~0 sources}

In the preceding sections we have identified Class~I or Stage~I sources as those that are dominated by their envelopes. 
We would like now to identify the youngest of them, the Class~0 sources. The lifetime of these YSOs is $\leq 10^5$~yr, 
according to Maury et al. (\cite{mau11}; assuming that the lifetime of Class~I YSOs is a few $10^5$~yr). (These lifetimes are still very uncertain 
and depend possibly of the final mass of the central source.) Different indicators 
can be used to identify Class~0 sources: i) Class~0 sources have  M$_{\rm env}$/M$_*$ larger than 1 (Andr\'e et al. \cite{and93} and references therein; 
M$_*$ is the mass of the central object); 
however if the source is not observed at near- and mid-IR wavelengths it is difficult to estimate M$_*$. It has been shown, in nearby star forming regions, 
that the envelope of Class~0 sources is extended (FWHM in the range 10$\arcsec$--14$\arcsec$ for two Class~0 in the $\rho$~Oph region at 160~pc, 
Andr\'e \& Montmerle \cite{and94}); 
however the resolution of {\it Herschel}-SPIRE does not allow us to resolve these sources in W5-E at 2~kpc; ii) for sources observed at sub-mm or mm wavelengths 
another condition is L$_{\rm bol}$/L$_{\lambda \geq 1.3mm} \leq$2$\times$10$^4$, 
or L$_{\lambda \geq 350\mu{\rm m}}$/L$_{\rm bol} \geq$~10$^{-2}$  (Andr\'e et al. \cite{and93}); a roughly equivalent criterium is  
M$_{\rm env}$/L$_{\rm bol} \geq$0.1~$\msol / \lsol$ (Andr\'e et al. \cite{and00}).  
One can also use evolutionary diagrams like  M$_{\rm env}$ versus L$_{\rm bol}$ where it has been shown that 
Class~0 and Class~I sources occupy different locations (Motte \& Andr\'e \cite{mot01}; Andr\'e et al. \cite{and08}; Hennemann et al. \cite{hen10}); 
iii) some authors use T$_{\rm bol}$, the bolometric temperature (which is the temperature of a blackbody having the same mean frequency as the observed YSO SED)  
to estimate the importance of the envelope. Enoch et al. (\cite{eno09} and references therein) define 
Class~0, and Class~I  sources as sources with respectively T$_{\rm bol} <$~70~K, and 70~K $\leq$T$_{\rm bol} \leq$ 650~K.

Table~\ref{classe0} gives these different indicators estimated for the {\it Herschel} point sources, when possible. The first column 
is the identification of the source and the second and third columns are the mass of the envelope using the modified blackbody fit (as in Table~\ref{temperatureYSOs}) 
and the mass of the central object estimated using the SED fitting-tool of ROB07. The resulting ratio 
M$_{\rm env}$/M$_*$ is given in column 4. Columns 5 and 8 give respectively the bolometric luminosity and temperature 
(according to Enoch et al. \cite{eno09}), obtained by a simple integration of the observed SED:

\begin{equation}
L_{\rm bol}=4 \pi d^2 \int S_{\nu}~dv
\end{equation}
and
\begin{equation}
T_{\rm bol}=1.25 \times 10^{-11} \frac{\int \nu\ S_{\nu}~dv}{\int S_{\nu}~dv}
\end{equation}

Columns 6 and 7 give respectively the ratios M$_{\rm env}$/L$_{\rm bol}$ and L$_{\lambda \geq 350\mu{\rm m}}$/L$_{\rm bol}$. Column 9 gives the spectral index  
 $\alpha$(24~$\mu$m--100~$\mu$m). In column 10  we list the presence or absence of a near- or mid-IR component and of a CO outflow (see Sect.~6.4).

\begin{table*}[tb]
\caption{Selection of Class~0 candidates}
\resizebox{18cm}{!}{
\begin{tabular}{rrrrrrrrrl}
\hline\hline
Identification & M$_{\rm env}$(BB) & M$_*$      &  M$_{\rm env}$/M$_*$ & L$_{\rm bol}$(BB) &  M$_{\rm env}$/L$_{\rm bol}$ & L$_{\lambda \geq 350\mu{\rm m}}$/L$_{\rm bol}$ & T$_{\rm bol}$(BB) 
& $\alpha$(24--100) & Counterpart \\
               & ($\msol$ )        & ($\msol$ ) &                       & ($\lsol$)         &                               &          ($\%$)                & (K) 
&                   &   \\
\hline
BRC12-b        & 30.9              & 2.8--6.9   & 11.0--4.5             & 210               & 0.15                          & 1.6                            & 82 
& 1.45              & 2MASS+{\it Spitzer}, CO outflow\\
BRC12-d        & 32.1              & 4.1--5.1   & 7.8--6.3             & 92                & 0.35                          & 2.6                            & 46 
& 1.58               & {\it Spitzer}, CO outflow\\
BRC12-e        & 12.3              & 2.5--5.3   & 4.9--2.3              & 74                & 0.17                          & 1.1                            & 93 
&0.52               & 2MASS+{\it Spitzer}\\
BRC14-a3       & 15.6              & 1.4        & 11.3                  & 21                & 0.74                          & 4.5                            & 45 
&1.27               & {\it Spitzer}\\
BRC14-b        & 20.2              & 2.2--4.15  & 9.1--4.9              & 93                & 0.22                          & 2.1                            & 78 
&1.22               & 2MASS+{\it Spitzer}\\
BRC14-c        & 22.5              & 1.2        & 18.8                  & 7                 & 3.21                          & 11.1                           & 36 
&1.46               & {\it Spitzer} \\
S201-b         & 1.3               & 0.7        & 1.9                   & 18                & 0.07                          & 1.1                            & 129 
&1.40               & 2MASS+{\it Spitzer}\\
S201-c         & 18.0              & 1.4        & 12.9                  & 13                & 1.38                          & 6.9                            & 50 
& 1.43              & {\it Spitzer}\\
S201-d         & 5.4               & 4.7        & 1.15                  & 118               & 0.05                          & 0.4                            & 789 
& $-$0.85           & 2MASS+{\it Spitzer}\\
S201-g         & 46.5              &            &                       & 40                & 1.16                          & 7.5                            &  
& $\geq$3.31        & no counterpart, CO outflow \\
S201-h         & 4.4               &            &                       & 4                 & 1.10                          & 5.2                            &   
& $\geq$1.98        & no counterpart \\
IB-a           & 14.1              &            &                       & 7.5               & 1.88                          & 8.4                            & 
& $\geq$3.19        & {\it Spitzer}-MIPS \\
IB-b           & 8.7               & 0.7--1.4   & 12.4--6.2             & 26                & 0.33                          & 2.3                            & 78 
& 0.9               & 2MASS+{\it Spitzer} \\
IB-c           & 8.5               &            &                       & 15                & 0.57                          & 3.4                            & 286 
& 1.6               & 2MASS+{\it Spitzer}\\
IB-d           & 20.3              & 1.9--3.25  & 10.7--6.2             & 63                & 0.32                          & 2.0                            & 383 
& 0.3               & 2MASS+{\it Spitzer} \\
I-5            & 1.7               & 0.45--0.9  & 3.8--1.9              & 7.5               & 0.23                          & 1.9                            & 161 
& 1.1               & 2MASS+{\it Spitzer}\\
I-11           & 11.1              & 4.6        & 2.4                   & 51                & 0.22                          &  1.7                           & 246 
& $-$0.1            & 2MASS+{\it Spitzer}\\
\hline
\label{classe0}
\end{tabular}
}
\end{table*}

Table~\ref{classe0} shows that most of these indicators are in rather good agreement, pointing to the same sources for the candidate Class~0 objects. We think that the most reliable indicators are these estimated directly from the observations, such as M$_{\rm env}$/L$_{\rm bol}$ or T$_{\rm bol}$.  Good candidate Class~0 sources are BRC14-c and S201-c which have  M$_{\rm env}$/L$_{\rm bol}\geq$1 and T$_{\rm bol}\leq$70~K; S201-g, S201-h and IB-a, which have M$_{\rm env}$/L$_{\rm bol}\geq$1 and a large value of  $\alpha$(24~$\mu$m--100~$\mu$m), are also good candidate Class~0 sources. All of these candidate Class~0 sources display high values for the 
ratio L$_{\lambda \geq 350\mu{\rm m}}$/L$_{\rm bol}$. BRC14-a3 appears to be at the limit between Class~0 and Class~I. And all the indicators confirm that S201-d is a Class~II~YSO.     
The indicator M$_{\rm env}$/M$_*$ is less reliable as a large range of values are often proposed for M$_*$ by the SED fitting-tool of ROB07. 
(Table~\ref{classe0} shows that all sources, except S201-d, should be Class~0 sources according to this indicator; see also our reservation concerning M$_*$ in Sect.~5.) 
Enoch et al. (\cite{eno09}) 
find that the spectral index $\alpha$(2~$\mu$m--10~$\mu$m) shows a large dispersion for the Class~I sources; we observe the same trend for the index $\alpha$(24~$\mu$m--100~$\mu$m). 
However this index is high, $\geq$1.2, for all our candidate Class~0 sources.

Several other sources for which we have less measurements are also possibly Class~0 sources. They are BRC14-d, BRC14-i, and BRC14-j, which have no 
{\it Spitzer} counterpart (even at 24~$\mu$m) and thus present a very high spectral index ($\geq$2); their masses, estimated from their 250~$\mu$m integrated flux assuming a cold envelope of temperature 15.7~K  are respectively 16.8~$\msol$, 10.6~$\msol$, and 11.1~$\msol$. Also the two bright 100~$\mu$m sources s1 and s2 at the waist of 
Sh~201 could be Class~0 sources. 

Most of these candidate Class~0 sources have massive envelopes and, by accretion, may become massive stars (S201-h is an exception).\\  

S201-a, a distant background source (Sect.~6.4; distance of 5.45~kpc), appears as a Class~I object, with  T$_{\rm bol}$=113~K, M$_{\rm env}$=32.6~$\msol$, 
L$_{\rm bol}$=2460~$\lsol$. Its high luminosity suggests that it harbors a cluster containing one or several class~I YSOs. It is similar to the condensation at the head of BRC13 which harbors a cluster with Class~I and Class~II sources: same T$_{\rm bol}$ (112~K for BRC13-a$^{***}$), same spectral index (0.30 for S201-a, 0.37 for BRC13-a$^{***}$) and high luminosity (1400~\lsol\ for BRC13-a$^{***}$).
 
\subsection{The CO outflows}

CO outflows are a good tracer of ongoing embedded star formation. Class~0/I objects are very often associated with CO outflows in 
star forming regions (for example Curtis et al. \cite{cur10} for the Perseus region). Using JMCT HARP CO(3-2) observations, 
Ginsburg et al. (\cite{gin11}) detected 18 candidate CO outflows in the field mapped by {\it Herschel}. Table~\ref{CO} lists 
some parameters of these outflows and their possible {\it Herschel} counterparts. Columns 1 to 6 give the identification, the center coordinates, 
the momentum, the energy, and the dynamical age of the components of the outflows. In column 7 we identify the central sources at the origin of the outflows as 
one of our point sources; this identification is based on the proximity between the center of the flow and this source; the distance separating these positions is 
given in column 8. (We consider that the association is probable if the distance YSO--center of the outflow is $\leq$12$\arcsec$). 
Columns 9 and 10 give the evolutionary stage of these sources.
Eleven of the 18 CO outflows are confidently bipolar; the BRC14 region has many blue and red lobes, and according to Ginsburg et al. (\cite{gin11}) confusion prevents pairing. Two or possibly three of the bipolar outflows are probably associated with a candidate Class~0 YSOs (S201-g, BRC14-a3, S201-s1?), six with a Class~I YSO (BRC12-b, BRC12-d, IB-b, IB-d, BRC14-c1, S201-a), and one with a cluster (BRC13-cluster; this cluster contains several Class~I YSOs).  Furthermore, we propose to pair outflows \#31 and \#32 (respectively blue and red components) and to associate them with BRC14-g or -h, as they are situated close to and 
on each side of these condensations associated with possible Class~0/I sources. We do not identify clear counterparts for outflows \#22, \#23, \#27, \#28, \#29, \#30, and \#34. Thus we propose a {\it Herschel} counterpart for 12 over 13 bipolar outflows (the one exception is outflow \#34).

Ginsburg et al. (\cite{gin11}) have detected a total of 40 candidate CO outflows associated with W5-E and W5-W. Outflow \#26, associated with BRC14-c1 
(also AFGL~4029-IRS1) is the most powerful of their sample (it has the highest momentum and highest energy, and a high momentum flux).
Outflows \#20 (BRC12-b) and \#21 (BRC12-d) are also powerful.\\

Some other CO outflows detected by Ginsburg et al. (\cite{gin11}) have velocities inconsistent with the W5 complex velocities. Outflow \#53 is one of them;  
its central velocity of $-$59.7~km~s$^{-1}$ indicates, according to the authors, a kinematic distance of 5.45~kpc.  This velocity placed it in the outer arm, 
far away from W5-E. The center of this outflow is offset from the S201-a {\it Herschel} point source by 9.2~$\arcsec$. They are probably associated, 
and this association can possibly explain the 
peculiar properties of S201-a (its relatively hot dust temperature and low envelope mass - if placed at a distance of 2.0~kpc).

\begin{table*}[tb]
\caption{CO outflows and their Herschel counterparts}
\begin{tabular}{lrrrrrlll}
\hline\hline
CO name & RA & Dec & Momentum            & Energy           & Age         & IR counterpart & d\tablefootmark{a}       & Stage  \\
        &    &     & (\msol~km~s$^{-1}$) & (10$^{42}$~ergs) & (10$^4$~yr) &                & (\arcsec) &       \\
\hline
20r	& 43.755678	& 60.597782 & 0.50 & 46.3 & 0.5 & BRC12-b       & 8         & Class I     \\
20b	& 43.759476	& 60.596599 & 0.33 & 26.6 & 0.5 & idem          & 7 (b)       &                \\
21b	& 43.967801	& 60.628583 & 0.58 & 41.4 & 1.7 & BRC12-d       & 12        & Class I      \\ 
21r	& 43.973007	& 60.631990 & 0.08 & 4.3  & 1.7 & idem          &           &                 \\ 
23b	& 44.171755	& 60.769192 & 0.03 & 0.9  & 4.5 & IB-b?         & 20        & Class I          \\
23r	& 44.177242	& 60.761364 & 0.03 & 1.0  & 4.5 & idem          & 7 (r)       &                  \\
24r	& 44.259049	& 60.698702 & 0.30 & 26.1 & 1.7 & IB-d          & 9.5       & Class I        \\
24b	& 44.267036	& 60.694962 & 0.28 & 18.3 & 1.7 & idem          & 6.5 (b)     &                    \\ 
25b	& 45.229989	& 60.675872 & 0.09 & 6.8  & 1.0 & BRC13         &           & ? (cluster)                    \\
25r	& 45.236467	& 60.677109 & 0.35 & 23.0 & 1.0 & idem          &           &                   \\
26b	& 45.372510	& 60.487571 & 0.24 & 26.1 & 1.1 & BRC14-c1      & 3.5       & Class~I         \\ 
26r	& 45.384479	& 60.487318 & 0.85 & 106.0 & 1.1 & idem         &           &                    \\
31	& 45.423500	& 60.477057 & 0.03 & 0.7  &     & BRC14-g or h  &           & ? (cluster)      \\
32	& 45.437614	& 60.483742 & 0.18 & 14.5 &     & BRC14-g or h  &           &                    \\
33b	& 45.452775	& 60.412993 & 0.14 & 10.1 & 3.0 & BRC14-a3      & 11        & Class 0/I        \\
33r	& 45.445968	& 60.401839 & 0.20 & 16.1 & 3.0 & idem          & 10.5 (r)    &                     \\ 
35r	& 45.836843	& 60.471159 & 0.12 & 11.0 & 1.3 & S201-s1       & 8.5       & ? (Class 0)        \\
35b	& 45.837186	& 60.466721 & 0.12 & 11.0 & 1.3 & idem          & 5 (b)       &                      \\
36b	& 45.858773	& 60.483234 & 0.19 & 15.8 & 2.7 & S201-g        & 7.5       & Class 0          \\
36r	& 45.855888	& 60.475103 & 0.10 & 6.8  & 2.7 & idem          &           &                      \\ 
37r	& 45.886094	& 60.471649 & 0.04 & 1.6  & 2.1 & S201-e        & 12(r)      & ?        \\   
37b	& 45.896772	& 60.474332 & 0.06 & 4.2 & 2.1  & idem          &           &                    \\
\hline
53b & 45.952487 & 60.338230 & 3.16 & 194.0 & 4.7 & S201-a       & 9.2       & Class~I  \\
53r & 45.957001 & 60.339161 & 0.31 & 8.5   & 4.7 & idem         &           &          \\
\hline
\label{CO}
\end{tabular}
\tablefoot{
\tablefoottext{a}{Distance from the YSO to the center of the CO outflow, or to the blue (b) or red (r) CO component.}
}
\end{table*}

\subsection{The YSOs color-selected IRAS sources of Karr \& Martin (\cite{kar03})} 

IRAS sources with colors of YSOs were selected by Karr \& Martin (\cite{kar03}) to discuss star formation in W5. Twenty six YSOs were proposed  
by these authors in the field covered by the {\it Herschel} observations (see their table 4). We find that most of these IRAS sources are not YSOs but extended 
structures, bright fragments of PDRs or filaments. Only 3 of their sources correspond to YSOs detected by {\it Herschel}: IRAS02511+6023, IRAS02551+6042, and IRAS 02598+6008 which lie respectively in the direction of BRC12-b, I12, and S201-a. The sources IRAS02570+6028, IRAS02576+6017, and IRAS02593+6016 correspond 
to the extended sources associated with BRC13, BRC14, and Sh~201. IRAS02531+6032 lies in the direction of a bright 100~$\mu$m structure present in the center of the A \HII\ region, just above the exciting star ex1 (see Appendix C).

\subsection{Candidate prestellar condensations}

\begin{figure*}[tb]
\sidecaption
 \includegraphics[width=12cm]{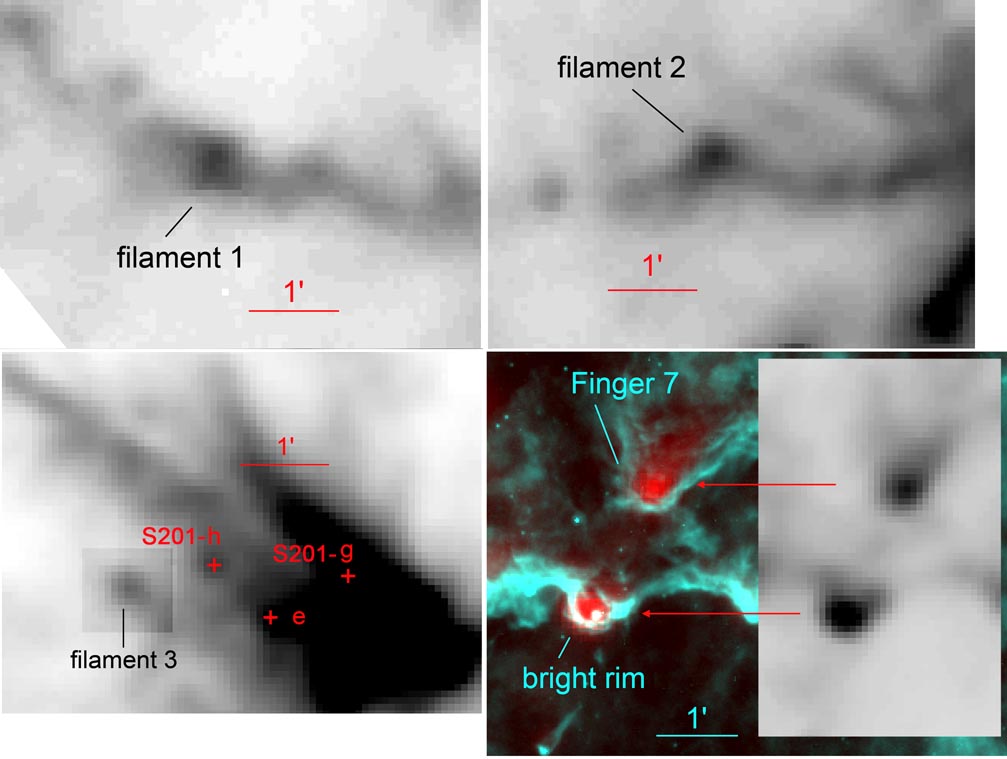}
  \caption{Identification of some of the candidate prestellar condensations discussed in the text. The grey images show the column density.
  In the colour image (bottom right), red is the column density map and turquoise is the {\it Spitzer} 8.0~$\mu$m image.  The ``bright rim'' feature is 
  associated with the {\it Herschel} source I-5 discussed in Appendix~E.}
  \label{compact}
\end{figure*}

Some compact condensations appear at {\it Herschel}-SPIRE wavelengths. They are resolved at 250~$\mu$m, with diameters in the range 0.2~pc -- 0.4~pc. 
They have no {\it Herschel}-PACS counterparts at 100~$\mu$m, and thus they may be prestellar. Some of these sources are located at the tip (or head) of fingers pointing toward 
exciting stars, some other are located along filaments.  We identify some of these sources in Fig.~\ref{compact}, and the others in Fig.~\ref{BRC12a} and Fig.~\ref{global}.  
In Table~\ref{divers} we give their coordinates, temperatures (from the temperature map), peak column densities, fluxes at 350~$\mu$m integrated in a 
circular aperture of radius 25$\arcsec$ or 30$\arcsec$ (column 6), and masses estimated from these temperatures and fluxes. 
These compact condensations have masses of a few solar masses. We do not know if they are gravitationally stable or not; they are potential 
sites of low-mass star formation.

\begin{table*}[tb]
\caption{Parameters of some compact condensations, candidate prestellar condensations. R is the radius of the aperture used to measure the flux.}
\begin{tabular}{llllllll}
\hline\hline
Name & RA(2000) & Dec(2000) & $T_{\rm dust}$ & N(H$_2$) & R & Flux(350~$\mu$m) & $M$ \\
     & ($\degr$) & ($\degr$) & (K)        & (cm$^{-2}$)  & ($\arcsec$) & (Jy)             & (\msol ) \\          
\hline
Finger \#1 & 44.7074 & +61.0495 & 20.3 & 1.6$\times$10$^{21}$ & 30 & 2.21 & 4.0 \\
Finger \#1bis & 44.7348 & +61.0753 & 20.0 & 1.5$\times$10$^{21}$ & 30 & 2.3 & 4.1 \\
Finger \#2 & 45.8231 & +60.9639 & 20.3 & 2.3$\times$10$^{21}$ & 30 & 3.00 & 5.5 \\
Finger \#5 & 45.5943 & +60.7624 & 21.0 & 2.0$\times$10$^{21}$ & 30 & 1.48 & 2.7 \\
Finger \#6 & 45.7089 & +60.6599 & 21.5 & 1.5$\times$10$^{21}$ & 30 & 0.85 & 1.4 \\
Finger \#7 & 44.6145 & +60.7526 & 19.8 & 3.2$\times$10$^{21}$ & 30 & 2.43 & 4.8 \\
\hline
Filament \#1 & 44.1416 & +60.4838 & 19.7 & 1.9$\times$10$^{21}$ & 30 & 2.03 & 4.1 \\
Filament \#2 & 45.9890 & +60.3777 & 19.7 & 2.6$\times$10$^{21}$ & 30 & 1.84 & 3.7 \\
filament \#3 & 45.9444 & +60.4745 & 18.3 & 2.2$\times$10$^{21}$ & 25 & 1.05 & 2.5 \\
\hline
BRC12-cond\#1 & 43.8066 & +60.7135 & 21.1 & 3.3$\times$10$^{21}$ & 25 & 3.08 & 5.2 \\
BRC12-cond\#2 & 43.8665 & +60.6674 & 22.4 & 3.7$\times$10$^{21}$ & 25 & 2.68 & 4.0 \\
BRC12-cond\#3 & 44.0554 & +60.6516 & 18.9 & 3.8$\times$10$^{21}$ & 25 & 2.14 & 4.7 \\
BRC12-cond\#4 & 43.8695 & +60.6050 & 20.1 & 3.5$\times$10$^{21}$ & 25 & 3.10 & 5.9 \\
\hline
\label{divers}
\end{tabular}\\
\end{table*} 

\subsection{The history of star formation in W5}

The radio-continuum flux at 1420~MHz of the W5-E \HII\ region has been measured by Karr \& Martin (\cite{kar03}) and found to be 30~Jy. 
Using Equation~(1) in Simpson and Rubin (\cite{sim90}) we estimate the ionizing photon flux to be $\sim$1.0$\times$10$^{49}$~s$^{-1}$ 
(for an electron temperature of 7000~K as assumed by Karr \& Martin). According to Martins et al. (\cite{mar05}) the ionizing 
flux of an O7V star, the exciting star of W5-E, is 4.27$\times$10$^{48}$~s$^{-1}$, which is too faint by a factor of two to account for the 
ionization of the gas.  Additionally, the observed extended 24~$\mu$m emission inside the central cavity may represent hot dust
that absorbs some of the ionizing photons. Thus we are probably missing some exciting stars\footnote{Karr \& Martin (\cite{kar03}) used the 
calibration of Vacca et al. (\cite{vac96}) for massive stars, which gives an 
ionizing flux of 1.3$\times$10$^{49}$~s$^{-1}$ for an O7V star. Thus they concluded that HD~18326 was the main exciting star of the W5-E \HII\ region. 
With the calibration of Martins et al. (\cite{mar05}) a O7V star does not provide enough ionizing photons. HD~18326 belongs to a cluster studied by 
Chauhan et al. (\cite{cha11a}). It is possible that this cluster, containing one O7V star, also contains several late O or early B stars 
contributing to the ionization of W5-E. For example BD+60$\degr$0606 located 1.1~pc away from HD~18326 has been classified as a B0e star 
(Voroshilov et al. \cite{vor85}).}. In the following we adopt a photon flux of 10$^{49}$~s$^{-1}$ for the ionizing flux of W5-E. Assuming 
that W5-E is a spherical \HII\ region of radius 12~pc, we estimate its electron density to be 13.5~cm$^{-3}$ and the ionized gas mass to be 3100~\msol. \\

The sound crossing time, assuming a sound speed  $\sim$10~km~s$^{-1}$ in the ionized gas, is $\sim$1.2~Myr, which gives a minimum limit for the age of W5-E.
The lifetime of an O7V star on the main-sequence gives the maximum limit, $\sim$8~Myr, according to the models of rotating stars by 
Meynet \& Maeder (\cite{mey03}, \cite{mey05}).  Estimating the age of a cluster is a difficult task because it requires accurate 
photometry and isochrones. As noted by Chauhan et al. (\cite{cha11a}) the use of different isochrones can introduce a systematic shift in 
the age. Nakano et al. (\cite{nak08}), using {\it g',i'} photometry and the isochrones of Palla and Stahler (\cite{pal99}), 
estimate that the mean age of the young H$\alpha$ emission stars located near the exciting star is $\sim$4~Myr. Chauhan et al. (\cite{cha11a}), 
using $V,I_{\rm c}$ photometry and the PMS isochrones of Siess et al. (\cite{sie00}), conclude that 
star formation in the cluster is non-coeval, extending over 3.5~Myr, with a mean age of 1.3~Myr. Thus no straightforward conclusions are 
obtained about the age of the exciting cluster of W5-E; it probably lies in the range 1~Myr -- 4~Myr. \\

AND12 estimate that about 8000~\msol\ of molecular material are associated with the W5-E \HII\ region.  
They estimate this mass from the 350~$\mu$m flux density integrated in an irregular aperture of area 2515~(\arcmin)$^2$, not very different from a circular 
aperture of radius 16.5~pc (the Sh~201 \HII\ region lies outside this aperture). We assume that this molecular material was collected around the 
ionized region during its expansion. If W5-E has evolved in a homogeneous medium of density 
$n_o$~atoms~cm$^{-3}$, we can estimate $n_o$ by equating the mass initially in the sphere of radius 16.5~pc to the mass now present in both the 
ionized region and its collected shell (3100~\msol + 8000~\msol). We obtain $n_o\sim20$~atoms~cm$^{-3}$. This very low value is unrealistic: 
such a large \HII\ region cannot have formed and evolved during all its life into such a low density medium. (See also fig. 14 in AND12 which shows that W5-E differs 
from the other regions of their sample, as it has collected little material.)

\begin{figure*}[tb]
\sidecaption
\includegraphics[width=12cm]{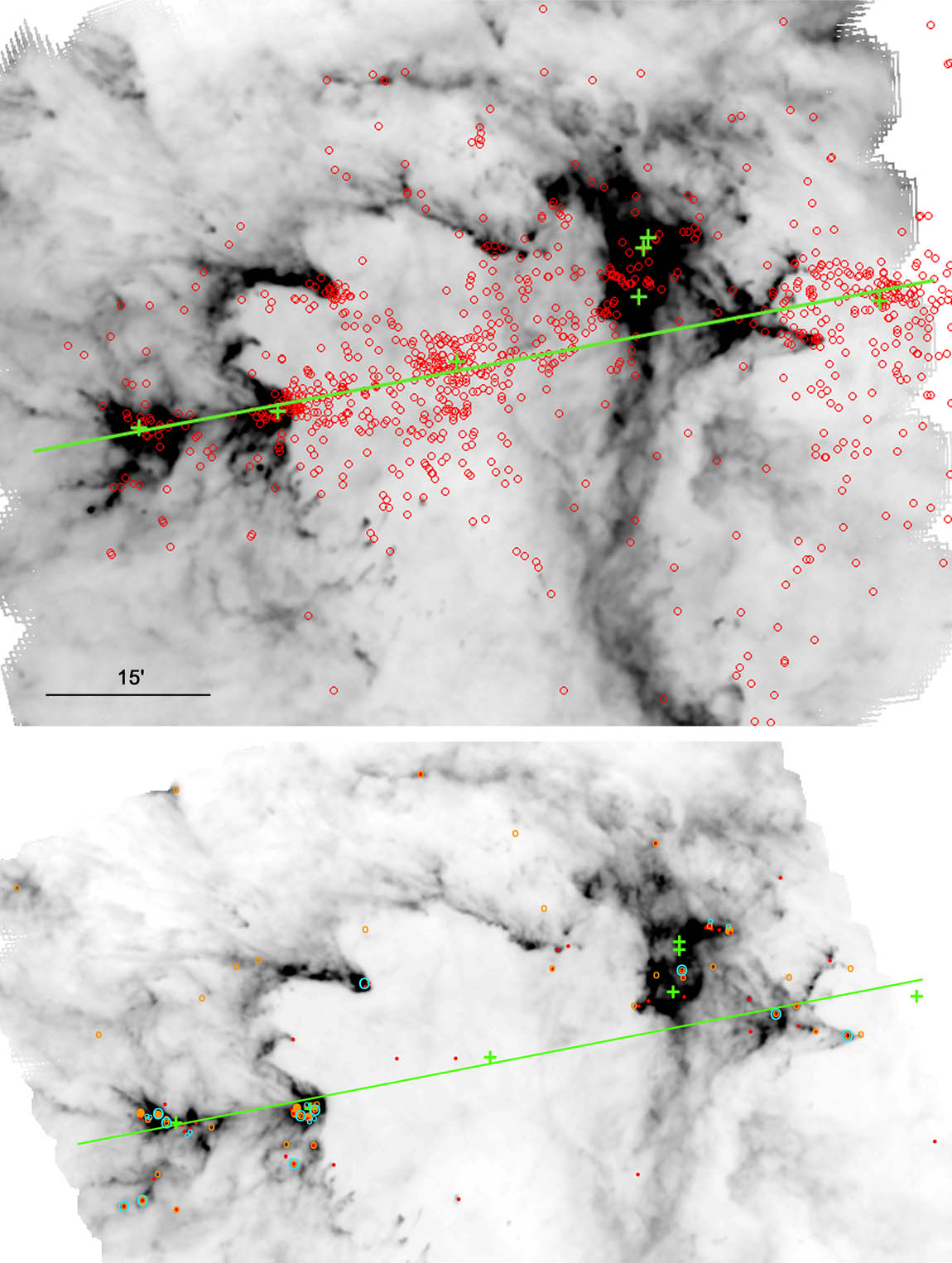}
\caption{Distribution of YSOs in W5-E. {\it Top:} Class~II YSOs and exciting stars. The red circles are the YSOs identified as Class~II sources 
  by KOE08. The green crosses are the massive exciting stars of the \HII\ regions (from east to west) Sh~201, AFGL~4029, W5-E, 
  A \& B between W5-E and W5-W, and W5-W. The green line shows the east-west filament inside which we suggest that the exciting stars of W5-E and W5-W formed first. 
  The underlying grey-scale image is the {\it Herschel}-SPIRE 350~$\mu$m map. {\it Bottom:} The youngest sources, Class~I and candidate Class~0 YSOs. 
  The red dots are the YSOs identified as Class~I sources by KOE08. The orange circles are the 100~$\mu$m point sources, mostly Stage~I YSOs. The large full orange 
  circles are our candidate Class~0 sources. The blue circles are for the CO outflows detected by Ginsburg et al. (\cite{gin11}). 
  The underlying grey-scale image is the column density map.}
  \label{pdr}
\end{figure*}

\begin{figure}[tb]
\centering
\includegraphics[width=85mm]{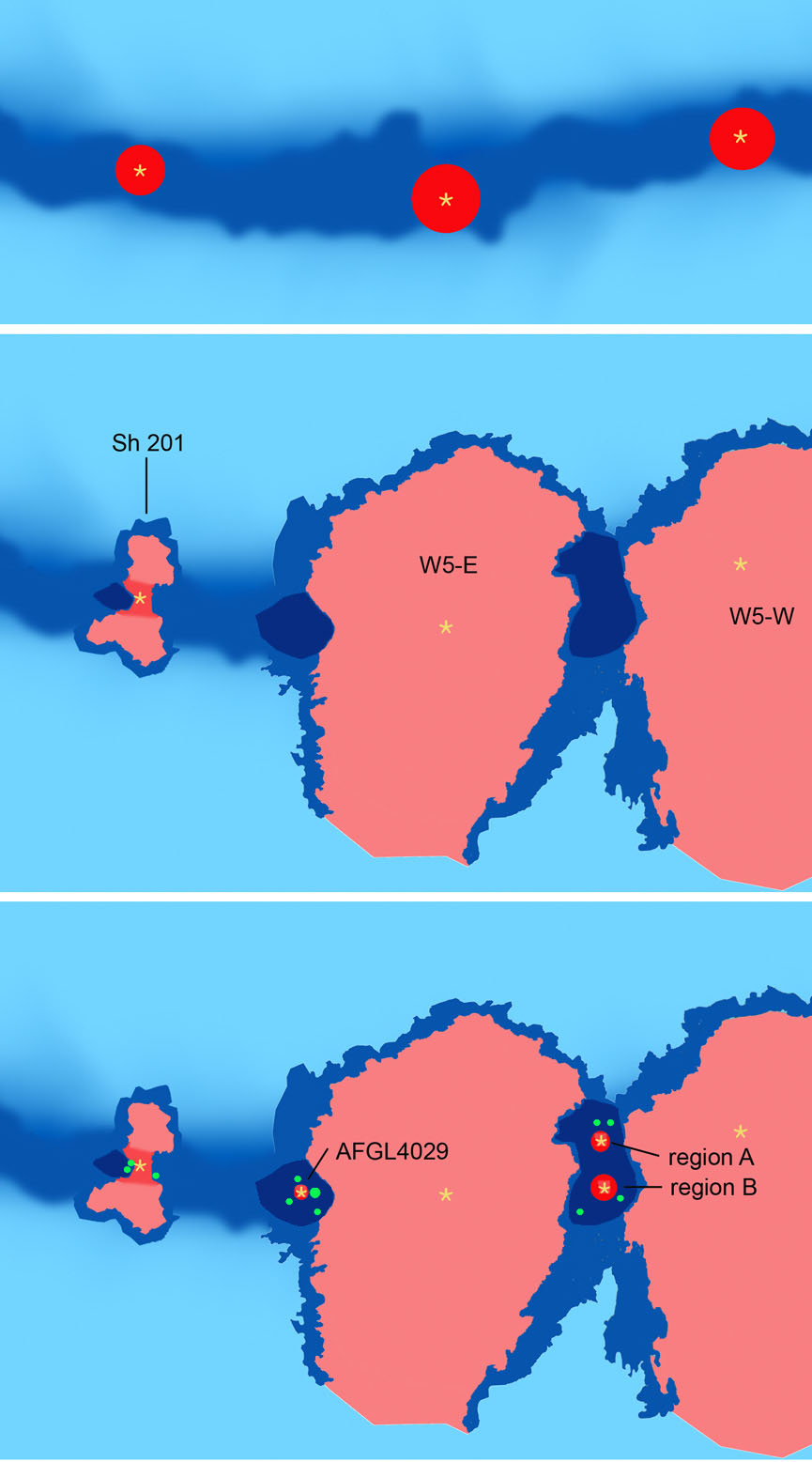}
 \caption{Scenario of star formation in the W5 region. {\it Top:} The exciting stars of Sh~201, W5-E and W5-W formed in a dense filament. The exciting star of Sh~201 formed in the middle of the filament whereas those of W5-E and W5-W formed on its border. {\it Middle:} The expanding \HII\ regions have bursted out of the filament. In Sh~201 the ionized gas flows away of the filament in two opposite directions, forming a bipolar \HII\ region. {\it Bottom:} A new generation of stars is forming in the densest condensations, adjacent to the 
  first generation \HII\ regions.}
   \label{scenario}
\end{figure}

An alternative situation is suggested by the distribution of the Class~II YSOs, as observed by KOE08.                                                                         
These authors found that the Class~II sources are clustered along filamentary structures. Their figures 7 and 12, and our Fig.~\ref{pdr} show clusters of Class~II YSOs 
forming an east-west filament going through Sh~201, BRC14, HD~18326 , the A and B \HII\ regions between W5-E and W5-W, BRC12, to BD+60$\degr$0586, 
and even possibly to HD~17505 and HD~17520 (also BD+59$\degr$0552 and BD+59$\degr$0553, other exciting stars of W5-W). This suggests that a massive and probably dense filament 
or sheet was originally present, and that the exciting clusters of W5-E and W5-W formed inside, 
a few megayears ago (see Fig.~\ref{scenario}). Their associated \HII\ regions subsequently
grew in size, first inside this dense filament, then opening toward the outside, on the low density southern side (figure~7 in KOE08 and our Fig.~\ref{radio} show 
that the general morphology of W5-E and W5-W are similar). During their expansion these \HII\ regions accumulated neutral material at their borders, 
remaining mostly ionization bounded on the north side (and density bounded on the south side for W5-E). 

The \HII\ region Sh~201 fits well with this scenario (Fig.~\ref{scenario}). Its morphology suggests that its exciting star formed near the axis of the filament. 
During the expansion of the \HII\ region the IF reached almost simultaneously the north and south low-density borders of the filament. The ionized gas 
expanding away from the dense regions, in two directions perpendicular to the filament, is at the origin of the bipolar structure of Sh~201 in agrement with 
the simulations of Bodenheimer et al. (\cite{bod79}). The two ionized lobes are surrounded by 
neutral material collected during the expansion. The dust located in this collected material, close to the IF, heated by the radiation leaking from the \HII\ region, 
is observed at 24~$\mu$m and 100~$\mu$m. Colder dust, located in the collected material but farther away from the IF, is observed 
at all {\it Herschel}-SPIRE wavelengths.\\

It is not clear presently if the late OB stars of the field, those exciting Sh~201, the AFGL~4029 cluster, and the A and B \HII\ regions,   
formed at the same time as the exciting clusters of W5-E and W5-W or later. Are these massive stars first- or second-generation stars? 
The associated \HII\ regions are much smaller than W5-E and W5-W, and thus are possibly younger, but their exciting stars are also less massive than those of W5-E   
(O9-B1V versus O7V). Also, they possibly
 formed at the center of the dense filament and/or inside dense condensations, which could explain their smaller sizes. 
The fact that the AFGL~4029 cluster lies in the core enclosed by BRC14, and that the A and B \HII\ regions lie in the region of possible compression  
between W5-E and W5-W could indicate that they are triggered second-generation massive stars. Only accurate age determinations could answer this question, but the  
required accuracy is presently far from being achieved\footnote{It is almost impossible to estimate the dynamical age of an \HII\ region. 
For this it is necessary to know at each time the physical conditions (especially the density) of the medium into which the \HII\ region evolves (and \HII\ regions never evolve into an homogeneous medium). The medium that we see presently has been highly modified by ionization. If we want to know the age of an \HII\ region we must estimate the age of its exciting star(s). With the presently available tools we need accurate photometry or deep spectra at UV or optical wavelengths; an accuracy better than 1~Myr is very difficult to achieve. It is why we are unable to determine if AFGL~4029 or the A and B \HII\ regions are younger than HD~18326, and thus if they are first- or second-generation \HII\ regions.}.\\

Fig.~\ref{pdr} shows the distribution of the youngest sources: Class~I YSOs according to KOE08 (small red dots), 
100~$\mu$m point-like sources, most of them Stage~I YSOs (small orange circles), our candidate Class~0 sources (large and full orange circles); it shows also the location of the  CO outflows (Ginsburg et al. \cite{gin11}; blue circles). These sources are all young, much younger than the exciting star of W5-E, and younger than the 
Class~II YSOs. A timelife of about 1--2$\times$10$^5$~yr is generally estimated for low-mass Class~I YSOs (Andr\'e et al. \cite{and00} and references therein); Class~0 are younger,  with a lifetime of a few 10$^4$~yr (Maury et al. \cite{mau11}). Fig.~\ref{pdr} shows that very few Class~I sources are observed in the direction of 
the ionized region ($\leq$10\%)\footnote{Two Class~I sources, \#16221 and \#15522, lie in the center of W5-E. They have no 100~$\mu$m counterparts. Source \#16221 has no 24~$\mu$m emission; thus, its classification is doubtful. Three Class~I sources lie in the center of Sh~201, sources \#5, \#6, and \#7; we have seen in Appendix~C that their classification is doubtful.}. 
Our study confirms the finding of KOE08 that the ratio of Class~II to Class~I YSOs is much higher in the direction of the ionized cavities than in the direction 
of the surrounding molecular material. We find that the youngest sources are mostly observed 
in the regions of higher column density, in condensations or filaments; they are preferentially located along the previously mentioned east-west filament, in the condensations adjacent to the \HII\ region Sh~201, in the condensations enclosed by the bright-rimmed clouds, and in the region of compression 
nearby the A and B \HII\ regions. 

Another characteristic of the young population is the lack of luminous sources. All the confirmed Stage~I sources have luminosities smaller than a few hundred solar luminosities, and central objects of only a few solar masses. The only locations of high-mass star formation are the massive condensation enclosed by BRC14 (with the AFGL4029 cluster 
already formed and c1), and the massive condensation at the east waist of Sh~201 (with the source s1, if confirmed). Presently we are in a phase of low- to intermediate-mass star formation. Is it the end of massive-star formation in this region? We cannot yet answer this question, as we note that there are two massive condensations that contain enough mass for further high-mass star formation.

\subsubsection{Processes of triggered star formation}

Several processes can trigger star formation on the borders of \HII\ regions. Four of them are described and illustrated in Deharveng et al. (\cite{deh10}; their fig.~4). 
They are: 1) small-scale instabilities of various origins in the collected layer; 2) large-scale gravitational instabilities along the collected layer leading to the formation of massive fragments (the collect and collapse process); 3) ionization radiation acting on a turbulent medium; 4) radiation-driven compression of pre-existing dense clumps (the RDI process). 
It is often difficult to prove that star formation has been triggered because the age difference between two stellar populations is difficult to estimate with accuracy 
(for example Martins et al. 2010), and we are often left with only morphological arguments. Furthermore it may be difficult to identify the processes at the origin of the 
triggering.  What can we learn from the {\it Herschel} observations of W5-E?

Neutral material has been collected around the W5-E \HII\ region, about 8000~$\msol$ according to AND12. It is observed for example in the direction of PDR1 and PDR2,  
at the south border of BRC13, and at the periphery of BRC12. The column density in these structures ($\leq$3$\times$10$^{21}$~cm$^{-2}$), however, is not high enough to 
allow triggered star formation via the collect and collapse process (according to Whitworth et al. \cite{whi94} who estimate that the collapse 
occurs when the column density is higher than about 6$\times$10$^{21}$~cm$^{-2}$). 
Collected neutral material is also observed around the bipolar nebula Sh~201. Around the bipolar lobes the column density is 
rather low,  $\leq$10$^{21}$~cm$^{-2}$, and no star formation is observed. The column density is much higher at the waist of the nebula. There, we hypothesize that it is the 
material of the dense parental filament that has been collected. 
Several candidate Class~0 and Class~I YSOs are detected in this dense collected material (S201-f, -g, -h, -s1, -s2; Fig.~\ref{S201c}, Fig.~\ref{taille}); CO outflows are present as well.
These YSOs are second-generation sources, probably younger than 10$^5$ yr. Some of these Class~0/I or candidate starless cores are massive 
(S201-g and possibly S201-s1-s2) and possibly susceptible to transform via accretion into massive stars. We are dealing with triggered star formation in 
the act, in the dense layer present at the waist of the Sh~201 \HII\ region. But, due to the poor resolution, the exact process operating is difficult 
to identify.

Triggered star formation is possibly observed in other locations:

$\bullet$ In the pillars, on the east and south-east border of W5-E. As discussed in Sect.~6.2 the best explanation for these numerous structures seems to be the interaction 
of the ionizing radiation of the central exciting star with a low-density turbulent medium. Some low-mass YSOs have already formed at the tip of these structures, 
which are observed as Class~I or Class~II sources. Low-mass condensations are also detected at the head of the pillars, which are potential sites of future low-mass star formation.  We suggest that this process may explain the dispersed population of low-mass Class ~II YSOs detected by KOE08. This process possibly leads to a continuous star formation: as shown by Gritschneder et al. (\cite{gri09a}) the ionizing radiation of an O star provides a good mechanism to substain and even drive turbulence in the parental molecular cloud (the turbulence does not decay with time), and thus, this star formation process may remain at work during a long period.

$\bullet$ In the vicinity and possibly inside the bright-rimmed clouds. BRC13 and BRC14 enclose massive condensations that harbor clusters and several Class~I YSOs. 
BRC12 is probably different: it is bordered by collected material, but does not enclose a massive condensation; it harbours several individual Class~I YSOs but no cluster.   
Star formation in the vicinity of bright rims has possibly been triggered by the ionizing radiation (i.e. by the expanding \HII\ region), 
as shown by several indicators:
\begin{enumerate}
\item The morphology of these BRCs, pointing towards the exciting star

\item The comparison of the pressure of the gas inside and outside the rims. Morgan et al. (\cite{mor04}), using radio-continuum observations, 
estimated the density and the pressure in the dense ionized layer bordering the BRCs. Morgan et al. (\cite{mor08}), using SCUBA observations, 
estimated the temperature and the density of the molecular condensations enclosed by the BRCs. They showed that the pressure of the ionized 
gas is higher than that of the molecular gas in BRC12, BRC13, and BRC14. As a 
consequence, an ionization front of D type and its associated shock front progress inside these structures, compressing material and thus providing favorable 
conditions for star formation.

\item The observed ``small-scale sequential star formation'' in the vicinity of the rims. This was first advocated by Sugitani et al. (\cite{sug95}) 
to explain the distribution of near-IR sources associated with BRCs.  They observed that the bluer sources, thus the older, were located closer to the
 exciting stars, sometimes outside the BRCs, while the redder sources, the younger, were located inside the BRC close to the enclosed IRAS source. This
 suggests a propagation of star formation from the side of the exciting star to the IRAS position. BRC12, BRC13, and BRC14 are 
 among the BRCs that have this configuration (Sugitani et al. \cite{sug95}). 
 The same effect was observed with the H$\alpha$ emission stars, which are pre-main-sequence stars of T Tauri type, and thus are low-mass Stage~II YSOs. H$\alpha$ 
emission stars have been identified in the vicinity of BRC12, BRC13, and BRC14 (Ogura et al. \cite{ogu02}). These stars are concentrated near the tip 
of the BRCs, most often just outside the rims, on the side of the exciting stars (see for example Appendix~A and Fig~A.1). Several photometric studies of these near-IR stars
 and/or these H$\alpha$ emission stars aim at estimating their ages (Ogura et al. \cite{ogu07}; Chauhan et al. \cite{cha09}; 
Chauhan et al. \cite{cha11a}). The recent study by Chauhan et al. (\cite{cha11a}), which also concerns mid-IR {\it Spitzer} sources, shows that in all the BRCs
 the YSOs lying on or inside the rim are younger (by several 10$^5$~yr; Appendix A \& B) than those located outside it. This sequential 
 small-scale star formation suggests that star formation occurs in the vicinity of the IF bordering the BRCs, and moves with it away from the central exciting star.
\end{enumerate}

The bright-rimmed clouds differ from the pillars in terms of size, mass enclosed, and stellar content (clusters versus isolated low-mass YSOs). Thus, they probably differ in origin, being due not to turbulence but more probably to pre-existing denses structures, condensations or filaments affected by the stellar radiation\footnote{To summarize we suggest that pillars are due to stellar radiation acting on a low-density turbulent medium, whereas bright-rimmed clouds result from the action of stellar radiation on pre-existing denses condensations}. The RDI process is generally proposed to explain star formation in their vicinity. But the whole picture remains unclear. 
Several simulations of the RDI process have recently been presented and discussed (references in Sect. 6.2). They all include self-gravity, but only the Haworth \& Harries' simulations take into account the diffuse 
radiation field. They also differ in their initial conditions: the mass (5~$\msol$ to 96~$\msol$), size (0.3~pc to 1.6~pc), density distribution in the pre-existing globule 
(uniform or centrally peaked), and the ionizing photon flux reaching the globule. 
Thus it is difficult to draw straightforward conclusions. All simulations (except those of Miao et al. \cite{mia09}) point to the importance of the ionizing photon flux. 
We estimate that these fluxes are $\leq$1.5$\times$10$^9$~s$^{-1}$~cm$^{-2}$ and $\leq$0.9$\times$10$^9$~s$^{-1}$~cm$^{-2}$ respectively at the center of the condensations 
enclosed by BRC13 and BRC14 (they may even be smaller if there is a projection effect). For all simulations this is a case of low ionizing photon flux, and the  
consequence for all simulations is that the evolution of the globule is slow, and that star formation is delayed (it occurs after $\sim$0.6~Myr 
in Gritschneder et al. \cite{gri09b}; this is compatible with the ages estimated by Chauhan et al. \cite{cha11a} for the enclosed clusters). The BRCs' morphology 
is more difficult to account for. BRC13 and BRC14 enclose massive and dense condensations (respectively $\sim$22~$\msol$ in a radius of 0.29~pc and more than 
200~$\msol$ in a radius of 0.48~pc); the associated clusters lie in the very centers of the condensations, far from the bordering IF (respectively at 0.29~pc and 0.81~pc 
from the IF in BRC13 and BRC14; see Fig.~\ref{BRC13} and \ref{AFGL4029}; this distance may be smaller if there is a projection effect). This is not predicted by  
the simulations which show star formation at the tip of a 
filamentary feature (the ultimate fate of a globule before dispersion). Also, no simulations consider a globule massive enough to allow the formation of a whole cluster, as is 
observed in BRC14. Simulations considering the action of a low ionizing photon flux falling upon a massive condensation or filament are missing. On the 
other hand, the {\it Herschel}-SPIRE images lack the resolution necessary to show the exact morphology of the enclosed condensations. As a consequence, if it is clear that 
the BRC13 and BRC14 structures are presently shaped by the UV radiation of HD~18326, it is difficult to determine if the formation of the enclosed clusters  
has been triggered or not by this radiation. 

The configuration of the east condensation at the waist of of Sh~201 has similarities and differences with the condensation enclosed by BRC14. Here again the condensation is massive and dense ($\sim$230~$\msol$ in a radius of 0.48~pc), and is bordered and compressed by ionized gas on one side. But the similarities stop there: the ionizing photon flux is much higher, 
of the order of 7$\times$10$^{10}$~s$^{-1}$~cm$^{-2}$, because the star lies much closer to the IF (if there is no strong projection effect). This condensation is experiencing high ionizing flux, 
leading to a rapid star formation (on a time scale of 0.1 -- 0.2~Myr according to Gritschneder et al. \cite{gri09b} or Bisbas et al. \cite{bis09}). The Class~0/I YSOs associated with 
this condensation (S201-g, S201-s1-s2) lie in the direction of the filaments separating the ionized and molecular material, and thus are very near to the IF (Fig.~\ref{taille}). They formed not 
at the center of the condensation but more probably in the compressed layer bordering the condensation. Triggering is more obvious than in the case of BRC14, 
possibly via the RDI process. But, here again, we do not have the resolution necessary to unravel the small scale morphology of the condensation and the exact location of the YSOs.

\section{Conclusions}

We have taken advantage of the very simple morphology of the W5-E \HII\ region to study star formation, and the possible impact 
of triggering. This region has various features interesting in terms of star formation: bright rims, pillars, 
and a nearby bipolar \HII\ region. For the first time, thanks to {\it Herschel}-PACS and SPIRE observations, we have access to the 
distribution of the cold dust over the whole region, tracing the dense molecular material inside which stars form. Also we have 
identified the youngest protostellar objects, the Class~0/I YSOs, via the emission of their cold envelopes. 

Some of our conclusions are still uncertain, for various reasons: i) the masses of the envelopes and condensations are 
uncertain because of uncertainties in the opacity of the dust. The nature of the dust varies with its location (or environment). 
How this affects its opacity is not yet well established; ii) we have shown that the SED fitting tool of Robitaille et al. (\cite{rob07}), 
which presently does not include dust colder than 30~K, does not allow us to accurately estimate the properties of the deeply embedded Class~I YSOs   
(especially those concerning their cold envelopes). New models will soon be available. Meanwhile it seems far better to use a 
modified blackbody model to fit the envelopes of Class~0/I YSOs; iii) some sources, such as S201-a, may not be part of the W5-E complex; 
only the knowledge of the source velocity (which requires high resolution molecular observations) could confirm the association. Furthermore 
a few associations between {\it Spitzer} and {\it Herschel} point sources are problematic due to the resolution of the 
{\it Herschel}-SPIRE observations; in some cases we may see with {\it Herschel} a compact condensation 
adjacent to a Class~I YSO rather than the YSO's envelope. This could explain why a number of SEDs are not well fitted 
(for example IB-c and IB-f in Appendix D, I-11 in Appendix E ).\\

Remembering these uncertainties our conclusions are the following:

$\bullet$ W5-E is a bubble \HII\ region: the ionized region is surrounded by a ring of PAHs emitting at 8.0~$\mu$m, observed by MSX and {\it Spitzer}. 
Several bubble \HII\ regions have been observed with {\it Herschel} (Anderson et al. \cite{and12}). W5-E differs from most of these regions, 
possibly because of its large size. As a whole its dust temperature is more uniform than that of the regions studied in Anderson et al. (\cite{and12}),
everywhere higher than 17.5~K, and its column density is rather 
low, everywhere lower than 10$^{23}$~cm$^{-2}$. Neutral material has been collected around the ionized region during its expansion. 
We observe this material at SPIRE wavelengths as a layer of cold dust adjacent to the IFs. The temperature in these PDRs is of the order of 21~K, and the 
column density is low, of the order of a few 10$^{21}$~cm$^{-2}$. According to the existing models, it is too low for star formation to proceed;  
very few YSOs are observed in these directions. 

The total mass of collected material is $\sim$8000~\msol\ (Anderson et al. \cite{and12}), which is low with respect to the volume 
of the \HII\ region, and differs from the other studied regions. We suggest that W5-E has not formed and evolved in a spherical homogeneous cloud, but in a dense 
filament (or sheet). One signature of this scenario is the distribution of Class~II sources (along a filamentary structure) according to Koenig et al. (\cite{koe08}).  
Another signature is the detection by {\it Herschel}-SPIRE of the remaining parental filament extending east-west over more than 15~pc, from 
BRC14 to regions east of Sh~201. 

$\bullet$ We have detected and measured 50 point-like sources and a few compact condensations. Their SEDs have been fitted using a modified blackbody model. 
The envelopes of the YSOs are cold, with a mean temperature of 15.7$\pm$1.8~K. The more extended condensations, harboring young clusters 
(enclosed by BRC13 and BRC14), are hotter, with a temperature around 26~K. The far away point source, S201-a, lying in the background 
of W5-E at 5.45~kpc, belongs to this last group.

$\bullet$ The masses of the envelopes of the YSOs detected by {\it Herschel} lie in the range 1.3~$\msol$ -- 47~$\msol$.

$\bullet$ The spectral index between 24~$\mu$m and 100~$\mu$m cannot, alone, be used to separate infallibly Class~I and Class~II sources. 

$\bullet$ We have used several indicators to identify the youngest YSOs, the Class~0 sources. The purely observational indicators, 
M$_{\rm env}$/L$_{\rm bol}$ and T$_{\rm bol}$, are probably the best ones in distant regions such as W5-E. Eight of the fifty 100~$\mu$m 
point sources are good candidate Class~0 YSOs; three more are possible candidate Class~0.  They are BRC14-c, BRC14-d, BRC14-i, BRC14-j, S201-c, S201-g, S201-i, 
IB-a (also possibly BRC14-a3, S201-s1, -s2).

$\bullet$ Eighteen CO outflows have been detected in the field observed by {\it Herschel} (Ginsburg et al. \cite{gin11}). Thirteen of them 
are bipolar. We found a {\it Herschel} point source counterpart for twelve of them (3 are possibly associated with a Class~0 YSO, 
6 with a Class~I, 2 with clusters containing Class~I YSOs, 1 with an YSO of uncertain evolutionary stage).

$\bullet$ Most of the Class~0/I YSOs are located in regions of high column density: 
near condensations enclosed by BRCs, near condensations at the waist of Sh~201, or in the vicinity of the compact A and B \HII\ regions.  
These YSOs are not very massive. We have used the SED fitting-tool of Robitaille et al. (\cite{rob07}) to estimate  
the parameters of the central object and of the disk in Class~I and Class~II sources (their masses, the disk and envelope accretion rates).  We find only central 
sources of low- or intermediate-mass (except for c1 and c2 in BRC14). But about fifteen sources have massive envelopes (10$\msol$ -- 47~$\msol$); 
some of these may have high envelope accretion rates and thus may become massive stars.

$\bullet$ We observe probable triggered star formation in several locations: 
\begin{enumerate}
\item Near the numerous pillars present south-east of W5-E, in a region of low density. 
Some Class~I and Class~II low-mass YSOs have been detected there by {\it Spitzer} (Koenig et al. \cite{koe08}). The {\it Herschel} 250~$\mu$m map 
shows the presence of low-mass condensations at the head of the pillars. We propose that the UV radiation 
of the central exciting star, irradiating a turbulent low-density medium, is responsible for these structures and the YSOs formation, in good 
agreement with the simulations of Gritschneder et al. (\cite{gri10}). 

\item Triggered star formation was already identified in the well known bright-rimmed clouds BRC12, BRC13, BRC14. {\it Herschel} images confirm the presence 
of condensations enclosed by BRC13 and BRC14, with masses respectively of a few tens to a few hundreds of solar masses. 
The massive BRC14 condensation harbors a cluster exciting a compact \HII\ region, several Class~I YSOs, at least four good candidate Class~0 sources, 
and several CO outflows. The RDI process, in the scope of a low ionizing photon flux, is possibly at work there. But {\it Herschel}-SPIRE images lack the 
resolution necessary to display the morphology of the condensations and allow a comparison  with the RDI simulations.

\item At the waist of the bipolar \HII\ region Sh~201.
\end{enumerate}

$\bullet$ The {\it Spitzer} images show many ``fingers'' (structures intermediate in size between the large BRCs and the narrow pillars) pointing to the central  
exciting star, and located in the outskirts of the W5-E region, far away from the main PDRs. {\it Herschel}-SPIRE images show the presence of 
condensations of a few solar masses (diameters in the range 0.2~pc -- 0.4~pc) at the head of these fingers; these condensations are good candidate 
prestellar cores as they are bordered by high pressure ionized gas. They are shaped by the UV radiation of the exciting star. This suggests that the main IF 
limiting the \HII\ region presents holes, allowing the ionizing radiation to leak and interact with the surrounding neutral medium, at large distances from the 
exciting star.

$\bullet$  The bipolar nature of the Sh~201 \HII\ region, well demonstrated
 by the {\it Spitzer} images (Koenig et al. \cite{koe08}), is confirmed by the {\it Herschel} observations. The 100~$\mu$m image shows the warm dust surrounding the two
 ionized lobes. The SPIRE images show a layer of cold dust of low column density surrounding the two lobes;  for the first time we see the material collected 
 during the expansion of this \HII\ region. The SPIRE images also show a cold filament of high column density on each side of the \HII\ region (the parental filament). 
 Two massives cold condensations are observed at the waist of the nebula, on each side of the ionized region. For the first time star formation is observed  
in the direction of the most massive one (mass $\geq$200~$\msol$); six 100~$\mu$m point sources are detected there, two of them are 
good candidate Class~0 sources; three more are possible candidate Class~0 YSOs. The fact that they are observed in the direction of filaments, 
in the zone of interaction beween the ionized gas and the molecular condensation, strongly suggests that their formation has been triggered 
by the expanding Sh~201 \HII\ region.\\

\begin{acknowledgements}

We thank T.P. Robitaille for helpful discussions about the envelopes of YSOs, and X.P. Koenig for providing us with beautiful {\it Spitzer} images of the W5 region.
We also thank the anonymous referee whose comments and suggestions helped improve the rigour and the clarity of the paper. This research has made use of the SIMBAD database operated at the CDS, strasbourg, France, and of the interactive sky atlas Aladin (Bonnarel et al. \cite{bon00}). 

This work is mainly based on observations obtained with the {\it Herschel}-PACS and 
{\it Herschel}-SPIRE photometers. PACS has been developed by a consortium of institutes led by MPE (Germany) and including UVIE (Austria); KU Leuven, CSL, IMEC (Belgium); 
CEA, LAM (France); MPIA (Germany); INAF-IFSI/OAA/OAP/OAT, LENS, SISSA (Italy); IAC (Spain). This development has been supported by the funding agencies BMVIT (Austria), ESA-PRODEX (Belgium), CEA/CNES (France), DLR (Germany), ASI/INAF (Italy), and CICYT/MCYT (Spain). SPIRE has been developed by a consortium of institutes led by 
Cardiff Univ. (UK) and including Univ. Lethbridge (Canada); NAOC (China); CEA, LAM (France); IFSI, Univ. Padua (Italy); 
IAC (Spain); Stockholm Observatory (Sweden); Imperial College London, RAL, UCL-MSSL, UKATC, Univ. Sussex (UK); Caltech, JPL, 
NHSC, Univ. colourado (USA). This development has been supported by national funding agencies: CSA (Canada); NAOC (China); CEA, 
CNES, CNRS (France); ASI (Italy); MCINN (Spain); SNSB (Sweden); STFC and UKSA (UK); and NASA (USA). We thank the French Space Agency (CNES) for financial support.
 
This work is based in part on observations made with the {\it Spitzer Space Telescope}, which is operated by the Jet Propulsion Laboratory, California 
Institute of Technology, under contract with NASA. We have made use of the NASA/IPAC Infrared Science Archive to obtain data products from 
the 2MASS, {\it Spitzer}-GLIMPSE, and {\it Spitzer}-MIPSGAL surveys.

This research used the facilities of the Canadian Astronomy Data Centre operated by the National Research Council of Canada with the support of the Canadian Space Agency.

Part of this work was supported by the ANR (\emph{Agence Nationale pour la Recherche}) project ``PROBeS", number ANR-08-BLAN-0241. 
L.D. Anderson acknowledges support by this ANR.

\end{acknowledgements}
 


\clearpage
\begin{appendices}
\appendix

\section{The bright-rim cloud BRC13 and vicinity}

\begin{figure*}[tb]
\sidecaption
\includegraphics[width=12cm]{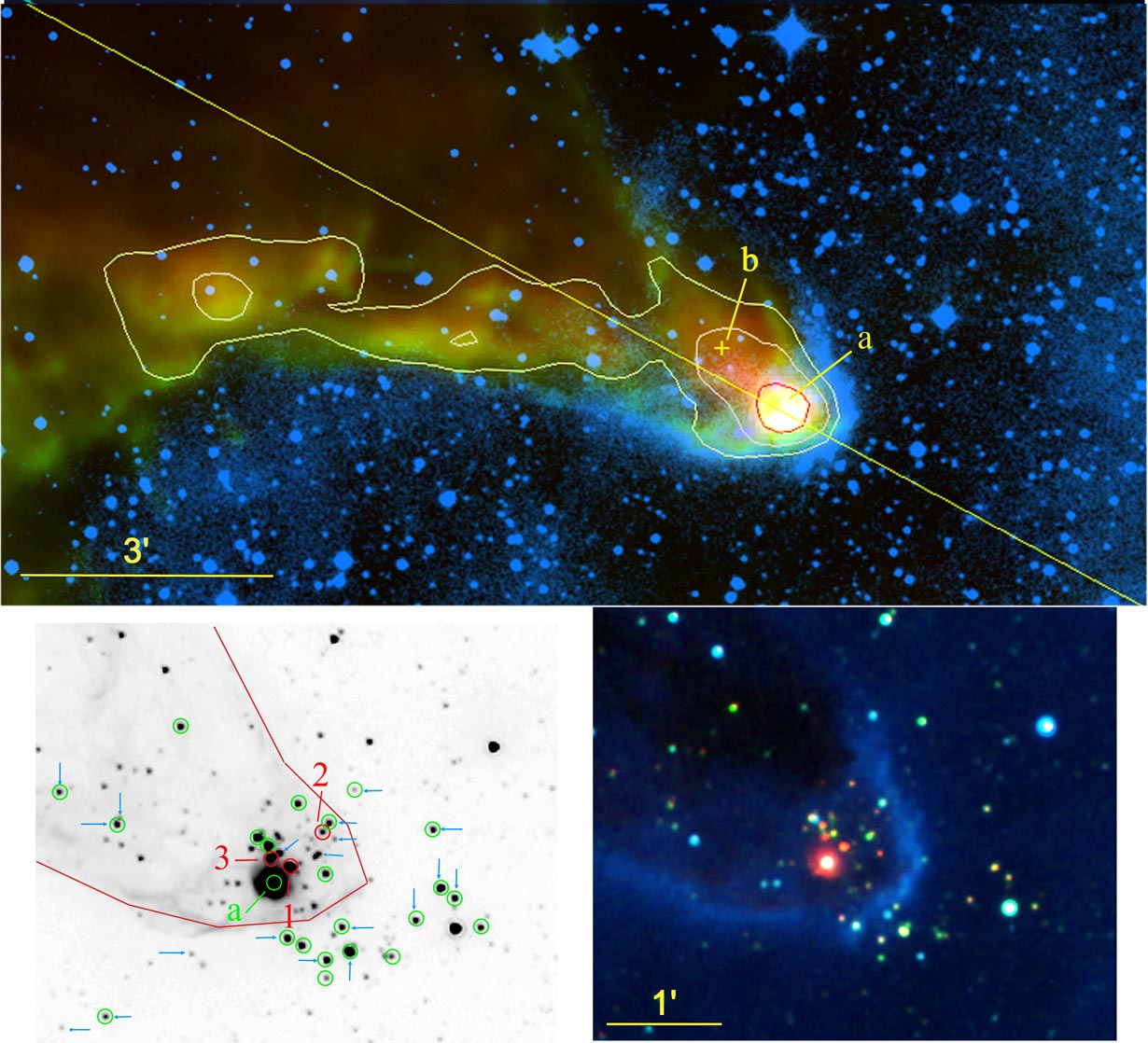}
  \caption{The bright-rimmed cloud BRC13. {\it Top:} Composite colour image: red is the column density map from {\it Herschel} data, green is the 
  {\it Herschel}-PACS image at 100~$\mu$m, and blue is the DSS2-red image. The peak column density is 1.7$\times$10$^{22}$~cm$^{-2}$. The contours 
  correspond to column densities of 1/2, 1/4, and 1/8 of the peak value. The yellow line shows the direction of the exciting star HD~18326. The field 
  size is 14.7\arcmin$\times$7.5\arcmin. North is up and east is left. {\it Bottom right:} 
  The cluster inside BRC13: red is the {\it Spitzer} emission at 4.5~$\mu$m, green is the 2MASS $K$ 
  stellar emission, and blue is the H$\alpha$ emission from the ionized boundary layer (image obtained at the 1.2-m telescope of the Observatoire de 
  Haute-Provence). {\it Bottom left:}  
  Identification of the YSOs discussed in the text and present in the KOE08 catalogue.  The grey-scale underlying image 
  is the {\it Spitzer} 4.5~$\mu$m data. The red and green circles are respectively the proposed Class~I and Class~II YSOs. The blue arrows point 
  to the H$\alpha$ emission stars (classical T Tauri-like stars) identified by Ogura et al. (\cite{ogu02}).}   
  \label{BRC13}
\end{figure*}

BRC13 (Sugitani et al. \cite{sug91}) is the textbook example of a bright-rimmed cloud: it resembles a finger 
pointing to HD~18326, the exciting star of W5-E. An IR source IRAS~02570+6028 lies at the tip of the finger, inside the rim; its luminosity 
is 1440 \lsol\ according to Sugitani et al. (\cite{sug91}; from IRAS fluxes) 
or 490~\lsol\ according to Morgan et al. (\cite{mor08}; from SCUBA fluxes); both estimates are for a distance of 2~kpc. 

Fig.~\ref{BRC13} shows the morphology of BRC13.  The head of BRC13 is symmetric (over a length of 1.7~pc) with respect to the direction of HD~18326, the exciting star of W5-E. 
This is strong evidence that the ionizing radiation of HD~18326 has shaped this structure. Fig.~\ref{BRC13} shows that a small IR cluster 
lies inside BRC13, at its tip. This cluster is likely the counterpart of IRAS~02570+6028. According to KOE08  
it contains several Class~II and Class~I YSOs. Sources \#1 and \#2 in Table~\ref{BRC13} have been identified by KOE08  
as Class~I YSOs. YSO \#2 has no detectable counterpart at {\it Herschel} wavelengths. Source \#3 (identified in Fig.~\ref{BRC13}) is not present 
in the catalogue of KOE08; however it is probably also a Class~I YSO:
it strongly resembles YSO \#1 as both have 24~$\mu$m and 100~$\mu$m counterparts.  The brightest source at 24~$\mu$m is 
a Class~II source, BRC13-a, located at the center of a region of extended emission visible in H$\alpha$ and at all {\it Spitzer} wavelengths. It is this YSO and 
the extended emission region (plus probably the nearby Class~I YSOs \#1 and \#3) that are seen in the PACS and SPIRE images. At these wavelengths we are unable to 
separate the central object from its associated extended emission zone.  We have measured the fluxes of this extended source using
apertures of diameter 30$\arcsec$, 45$\arcsec$, and 60$\arcsec$ (respectively one, two, and three asterisks in Table~\ref{BRC13}); the first aperture 
is that used by Morgan et al. (\cite{mor08}) to measure the SCUBA 
450~$\mu$m and 850~$\mu$m fluxes, but only the last aperture covers the entire extended emission zone. 
We have used the same apertures to measure the fluxes in the {\it Spitzer} bands (asterisks).  We list these measurements in Table~\ref{BRC13}.

Fig.~\ref{BRC13} shows the dense molecular material present at the head of BRC13, around BRC13-a. The diameter at half intensity of the molecular condensation is 0.31~pc. 
Its peak column density is N(H$_2$)$=$1.7$\times$10$^{22}$~cm$^{-2}$, thus a density n(H$_2$)$\sim$1.7$\times$10$^4$~cm$^{-3}$. 
Its temperature, estimated from the {\it Herschel} 
fluxes, is in the range 24.4~K to 27.8~K (the smallest aperture giving the hotter temperature; Table \ref{temperatureYSOs}).  This differs slightly  
from the temperature of 23~K obtained by Morgan et al. (\cite{mor08}). The mass of this condensation is estimated to be $\sim$6.1\msol\ in the 30$\arcsec$ diameter 
aperture and increases to 21.8$\msol$ in the 60$\arcsec$ diameter aperture (considering only the {\it Herschel} data; Table \ref{temperatureYSOs}). (from SCUBA data Morgan et al. estimate a mass of 35~\msol\ in the 30$\arcsec$ central aperture, corrected for a distance of 2~kpc and a dust opacity at 850~$\mu$m of 1.25~cm$^2$~g$^{-1}$). 
We suggest that this condensation is heated from the inside by the nearby 
Class~II object BRC13-a; this YSO is possibly partly embedded and appears to have carved a cavity open in our direction.  This has created a path of low extinction allowing us to see BRC13-a and the light reflected by the borders of the cavity, even at optical wavelengths.

Another fainter condensation (called b) is also present along the axis of BRC13, at the back of the brightest condensation, 
and thus farther away from the exciting star (Fig.~\ref{BRC13}). We measured its {\it Herschel} fluxes using an aperture of 
diameter 90$\arcsec$ (measurements in Table~\ref{BRC13}). The mass of this condensation is estimated to be 55\msol\ for a temperature of 19.5~K 
found in the temperature map (Fig.~\ref{temperature}). The column density decreases along the axis when going away from the tip of the rim. Dense material 
is also present on the south border of BRC13 (and not at its north border, a deviation from axial symmetry). The observed column density is not very high, 
of the order of 3$\times$10$^{21}$~cm$^{-2}$. This material is probably made of material collected during the expansion of W5-E. 
Alternatively it may be part of a east-west pre-existing filament as shown by Fig.~\ref{column}. The column density is not high enough 
for star formation to proceed via the collect and collapse process, and indeed no YSOs are detected there. 

At H$\alpha$ wavelength BRC13 appears as an extinction region surrounded by a bright rim.  Therefore the molecular condensation inside the rim,  
responsible for the extinction, is located slightly in front of the ionized region. The NVSS image shown by Morgan et al. (\cite{mor04}) 
shows faint radio-continuum emission from the very head of BRC13 (the ionized boundary layer seen also in H$\alpha$); 
they estimated the electron density there to be $\sim$250~cm$^{-3}$. 
Morgan et al. (\cite{mor04}) have shown that the pressure in the ionized boundary layer is much higher than the pressure in the enclosed molecular material. 
(This is confirmed by the {\it Herschel} observations.) Thus a D type ionization front progresses inside the BRC, compressing the molecular material 
and potentially triggering star formation. The observed IR cluster possibly results from this triggering. 

Another signature of triggered star formation in BRC13 is the small-scale sequential star formation observed in this object. Numerous H$\alpha$ emission line stars 
have been detected in the vicinity of BRC13 by Ogura et al. (\cite{ogu02}). These stars are YSOs, pre-main sequence stars in the T Tauri/Herbig AeBe phase. 
They are identified on Fig.~\ref{BRC13}. They are located near the tip of the bright rim, at the head of the rim, or just outside the rim. Most of these emission line stars 
are also Class~II YSOs in KOE08. The youngest Class~I YSOs are located inside the bright rim. This suggests  
small-scale sequential star formation, with star formation proceeding from the side of the exciting star outward away from the central \HII\ region. 
Chauhan et al. (\cite{cha11a}) used deep $V,I$ photometry to estimate the age of the H$\alpha$ emission stars; they obtained different ages, 
2.44$\pm$1.37Myr and 1.61$\pm$1.41~Myr respectively for 24 stars outside the rim and 10 stars inside the rim. This confirms the 
small-scale sequential star formation (however the large scatter in the stellar ages is worrisome).

\begin{table*}[tb]
\caption{Photometry of point sources: the bright-rim cloud BRC13 and vicinity}
\resizebox{18cm}{!}{
\begin{tabular}{llllllllllllrrrrrr}
\hline\hline \\ [0.5ex]
Name & \# & RA(2000) & Dec(2000) & $J$    & $H$   & $K$   & [3.6] & [4.5]  & [5.8]  & [8.0] & [24]     & Class & S(100)  & S(160)   & S(250)  & S(350) & S(500) \\
     &         & ($\degr$) & ($\degr$) & mag & mag & mag & mag & mag & mag & mag & mag & & Jy & Jy & Jy & Jy & Jy \\          
\hline \\ [0.5ex]
a & 16614 & 45.235304 & 60.669934 & 12.36 & 11.68 & 11.23 & 10.50 & 9.99 & &  & 0.96 &  II & & & & & \\ 
a* &      &           &           &        &       &       &        &        &        &        &        &             & 78.95 & 67.42 &  20.07 & 6.99 & 2.55 \\
a** &     &           &           &       &       &       & 7.57**  & 7.36**   &  4.65** &  2.96** & 0.18**     & & 112.08 & 113.54 & 35.83 & 13.36 & 5.02 \\
a***&     &           &           &       &       &       & 7.33***  & 7.16***   & 4.48***  & 2.80*** & 0.033*** & & 137.22 & 152.19 * & 49.55 & 19.25 & 7.48 \\
b   &     & 45.2615   & 60.6843  &        &       &      &       &       &       &      &      &   &                            &          & 56.2  & 27.1 & 12.8 \\  
1 & 16594 & 45.230365 & 60.672069 &       & 15.08 & 13.51 & 11.18 & 10.05 & 8.82  & 7.57 & 3.32 & I & & & & & \\
2 & 16555 & 45.221034 & 60.677145 &       & 15.54 & 14.62 & 12.89 & 12.30 & 11.29 & 9.63 & 5.46 & I & & & & & \\
3* &      & 45.2359*  & 60.6735*  &       &       &       &       &       &        &     &      &    & & & & & \\  [0.5ex]        
\hline
\label{BRC13}
\end{tabular}\\
}
\end{table*}

\section{The bright-rim cloud BRC14 and vicinity}

\begin{figure}[tb]
\centering
\includegraphics[width=8cm]{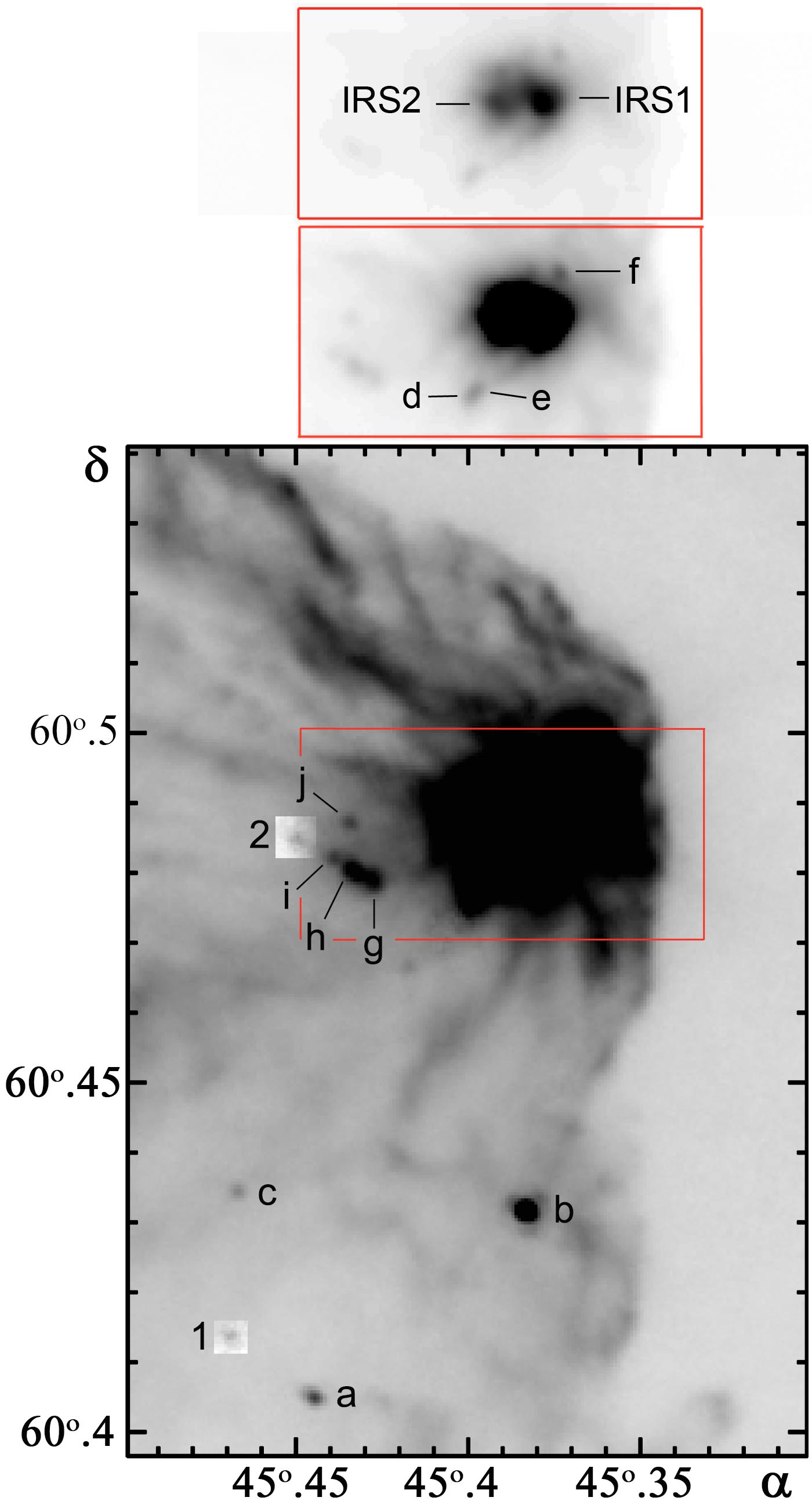}
  \caption{Identification of the YSOs in the BRC14 field. The grey-scale image shows the {\it Herschel}-PACS 100~$\mu$m emission. The three red boxes 
  show the same zone, displayed with different intensity cuts to allow the identification of all the central sources.}
  \label{BRC14a}
\end{figure}

BRC14 is another bright-rimmed cloud catalogued by Sugitani et al. (\cite{sug91}). It harbours the bright IR source IRAS 02575+6017 (also AFGL~4029,  
Price and Walker, \cite{pri76}). The luminosity of this IRAS source is 
$\sim$1.9$\times$10$^4$~\lsol\ (Snell et al. \cite{sne88}; corrected for a distance of 2~kpc). Beichman (\cite{bei79}) has shown that AFGL~4029 is composed of
two mid-IR sources, IRS1 and IRS2, separated by 22\arcsec. 

IRS1 and IRS2 are radio sources, respectively G138.295+1.555 and G138.300+1.558 (Kurtz et al. \cite{kur94}, Zapata et al. \cite{zap01}). 
Both are also small H$\alpha$ emission regions (see Figure 1 in Deharveng et al. \cite{deh97}). G138.300+1.558 is a thermal \HII\ region harbouring 
a small cluster dominated by a B1V star (Deharveng et al.~\cite{deh97}; their star \#26, affected by a visual extinction $\sim$8~mag). 
IRS1 is associated with a YSO and its reflection nebula  (Deharveng et al.; their star \#25, affected by a visual extinction $\sim$30~mag). 
A high velocity optical jet originates from IRS1 (Ray et al. \cite{ray90}). 
The {\it Spitzer} images confirm the presence of two bright IR components, a cluster around a bright star in the direction of IRS2, and a bright 
YSO associated with a reflection nebulosity in the direction of IRS1. IRS1 has been observed at high resolution in the 8~$\mu$m--13~$\mu$m range 
by Zavagno et al. (\cite{zav99}) and at 24.5~$\mu$m by de Wit et al. (\cite{dew09}).  
De Wit et al. estimate a flux density of 680~Jy for IRS1 from the {\it Spitzer}-MIPS observations at 24~$\mu$m. 

The bright-rimmed cloud BRC14 borders a massive molecular cloud. This cloud has been mapped in $^{12}$CO(1-0) by Loren and Wootten (\cite{lor78}), 
Snell et al. (\cite{sne88}), Carpenter et al. (\cite{car90}; \cite{car00}), Karr and Martin (\cite{kar03}). High resolution maps have been obtained by 
Niwa et al. (\cite{niw09}; HPBW=15.6\arcsec) 
in $^{13}$CO(1-0) and C$^{18}$O(1-0). The condensation enclosed by BRC14 is their clump~7, the brightest one, with a peak column density in the 
range 5.8--9.4$\times$10$^{22}$~cm$^{-2}$ (depending on the transition), and a mass of 740~\msol.

A CO outflow was detected by Snell et al. (\cite{sne88}) in the direction of AFGL~4029 (size $\sim$0.55~pc; age $\sim$1.7$\times$10$^5$~yr). 
Their resolution (HPBW$\sim$45\arcsec) is not sufficient to associate it definitively with IRS1 or IRS2 (see figure 2 in Deharveng et al. \cite{deh97}). 
An H$_2$O maser has been detected in the direction of IRAS~02575+6017 by Churchwell et al. (\cite{chu90}); its velocity, V$_{\rm LSR}=-33.3$~km~s$^{-1}$, 
indicates that it is associated with the region.  Once again the resolution of the observations do not allow a direct association with IRS1. 
Methanol maser emission at 6.7~GHz has been searched for but not detected in the direction of the IRAS source (Slysh et al. \cite{sly99}).

Small-scale sequential star formation is also observed in the direction of BRC14 (Matsuyanagi et al. \cite{mat06}, 
Chauhan  et al. \cite{cha11a} and references therein). Using deep photometry Chauhan et al. (\cite{cha11a}) estimated a mean age of 
1.01$\pm$0.73~Myr for 18 YSOs on or inside the rim, and an age of 2.32$\pm$1.22~Myr for 58 sources outside the rim.

\subsection{YSOs in the field of BRC14}

\begin{table*}[tb]
\caption{Photometry of point sources: the bright-rim cloud BRC14 and vicinity}
\resizebox{18cm}{!}{
\begin{tabular}{llllrrrrrrrrrrrrrr}
\hline\hline \\ [0.5ex]
Name & \# & RA(2000) & Dec(2000) & $J$    & $H$   & $K$   & [3.6] & [4.5]  & [5.8]  & [8.0] & [24]     & Class & S(100)  & S(160)   & S(250)  & S(350) & S(500) \\
     &         & ($\degr$) & ($\degr$) & mag & mag & mag & mag & mag & mag & mag & mag & & Jy & Jy & Jy & Jy & Jy \\          
\hline \\ [0.5ex]
a1 & 17120 & 45.443647 & 60.406051 &      &       &        & 13.91 & 13.20 & 12.39  & 11.40 &          & I     &   &   &   &   &   \\
a2 & 17121 & 45.444053 & 60.404731 &      &       &        & 14.48 & 13.16 & 12.09  & 11.00 &          & I     &   &   &   &   &   \\
a3 & 17125 & 45.447332 & 60.405626 &      &       &        & 14.30 & 12.72 & 11.37  & 10.14 & 4.97     & I     & 1.88  & 4.22  & 4.84  & 3.59 & 1.63 \\
b  & 17033 & 45.384975 & 60.431850 &      & 14.95 & 13.32  & 10.72 & 9.44  & 8.37   & 7.33  & 3.17     & I     & 9.21  & 14.82 & 11.08 & 7.05 & 4.86 \\
c  & 17157 & 45.468562 & 60.434342 &      &       &        & 14.96 & 13.64 & 12.77  & 11.44 & 6.67     & I     & 0.51  & 1.66  & 2.83  & 2.71 & 2.77 \\
d  &       & 45.40186  & 60.47630  &      &      &        &        &       &       &        &  no     &        & 4.36  & 8.38  & 7.84:  &   &   \\
e  & 17052 & 45.400109 & 60.477348 &      & 15.08 & 13.63  & 12.09 & 10.94 & 10.02  & 9.19  & 3.71    & I      & 4.72  & 13.14 & 7.84:  &   &   \\
f  & 17012 & 45.375738 & 60.493892 &      & 14.06 & 12.42  & 10.87 & 10.20 & 9.85  & 9.27   &         & II     & 4.8: &       &       &   &   \\ 
g  &       & 45.43024  & 60.47904  &      &       &        &       &       &       &        &         &        & 1.74:  & 7.71  & 7.15: &   &   \\
g1 & 17094 & 45.430795 & 60.480435 &       &       &       & 11.37 & 10.63 & 9.96  & 9.26   & 4.67    & II     &        &       &       &   &   \\
g2 & 17091 & 45.429855 & 60.478460 & 15.88 & 14.02 & 12.99 & 12.35 & 11.88 & 11.47 & 11.01  &         & II     &        &       &       &   &   \\
h  &       & 45.43635  & 60.48052  &       &      &        &       &       &       &        &         &        & 2.22:  & 6.81  & 5.43: & 9.61 (g+h)  &   \\
h1  & 17111 & 45.436898 & 60.480450 & 15.88 & 13.86 & 12.48 & 10.76 & 10.12 & 9.58  & 8.98   &         & II    &        &       &       &        &   \\
i  &       & 45.44198  & 60.48221  &      &       &        &       &       &       &        & no      &        & 1.07  & 4.48   & 4.95:  &   &   \\
j  &       & 45.43740 & 60.48739   &      &       &        &       &       &       &        & no      &        & 0.94  & 2.03  &       &   &   \\
1 & 17160 & 45.470193 & 60.413711 &       &       & 14.78 & 12.35 & 11.24 & 10.24 & 9.35   & 5.49    & I      & 0.13:  &       &       &    &  \\
2 & 17132 & 45.452133 & 60.485071 &       & 15.97 & 14.88 & 11.04 & 10.18 & 9.27  & 8.42   & 5.04    & I      & 0.16:  &       &       &    &  \\
\hline
c1 & 17020 & 45.380310 & 60.486940 &       & 9.90  & 7.25  & 5.27  & 4.45  & 3.27  & 2.56   &        & I      &        &       &       &     &   \\
c2 & 17044 & 45.393692 & 60.486907 &       &       &       & 8.63  & 8.10  & 6.18  & 4.34   &        & I      &        &       &       &     &   \\
c*&       &           &           &        &       &       &       &       &       &        &        &        &  1503 &  1319  &  407.7 & 157.1   & 60.4 \\
c** &     &           &           &        &       &       &       &       &       &        &        &        &  1628 &  1508  &  476.5 &  189.3  & 77.2 \\
c*** &    &           &           &        &       &       &       &       &       &        &        &        &  1731 &  1661  &  534.1 &  216.0  & 86.9 \\ [0.5ex]
\hline
\label{BRC14tablea}
\end{tabular}\\
}
\end{table*}

\begin{table*}[tb]
\caption{Characteristics of the YSOs: the bright-rim cloud BRC14 and vicinity}
\resizebox{18cm}{!}{
\begin{tabular}{llllllllllll}
\hline\hline \\ [0.5ex]
Name & \# & $\chi ^2$ & $M_*$   & $T_*$ & $M_{\rm disk}$  & $\dot{M}_{\rm disk}$  &  $\dot{M}_{\rm env}$  & $L$     & $A_{\rm V}(ext)$ & $i$     & Stage \\
     &    &           & (\msol )& (K)   &(\msol)          &(\msol\ yr$^{-1}$)     & (\msol\ yr$^{-1}$)    & (\lsol) & (mag)            & (\degr) &       \\          
\hline \\ [0.5ex]
a3 & 17125 & 149 & 1.4             & 4218   & 1.1e-2                  & 1.0e-7                   & 4.7e-4  & 21            & 1.96         & 49  &  I \\
b  & 17033 & 151 (--187) & 4.2 (1.7--5.5) & 4420  & 4.5e-2 (8.4e-4--2.2e-1) & 5.3e-7 (5.6e-10--4.2e-6) & 3.1e-4 (1.5e-4--1.0e-3) & 133 (--60) & 3.54 (--8.0) &49 (--18) & I \\
c  & 17157 & 51 & 1.3              & 4187  & 5.7e-4                  & 4.9e-11                  & 4.4e-4  & 20          & 8.0           & 49  & I \\
e  & 17052 & 76 (--103) & 5.1 (--1.2)  & 4772  & 3.9e-4 (--2.3e-2)   & 5.3e-10 (--5.2e-7)   & 1.0e-4 (0.8e-4--5.4e-4) & 97 (--31) & 5.51 (--8.0) & 76 (--18) & I \\ 
2  & 17132 & 17 (--34) & 2.8 (--6.4) & 11428 & 4.2e-2 (7.2e-5--1.2e-1) & 5.5e-7 (--1.5e-10) & 0       & 79 (50--851) & 8.0 (--1.96) & 81 (--87) & II \\ [0.5ex]
\hline
\label{BRC14tableb}
\end{tabular}\\
}
\end{table*}

\begin{figure*}[tb]
\centering
\includegraphics[width=150mm]{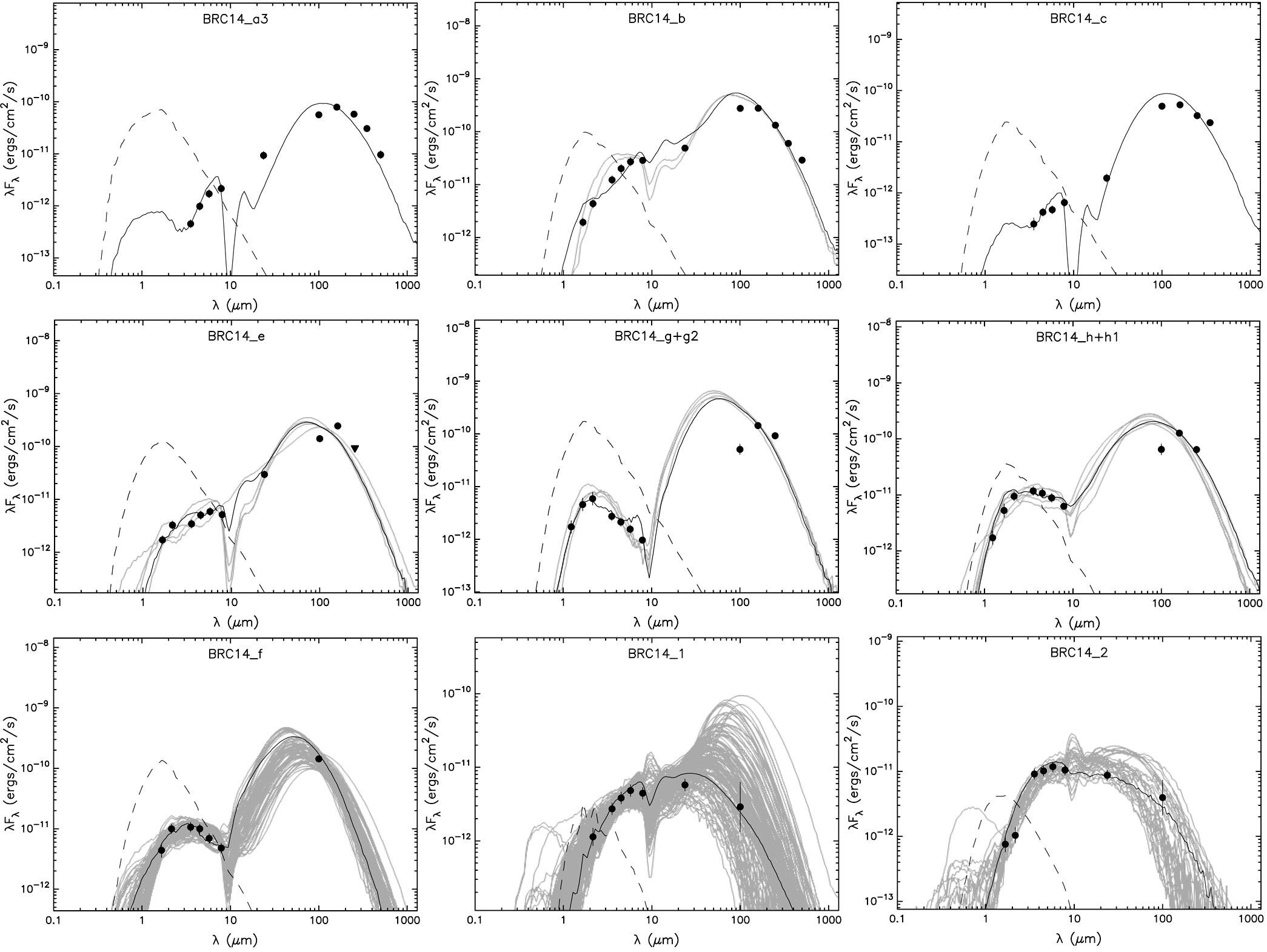}
  \caption{Spectral energy distributions of a few point sources in the BRC14 field. We used the SED fitting tool 
  of ROB07. }
  \label{BRC14SED}
\end{figure*}

\begin{figure*}[tb]
\sidecaption
\includegraphics[width=12cm]{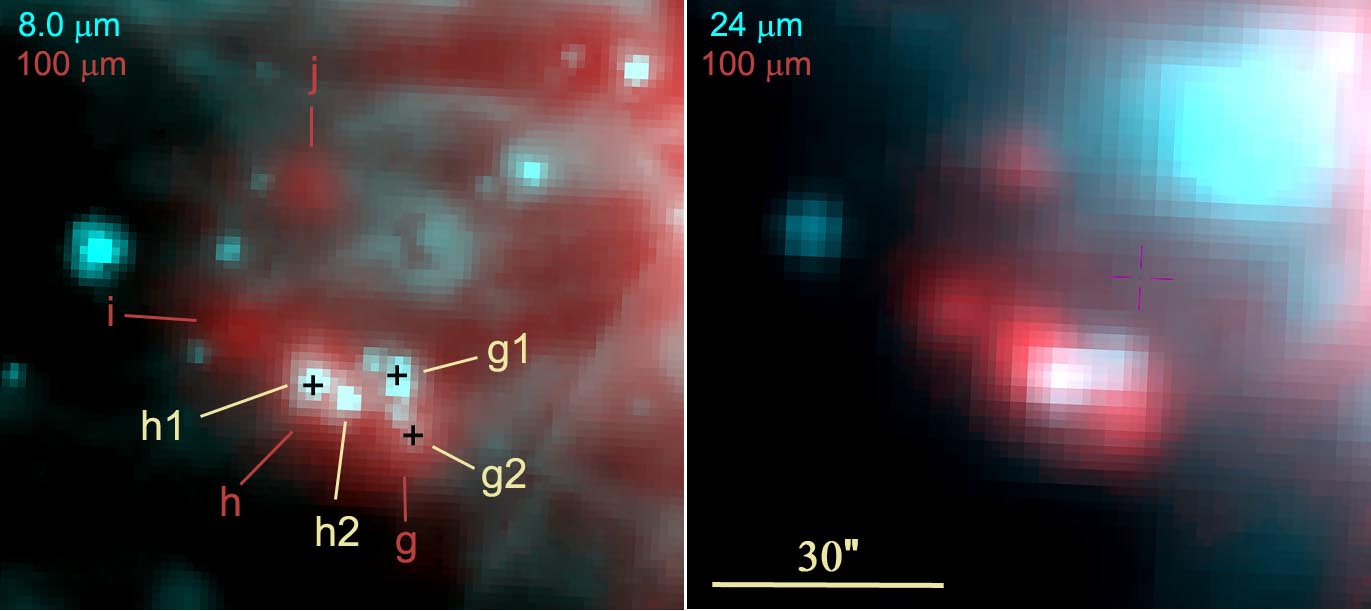}
  \caption{The BRC14-g, -h, -i, and -j 100~$\mu$m point sources and their associated mid-IR cluster. The black crosses show the position of the three Class~II YSOs identified by KOE08.}
  \label{BRC14h}
\end{figure*}

The {\it Herschel} observations show a bright massive condensation enclosed by BRC14, and numerous point-like sources. These objects are 
identified in Fig~\ref{BRC14a}, listed in Table~\ref{BRC14tablea}, 
and commented on below.  We show a few SEDs in Fig.~\ref{BRC14SED}. The SEDs have been fitted allowing the interstellar (external) extinction to range between 1.96~mag and 8~mag.

$\bullet$ BRC14-a: This source is observed in the direction of three nearby stellar sources (separated by some 6\arcsec), all classified as Class~I by KOE08. 
At least one of them (a3) is a 24~$\mu$m source. BRC14-a is likely associated with it. If true, only one model satisfies 
$\chi ^2-\chi^2$(best) per data point $\leq$3. This model corresponds to an intermediate mass Stage~I YSO. But the SED is not well constrained, perhaps because 
of the multiplicity of the central object.

$\bullet$ BRC14-b: A bright 24~$\mu$m source lies in this direction. It belongs to a small group of stars, observed at {\it Spitzer}-IRAC wavelengths; an object is clearly dominant  at 8.0~$\mu$m and 24~$\mu$m. The SED is that of a Stage~I YSO of intermediate mass. The parameters of the disk are not well constrained.

$\bullet$ BRC14-c: This source is faint and observed in the direction of a rather faint 24~$\mu$m source. This source was identified by KOE08 as a ``deeply embedded protostar''. We confirm here that its SED is that of a Stage~I YSO of intermediate mass.

$\bullet$ BRC14-d and -e: these sources, separated by some 5\arcsec, lie near the bright central IRS sources.  They cannot be separated at SPIRE wavelengths. 
Only one 24~$\mu$m source is observed in the direction of BRC14-e; BRC14-d does not have any 24~$\mu$m counterpart and thus is a candidate Class~0 object. 
For BRC14-e, six models satisfy $\chi ^2-\chi^2$(best) per data point $\leq$3.  They point to a Stage~I YSO of intermediate mass; but here again the parameters of the 
disk are not well constrained. 

$\bullet$ BRC14-f: This source is observed in the direction of a Class~II YSO according to KOE08. Because it lies near IRS1, its fluxes are difficult to measure. Its  
24~$\mu$m counterpart has not been measured by KOE08, and our measurement at 100~$\mu$m is very uncertain. 
At {\it Spitzer}-IRAC wavelengths this YSO is the brightest source of a small group of stars. The SED is not well constrained; the 
{\it Herschel} source is possibly not linked to the central YSO (but adjacent to it).

$\bullet$ BRC14-g, -h, -i: These sources lie near each other on the sky. BRC14-g and -h are possibly slightly extended. This region is complicated, as a small  
cluster is present in its direction. One IRAC Class~II source is observed in the direction of BRC14-h (called h1 in Fig.~\ref{BRC14h}), which 
has no 24~$\mu$m counterpart. Two Class~II sources are observed in the direction of BRC14-g (called g1 and g2 in Fig.~\ref{BRC14h}), and only 
g1 has a 24~$\mu$m counterpart. Another 24~$\mu$m source lies between -g and -h, that is not in KOE08 catalogue. It is difficult to 
associate -g and -h with any of these sources. For example, Fig.~\ref{BRC14SED} shows that if we associate BRC14-h to the 
Class~II YSO h1 the SED is not well fitted at {\it Herschel} wavelengths.
 The situation is the same for BRC14-g.  This and the fact that the {\it Herschel} sources are 
slightly extended suggest either that the cluster with its three Class~II YSOs is embedded in (or lie adjacent to) the  extended g and h condensations,
or that it contains several Class~0 sources. BRC14-i has no IRAC 
counterpart, and has possibly a very faint 24~$\mu$m extended counterpart; it is a candidate Class~0 source.

$\bullet$ BRC14-j: This 100~$\mu$m source has no IRAC or 24~$\mu$m counterpart. It is a candidate Class~0 source. 

Several other sources are Class~I YSOs according to KOE08. YSOs \#1 and \#2 have very faint 100~$\mu$m counterparts, at the limit of detection (Fig.~\ref{BRC14a}). 
Their SEDs are not at all well constrained, as shown by Fig.~\ref{BRC14SED}. However, the first 20 best fits for YSO \#2 have all an envelope accretion rate equal to zero. Thus, 
YSO~\#2 is probably a Class~II YSO, of intermediate mass; the parameters of its disk are not constrained. The SED of YSO \#1 resembles that of YSO \#2; it is also 
probably a Class~II YSO (in contrast to the finding of KOE08). The four other Class~I YSOs (according to KOE08; \#17122 close to BRC14-i, \#17029, \#17037, and \#17076, close to the central IR sources) have no detectable or measurable 100~$\mu$m counterpart. We cannot estimate their evolutionary stage.

The cloud enclosed by BRC14 contains in its center two very bright {\it Herschel} sources, c1 and c2, observed in the direction of IRS1 and IRS2. They are discussed in the following section.

\subsection{The dense condensation enclosed by BRC14}

\begin{figure}[tb]
\centering
\includegraphics[width=85mm]{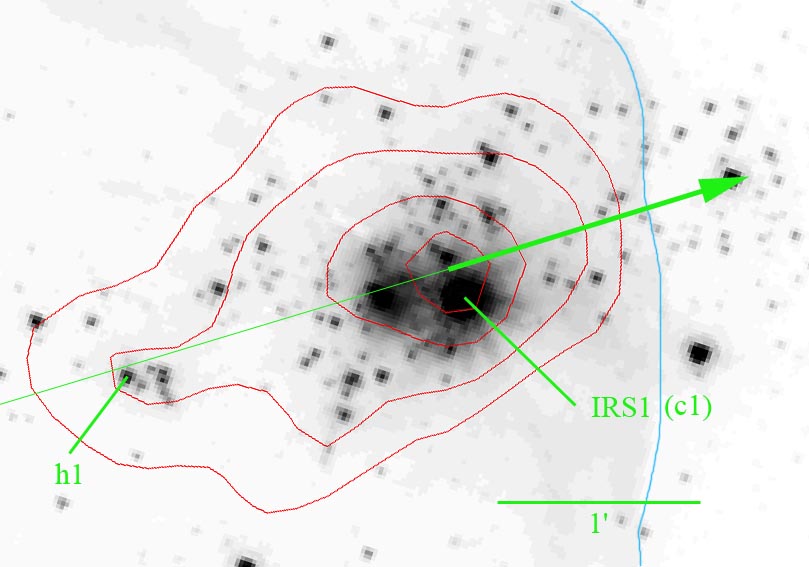}
\caption{The bright-rimmed cloud BRC14. The underlying greyscale image is the {\it Spitzer} 3.6~$\mu$m emission showing the 
cluster AFGL4029 (IRS2) and the YSOs BRC14-c1 (IRS1; a Class~I source according to KOE08)). The red contours correspond to column densities of 0.5, 1, 2.5, 
and 5~$\times$10$^{22}$~cm$^{-2}$. The peak column density is $\sim$1.0~$\times$10$^{23}$~cm$^{-2}$. The arrow points to the 
exciting star HD~18326. The blue line shows the position of the IF. North is up and east is left.}
\label{AFGL4029}
\end{figure}

The molecular condensation enclosed by BRC14 is conspicuous in the {\it Herschel} images. It is the brightest and  most massive condensation of the observed field (Fig.~\ref{AFGL4029}). 
It belongs to the east-west filament extending toward Sh~201. It is symmetric with respect to the direction of HD~18326. It contains the two bright IR sources IRS1 and IRS2.

The morphology of this condensation varies with the wavelength. At 100~$\mu$m it contains two slightly extended sources, observed in the direction of IRS1 and IRS2 (Fig.~\ref{BRC14a}); we call them c1 and c2 in Table~\ref{BRC14tablea}.  The western one (IRS1) is the brightest (there is a factor of 3.5 in intensity between their peak emission values). At 160~$\mu$m, IRS2 becomes fainter; the ratio of the peak intensities is 2.85. 
At SPIRE wavelengths IRS2 is barely detected, IRS1 is dominant, but the peak emission is shifted toward the north-east; the column density peaks some 8$\arcsec$ (0.08~pc) north east of c1. The IR sources c1 and c2 are very bright 24~$\mu$m sources. They have been classified as Class~I YSOs by KOE08, but no 24~$\mu$m measurements are given in the catalogue;   
c2 is the brightest at 24~$\mu$m, the contrary of what is observed at 100~$\mu$m and at longer wavelengths. This indicates that IRS2 is the hottest of the two sources. This is expected as IRS2 is not a single source, but corresponds to dust heated by the cluster exciting a compact \HII\ region (Deharveng et al. \cite{deh97}).

This condensation has a peak column density N(H$_2$)$=$1.0$\times$10$^{23}$~cm$^{-2}$, and a half-intensity diameter of 0.27~pc; this indicates a density 
n(H$_2$)$\sim$10$^5$~cm$^{-3}$. We have measured the flux of this condensation through circular apertures of radius 40\arcsec, 50\arcsec, and 60\arcsec, centered on the column density peak at  $\alpha$(2000)$=03^{\rm h} 01^{\rm m} 31^{\rm s}79$, $\delta$(2000)$=+60\degr$ 29$\arcmin$ 21$\arcsec$.  We give these measurements in 
Table~\ref{BRC14tablea}, under the labels c*, c**, and c***, respectively for the three apertures. The temperatures we derive from the {\it Herschel} fluxes 
are in the range 25.0~K to 26.4~K (the smallest aperture is the hottest), and the masses are in the range 170--250~$\msol$ respectively for the smallest to 
the largest aperture (see table \ref{temperatureYSOs}). This can be compared to the results obtained by Morgan et al. (\cite{mor08}) from SCUBA data; a temperature of 27~K, 
a diameter of 0.4~pc, and a mass of 94~$\msol$ for integration in a 30\arcsec\ aperture (if corrected for a distance of 2~kpc and an opacity at 850~$\mu$m of 1.25~cm$^2$~g$^{-1}$).

\section{The Sh~201 bipolar \HII\ region and vicinity}

\subsection{Presentation of the region}

\begin{figure}[tb]
\centering
\includegraphics[width=85mm]{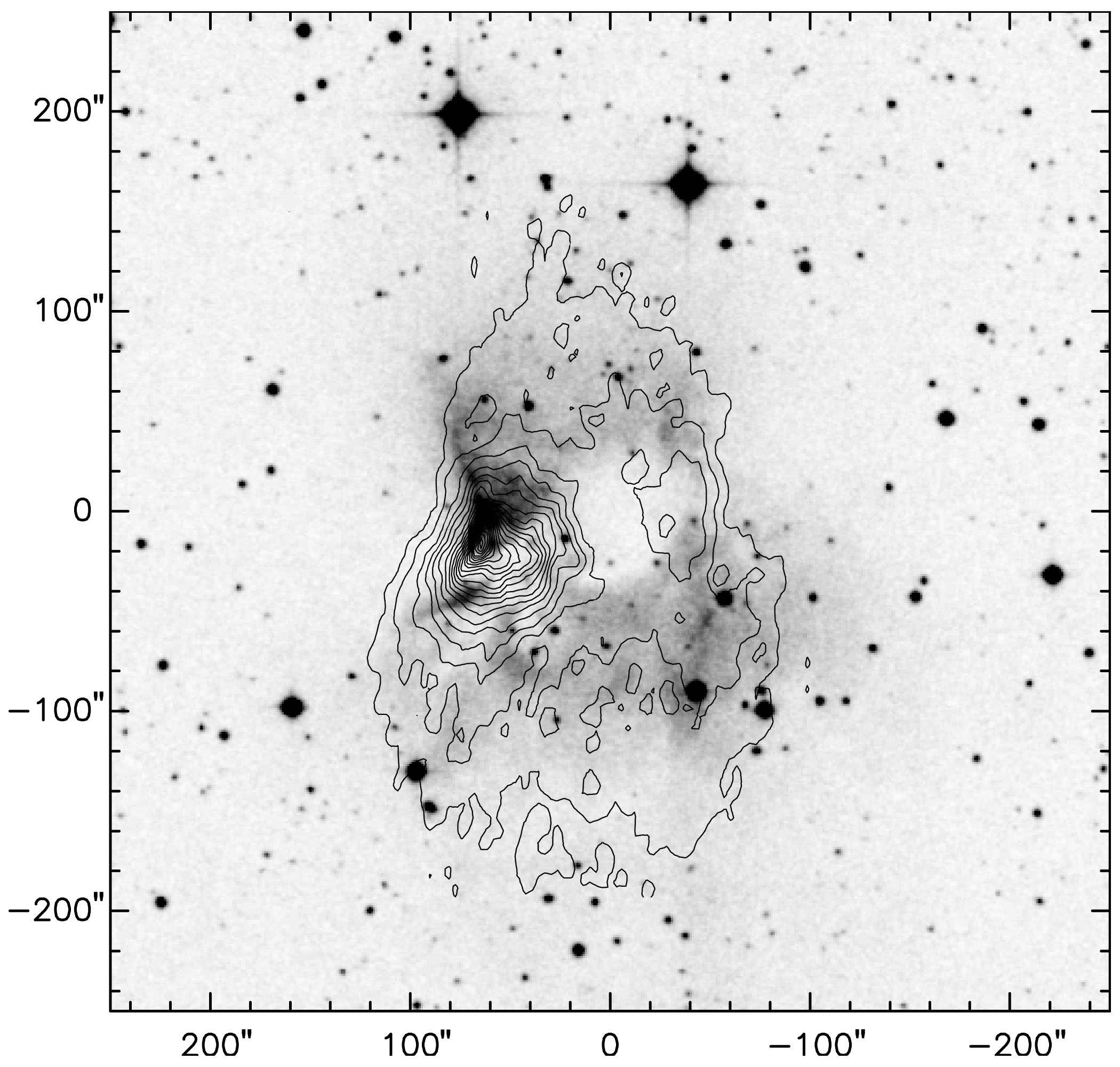}
  \caption{Emission of the ionized gas in Sh~201. The contours of the radio-continuum emission, as given by Ojha et al. (\cite{ojh04}; their figure 6), are superimposed 
 on a grey image of the H$\alpha$ emission (DSS-2 red image). The brightest emission comes from the dense ionized layer bordering a molecular condensation. North is up and east is left. } 
\label{S201aa}
\end{figure}

Sh~201 is a small \HII\ region located to the east of W5-E, about 7.6~pc away from the IF bordering BRC14. It is coincident with the 
IRAS source IRAS~02593+6016, which has a luminosity of $\sim$1.1 $\times$ 10$^4$~\lsol. It probably belongs to the same complex 
as W5-E since its velocity (V(RRLs)=$-$34.6~km~s$^{-1}$, Lockman \cite{loc89}; V(H$\alpha$)=$-$35.5~km~s$^{-1}$, Fich \cite{fic90})
is very similar to that of W5-E (V(H158$\alpha$)=$-$37.0~km~s$^{-1}$, Dieter \cite{die67}; V(H$\alpha$)=$-$39.0~km~s$^{-1}$; 
V(H$\alpha$)=$-$34.3~km~s$^{-1}$, Mampaso et al. \cite{mam87}).

Sh~201 is a bipolar optical \HII\ region of low excitation and low density, except for a bright H$\alpha$ knot on the east side 
(Mampaso et al. \cite{mam87} and Fig.~\ref{S201aa}). These authors suggest that this bright optical structure 
results from the interaction of the ionized gas with an adjacent dense molecular cloud.  
Sh~201 is a thermal radio source with an integrated flux density  
of 1.2$\pm$0.2~Jy at 6-cm (Felli et al. \cite{fel87} and references therein). At a distance of 2~kpc this flux density indicates 
an ionizing photons flux of 4.25 $\times$ 10$^{47}$~s$^{-1}$ (using equation 1 in Simpson \& Rubin \cite{sim90}).  This ionized flux points to an 
ionizing star of spectral type O9--O9.5V, if single, according to the calibration of Martins et al. (\cite{mar05}).  Felli et al. (\cite{fel87}), 
Fich (\cite{fic93}), and Ojha et al. (\cite{ojh04}) all show high-resolution radio continuum maps of Sh~201.
These maps, obtained at different frequencies, are very similar to one another; this  confirms the thermal nature of the radio-continuum emission. 
The radio source has a bright 
arc shaped edge on the east side, and a smoothly decreasing surface brightness on the other side (Fig.~\ref{S201aa}).  All aforementioned authors interpret this
arc-shaped structure as a result of the interaction between the \HII\ region and an adjacent molecular cloud (on the east side). 
Felli et al. (\cite{fel87}) have shown that an O9 ZAMS star (the exciting star of the \HII\ region) placed at a distance of 0.26~pc (for d=2~kpc) of the 
border of the cloud is consistent with the available data. This model does not attempt to explain the bipolar nature of Sh~201. 
Fig.~\ref{S201aa} summarises the situation for the ionized gas: 1) the regions of radio and H$\alpha$ emission have a similar extent, except for the central absorption 
zone; 2) the arc-shaped bright radio feature corresponds to the bright H$\alpha$ knot; they are bordering a region of high extinction to the east of Sh~201; 3) a faint secondary radio component is present about 1~pc west of the brightest radio component (on the opposite side at the waist of the bipolar nebula).

\begin{figure}[tb]
\centering
\includegraphics[width=85mm]{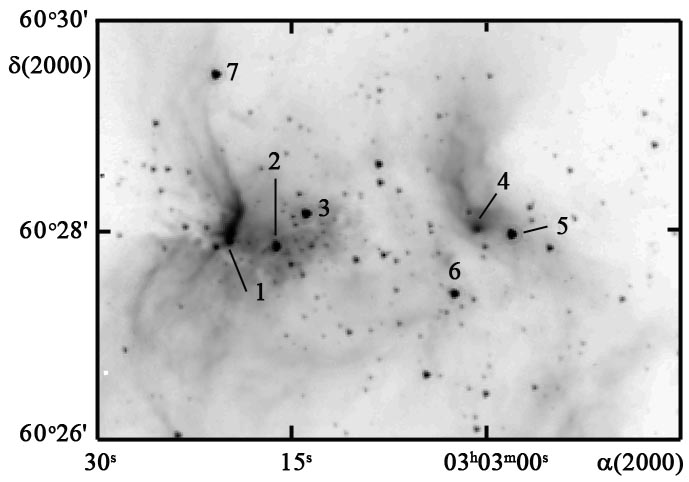}
  \caption{Young stellar objects and stars located at the waist of Sh~201. They are identified on the {\it Spitzer}-IRAC 4.5~$\mu$m image. 
  Star \#2 is the exciting star.}
  \label{S201d}
\end{figure}

Ojha et al. (\cite{ojh04}) discuss the stellar content of Sh~201, based on near-IR observations (see also Carpenter et al. \cite{car93}). A cluster, 
containing more than a hundred stars, is present in the direction of IRAS~02593+6016; it contains YSOs identified by their near-IR excess. 
The brightest and most reddened source is, according to these authors, an O6--O8 type star. Two massive stars with spectral type probably earlier than B2 are also present, which display a near-IR excess. These stars are identified on Fig.~\ref{S201d}; they are respectively \#2 (the exciting star), \#1 and \#3. 

The molecular content of the region has been the subject of various studies. Martin and Barrett (\cite{mar78}) have detected and mapped 
the  $^{13}$CO emission of a flat molecular cloud, located at the waist of the bipolar nebula, that is probably responsible for the optical absorption 
observed there (see also figure 3 in Felli et al. \cite{fel87}). This cloud extends east-west, with a size of more than 18~pc $\times$ 6~pc,   
and has a velocity similar to that of the \HII\ region (V(CO)=$-$39.1~km~s$^{-1}$, Blair et al. \cite{bla75}; V(CO)=$-$40$~km~s^{-1}$, 
Blitz and Fich \cite{bli82}; V(CO)=$-$38~km~s$^{-1}$, Wooterloot and Habing \cite{woo85}). Large-scale CO maps show the extent of this filament 
which goes from BRC14, through Sh~201, to eastern regions far away from Sh~201; its total east-west extent is $\sim$ 20~pc (Carpenter et al. \cite{car00}; 
Karr and Martin \cite{kar03}; Niwa et al. \cite{niw09}, their cloud 7). This filament has two condensations on each side of Sh~201 
(clumps 8 and 9 in Niwa et al. \cite{niw09}). The CS map of Carpenter et al. (\cite{car93}) shows that the east clump is   
by far the brightest. It is also the densest of all the CO clumps located in the periphery of W5, with a density 
of 2.6$\times$10$^{4}$~cm$^{-3}$ (Niwa et al. \cite{niw09}). It is the dense ionized layer bordering this clump that appears as 
the bright arc-shaped structure in the radio-continuum map and in H$\alpha$.

\begin{figure}[tb]
\includegraphics[width=8cm]{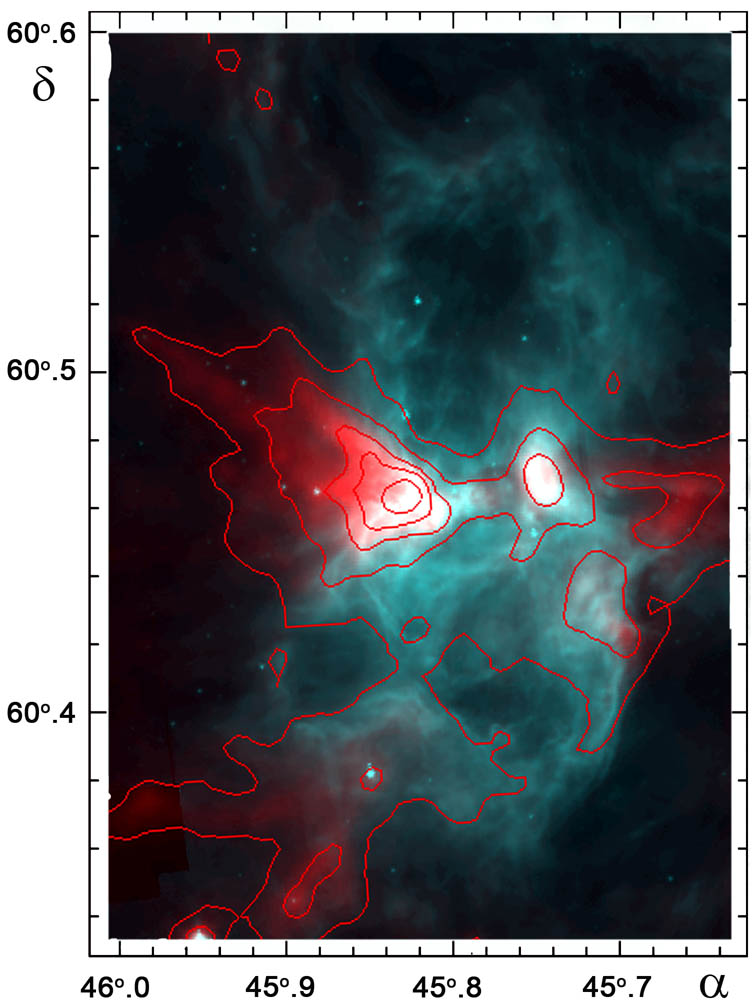}
  \caption{The bipolar nature of Sh~201. Red is for the column density map; the red contours are of column density; the peak column density is 6.2$\times$10$^{22}$~cm$^{-2}$ (east    condensation) and the contours levels  are 0.1, 0.25, 0.5, 1, 2.5, and 5$\times$10$^{22}$~cm$^{-2}$. Turquoise is for the {\it Spitzer} 8.0~$\mu$m image 
  showing the PAH emission from the vicinity of the ionization front. The coordinates are J2000 in degrees.}
  \label{S201bb}
\end{figure}

The Sh~201 region has been observed by {\it Spitzer} (KOE08). The 8.0~$\mu$m IRAC image, dominated by the PAH emission bands 
from the PDR, gives the best illustration of the bipolar nature of Sh~201. Fig.~\ref{S201bb} shows the column density contour levels (obtained by {\it Herschel}) superimposed on the 8.0~$\mu$m image. It confirms: 1) the presence of a dense neutral filament extending east-west, with two 
bright condensations at the waist of the bipolar nebula and a hole of emission in the central region (occupied by ionized material); 2) the bipolar nature of Sh~201: 
two lobes extend perpendicular to the filament, up to 3.5~pc from the central exciting cluster.\\

\begin{figure*}[tb]
\sidecaption
\includegraphics[width=12cm]{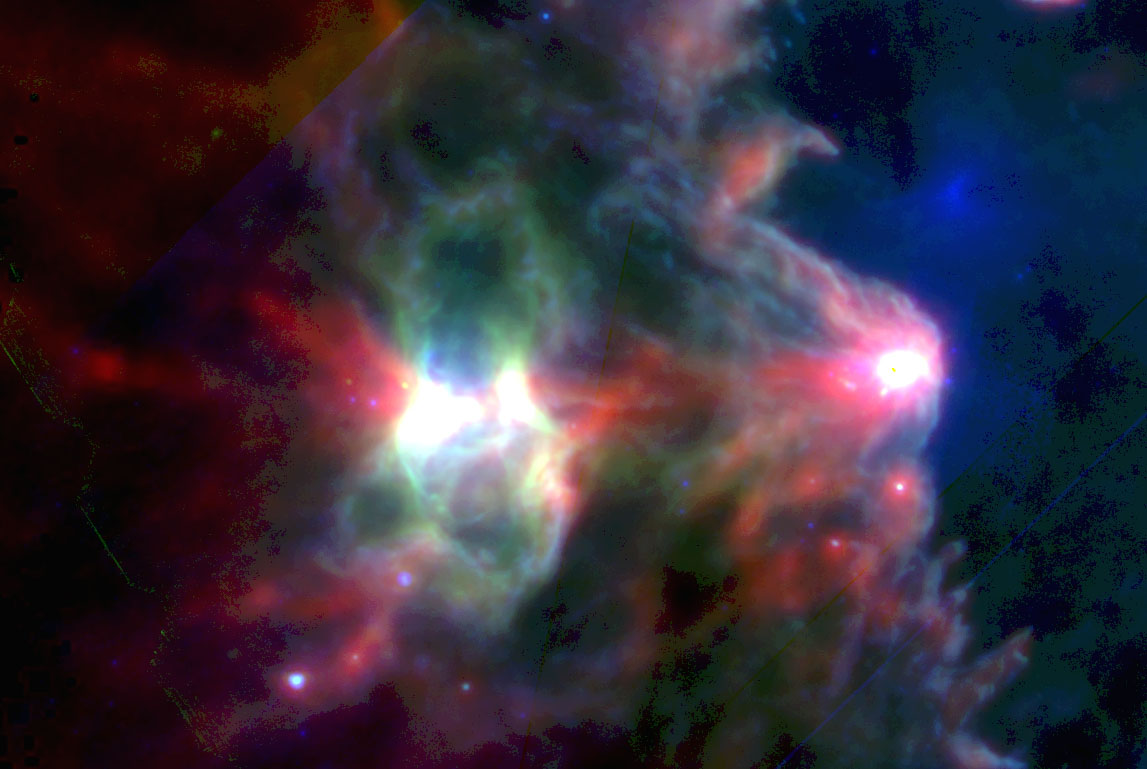}
  \caption{The vicinity of Sh~201 and BRC14. Composite colour image with the {\it Herschel}-SPIRE 500~$\mu$m image in red, the 
  {\it Herschel}-PACS 100~$\mu$m image in green and the {\it Spitzer}-MIPS 24~$\mu$m image in blue (all in logarithmic units). Red is the emission 
  of the cold dust; it shows the dense material in this region. Green traces the emission from warmer dust in the PDRs 
  of the \HII\ regions. Blue traces the emission from even hotter dust, located in the PDRs but also inside the ionized regions. YSOs appear as point sources 
  at 24~$\mu$m and 100~$\mu$m.  The field is $32.5 \arcmin \times 21.8 \arcmin$. North is up and east is left.}
  \label{S201b}
\end{figure*}

Figure \ref{S201b} displays the distribution and temperature of the dust observed by {\it Herschel}. Red is 500~$\mu$m band that traces 
the emission of cold dust.  As such, this channel
shows the numerous cold filaments containing most of the mass, especially the east-west filament mentioned earlier that extends from BRC14 to Sh~201.
Green is the 100~$\mu$m band, which traces the emission of warmer dust  
along the PDRs of the ionized regions. We see very clearly the two lobes perpendicular to the east-west filament, and a brighter emission at the waist 
of the bipolar nebula. Blue is the {\it Spitzer} 24~$\mu$m band, which traces even hotter dust.  Its emission is located
predominantly either in the PDRs or in the ionized region; we find 24~$\mu$m extended emission in the direction of the ``interior'' of W5-E and 
in the central region of Sh~201, close to the exciting cluster. Several cold dust condensations (red colour) are found along the east-west filament, 
one enclosed by BRC14, two on each sides of the waist of Sh~201 (the eastern one the brightest), and a small one at the extreme east, $\sim$5.5~pc from Sh~201 
(called filament~\#1 in Sect.~6.6 and Fig.~\ref{compact}). 
The eastern condensation at the waist of Sh~201 has a peak column density of 6.2$\times$10$^{22}$~cm$^{-2}$, corresponding to a visual extinction of 66~mag. 
The mass of this condensation, integrated over an area limited by the contour of 0.6$\times$10$^{22}$~cm$^{-2}$ (10\% of the peak intensity) is 235~\msol. 
This indicates a mean density of 3.5$\times$10$^4$~H$_2$~cm$^{-3}$, in rather good agreement 
with the results of Niwa et al. (\cite{niw09});  they find from C$^{18}$O observations of their clump 9 (corresponding to our eastern condensation) a column 
density of 4.3$\times$10$^{22}$~cm$^{-2}$, and a density of 2.6$\times$10$^4$~cm$^{-3}$. 
If we consider the contour at half-intensity as the integration limit, we measure a mass of 70~\msol\ for a size (diameter) of 0.25~pc. 
This indicates a core density of the order of 9$\times$10$^4$ H$_2$~cm$^{-3}$, in good agreement with the results of Carpenter et al. 
(\cite{car93}) who showed with their CS\,(2-1) observations that this condensation has a dense core. 

The western condensation at the waist of the bipolar nebula has a smaller column density of 1.1$\times$10$^{22}$~cm$^{-2}$ at the emission peak. 
It is slightly elongated north-south along the ionization front, with an equivalent diameter of 0.36~pc at half intensity; thus a smaller density of the order 
of 10$^4$~cm$^{-3}$. Note that this condensation is not clump 8 in Niwa et al. (\cite{niw09}).

\subsection{Young stellar objects in the vicinity of Sh~201}
\begin{figure}[tb]
\includegraphics[width=85mm]{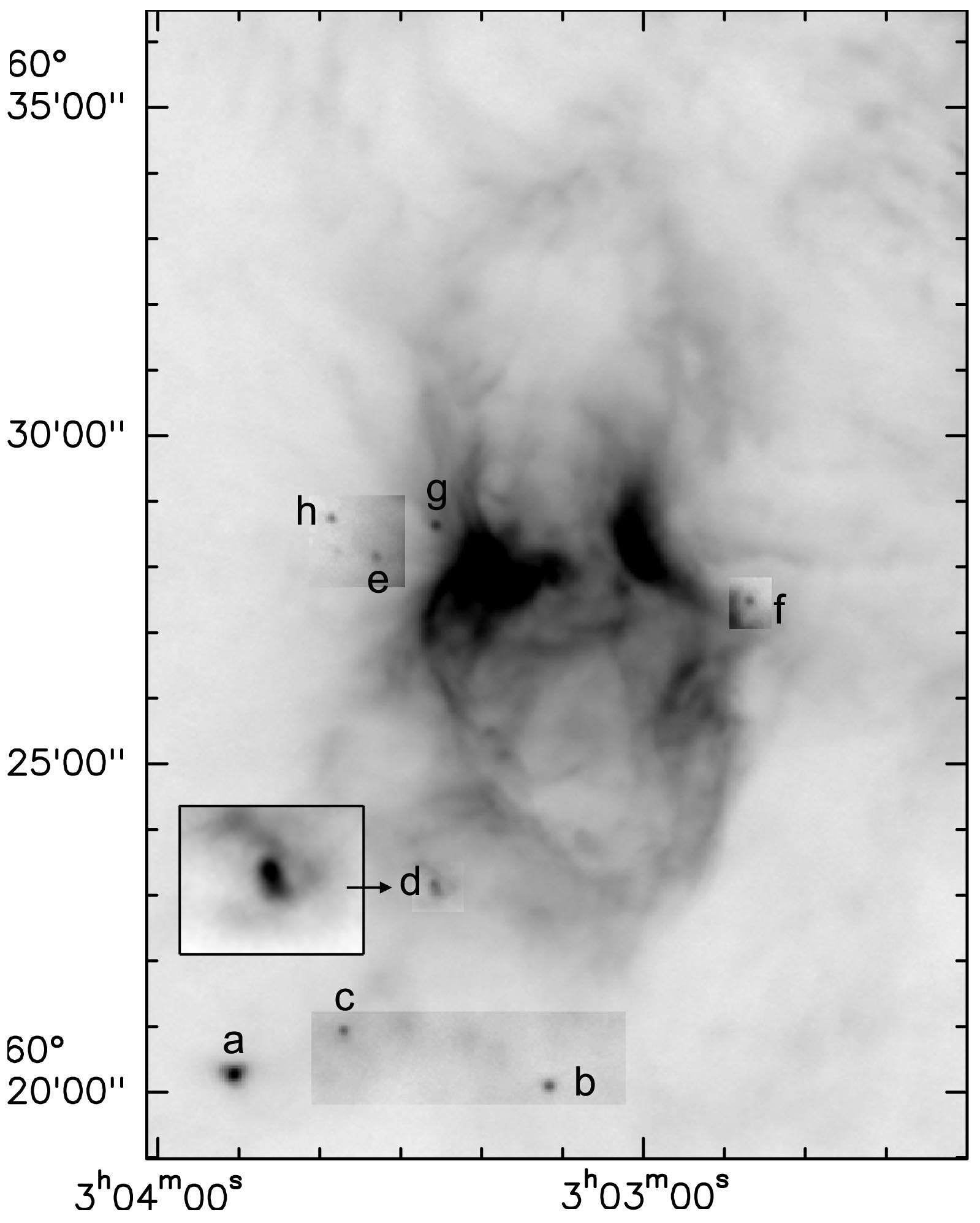}
  \caption{Young stellar objects detected at 100~$\mu$m in the vicinity of Sh~201. They are identified on the {\it Herschel}-PACS 100~$\mu$m image.
	(The contrast has been adjusted to better identify the sources.) The sources s1 and s2 are identified on Fig.~\ref{taille}. The coordinates are J2000.}
  \label{S201c}
\end{figure}

\begin{table*}[tb]
\caption{Photometry of point sources in the vicinity of Sh~201. The asterisks indicate our 24~$\mu$m measurements.}
\resizebox{18cm}{!}{
\begin{tabular}{llllrrrrrrrrrrrrrr}
\hline\hline \\ [0.5ex]
Name & \# & RA(2000) & Dec(2000) & $J$    & $H$   & $K$   & [3.6] & [4.5]  & [5.8]  & [8.0] & [24]     & Class & S(100)  & S(160)   & S(250)  & S(350) & S(500) \\
     &         & ($\degr$) & ($\degr$) & mag & mag & mag & mag & mag & mag & mag & mag & & Jy & Jy & Jy & Jy & Jy \\          
\hline \\ [0.5ex]
a & 17719 & 45.958974 & 60.336970 & 14.86 & 13.50 & 11.61 & 8.94  & 7.99  & 6.55  & 5.24 & 0.97*: & I & 30.00  & 40.00 & 9.32  & 3.91   & 2.48 \\
b & 17549 & 45.797180 & 60.335376 & 16.20 & 14.87 & 14.15 & 12.17 & 11.68 & 9.46  & 7.87 & 5.32*  & I &  1.64 & 2.85  & 1.62  & 0.66    & \\
c & 17660 & 45.903246 & 60.348649 &       &       &       & 14.07 & 12.22 & 10.77 & 9.85 & 5.76*  & I &  1.14 & 2.62  & 4.06  & 2.95:   & 2.31: \\
d & 17609 & 45.857765 & 60.385978 & 11.67 & 10.06 & 8.83  & 7.26  & 6.70  & 6.20  & 5.25 & 1.81*  & II&  1.66 & 3.75  & 2.90: & 1.65    & 1.27: \\
e & 17644 & 45.889840 & 60.468891 & 15.39 & 13.30 & 11.82 & 9.69  & 8.88  & 8.17  & 7.24 & 3.81*  & I &  0.55 & 1.01  & 1.2:   &         &  \\ 
f &       & 45.6981   & 60.4590   &       &       &       &       &       &       &      & 6.12*: &   &  1.03 & 3.16  & 5.19:  &         &  \\ 
g &       & 45.8594   & 60.4770   &       &       &       &       &       &       &      &    no  &   &  5.34 & 15.44 & 17.24  & 11.02:  &  \\
h &       & 45.9134   & 60.4785   &       &       &       &       &       &       &      &    no  &   &  0.79 & 1.44  & 1.21   & 1.39    &  \\
\hline
5 & 17484 & 45.741884 & 60.465859 & 13.75 & 11.84 & 10.40 &  9.01 &  8.23 & 7.36  & 5.95 & 1.86*  &  I &        &       &         &        & \\  
6 & 17506 & 45.760247 & 60.456447 & 11.71 & 11.20 & 10.85 & 10.00 &  9.09 & 7.96  & 6.20 & 2.01*  &  I &        &       &         &        & \\ 
7 & 17595 & 45.836823 & 60.491218 & 14.23 & 13.05 & 11.85 &  9.72 &  8.83 & 7.89  & 6.15 & 2.46*  &  I &        &       &         &        & \\ 
\hline
s1 &      & 45.83351 & 60.46711   &        &       &      &       &        &      &      &        &     &  80  &        &         &         &  \\
s2 &      & 45.83082 & 60.46896   &        &       &      &       &        &      &      &        &     &  36  &        &         &         &  \\                  
\hline
\label{S201}
\end{tabular}\\
}
\end{table*}

\begin{table*}[tb]
\caption{Characteristics of the YSOs in the vicinity of Sh~201}
\resizebox{18cm}{!}{
\begin{tabular}{lllllllllll}
\hline\hline \\ [0.5ex]
Name &  $\chi^2$ & $M_*$   & $T_*$ & $M_{\rm disk}$  & $\dot{M}_{\rm disk}$    & $\dot{M}_{\rm env}$  & $L$     & $A_V(ext)$ & $i$     & Stage \\
     &          & (\msol) & (K)   &(\msol)          &(\msol~yr$^{-1}$)        &(\msol~yr$^{-1}$)     & (\lsol) & (mag)            & (\degr) &       \\          
\hline \\ [0.5ex]
a (5.45kpc) & 74 (--110) & 9.6 (5.1--17.8) & 4450 (--9500) & 1.1e-1 (3.3e-3--9.3e-1) & 4.4e-6 (3.2e-9--1.7e-4) & 1.6e-3 (1.9e-4--3.0e-3) & 1540 (1060--9300) & 9.5 (1.95--10.0) & 18 (--63) & I \\
b &  72 (--108) & 0.7 (0.5--1.0) & 3800               & 8.6e-4 (5.0e-4--6.9e-2) & 5.8e-7 (4.2e-8--1.6e-6)  & 8.4e-5 (--1.2e-4) & 14 (12--17) & 2.1 (--4.9)  & 18 & I \\  
c & 104 (--137) & 1.4            & 4218               & 1.1e-2                  & 1.1e-7                   & 4.7e-4            & 21          & 1.96         & 49 & I  \\
d & 121 (--160 )& 4.7 (2.9--5.7) & 13170              & 1.6e-4 (--1.8e-3)       & 5.5e-9 (--6.4e-10)       & 0 (1.7e-7)        & 560 (--740) & 3.53 (--5.0) & 18 (--81) & II \\
e & 55 (--88)   & 1.9 (0.2--4.7) & 4317 (2970--15800) & 1.4e-1 (--1.2e-5)       & 9.1e-8 (1.5e-10--6.4e-6) & 1.7e-5 (--0)      & 34 (11--410) & 3.62 (--5.0)& 69 (18--87) & ? \\
\hline
5 &  16 (--40) & 0.4 (--5.4) & 3512 (--17760)  & 3.9e-2 (--1.9e-4) & 2.0e-5 (--7.4e-11)  & 3.0e-6 (--0) & 42 (--730)   & 5.0 (--3.68) & 18 (87)     & II? \\ 
6 &  48(--72)  & 7.4 (--3.6) & 21340 (--5810)  & 3.5e-8 (--1.4e-3) & 1.7e-13 (--3.2e-9)  & 0 (--1.9e-7) & 2070 (--145) & 4.0 (--1.96) & 69 (32--87) & II? \\
7 &  18 (--42) & 1.2 (--3.4) & 4034 (--12800)  & 3.4e-4 (--3.5e-2) & 1.0e-8 (--5.1e-10)  & 6.4e-6 (--0) & 30 (--110)   & 3.17 --(5.0) & 18 (--81)   & II? \\ 
\hline
\label{S201bis}
\end{tabular}\\ 
}
\end{table*}

\begin{figure*}[tb]
\centering
 \includegraphics[width=150mm]{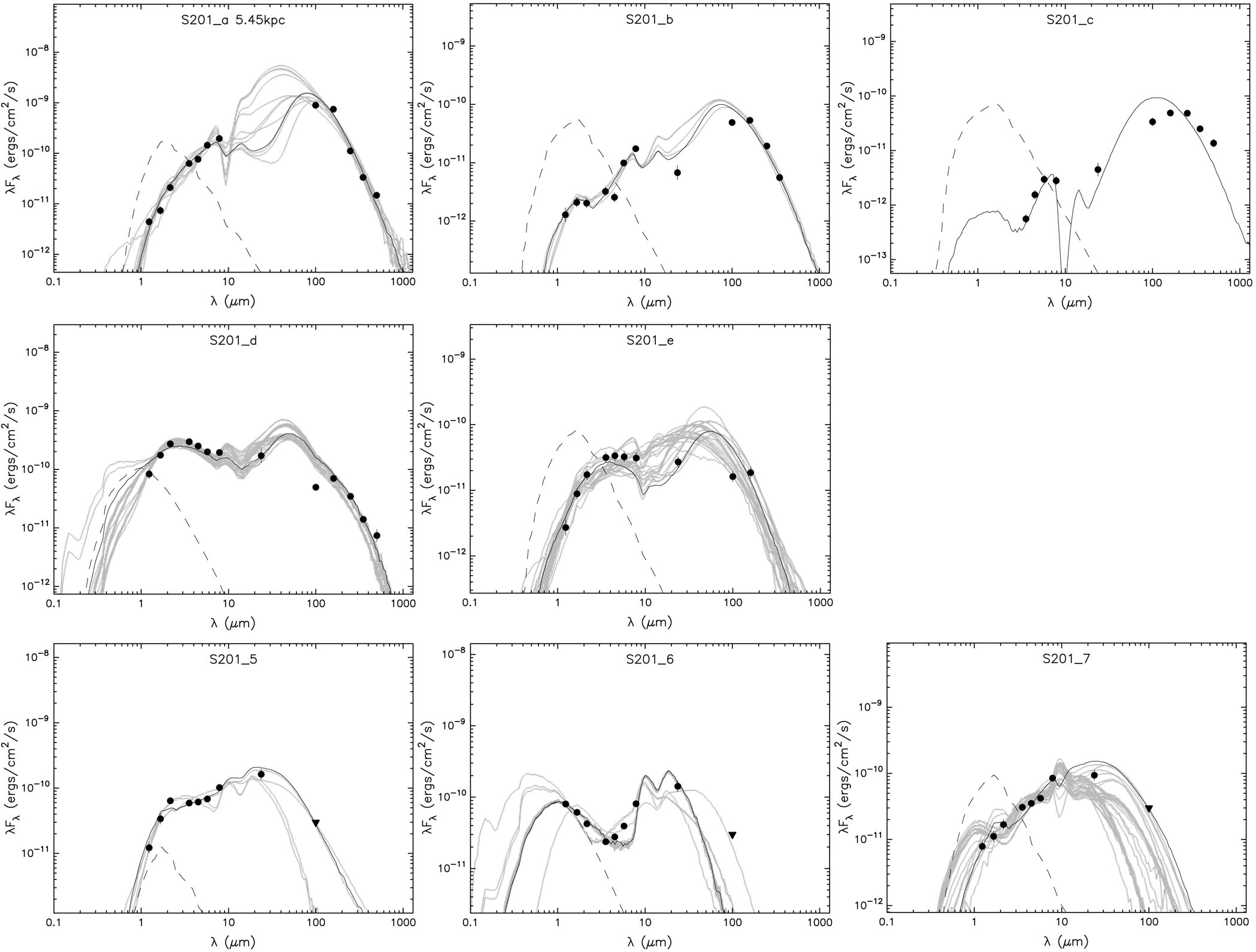}
  \caption{Spectral energy distributions of point sources in the Sh~201 field. We used the SED fitting-tool 
  of ROB07.}
  \label{S201SED}
\end{figure*}

Eight point sources, identified in Fig.~\ref{S201c}, are detected at 100~$\mu$m around Sh~201. Five of these point sources have {\it Spitzer} counterparts 
in the catalogue of KOE08.  Of these, four are Class~I sources and one is Class~II. Six of the point sources have 24~$\mu$m counterparts that
are not included in KOE08 catalogue.  We have measured the 24~$\mu$m fluxes of these six sources, as well as the {\it Herschel} fluxes. We give these data  
in Table~\ref{S201} (the values with an asterisk are our 24~$\mu$m measurements).

We fit the SEDs using the SED fitting-tool of ROB07, assuming an interstellar 
(external) visual extinction in the range 1.96~mag -- 5~mag; this last value is the mean extinction found for the whole associated cluster 
by Carpenter et al. (\cite{car93}). We list the parameters of the YSOs, as well as the external extinction for the best models, in Table~\ref{S201bis}. 
We show a few SEDs in Fig. \ref{S201SED}, and comment on individual sources below:

$\bullet$ S201-a: this strong source is located in the direction of a bright filament, south of Sh~201, and far from it. We show in Sect.~6.4 that, 
based on velocity arguments, this source is not associated with the W5 region, but lies far in the background at about 5.45~kpc. (The parameters given in Table~\ref{S201bis} 
are obtained for this distance, allowing an external extinction free in the range 1.95~mag to 10.0~mag, and not taking into account the saturated 24~$\mu$m emission.) Twelve models satisfy  $\chi ^2-\chi^2$(best) per data point $\leq$3; all point to a massive Stage~I YSO. Based on its relatively high temperature (Sect.~6.3.1) and luminosity  we suggest that this source contains a small cluster, and is similar to BRC13-a or BRC14-c1+c2. 

$\bullet$ S201-b: this source is located to the south of Sh~201, in the direction of a low column density region. The {\it Spitzer}-IRAC images show a small group of stars in this direction; one is dominant at  8.0~$\mu$m and at longer wavelengths. It is a rather low mass Stage~I YSO.

$\bullet$ S201-c: the fit for this source is not very good, and only one model satisfies $\chi ^2-\chi^2$(best) per data point $\leq$3. The model points to 
a Stage~I YSO.

$\bullet$ S201-d: this source is possibly located at the intersection of two filaments, to the south of Sh~201. This location is also a region of interaction between these  
filaments and the southern lobe of Sh~201. Many models satisfy $\chi ^2-\chi^2$(best) 
per data point $\leq$3; they are all rather similar, but the fit is not good. S201-d is probably an intermediate mass Stage~II YSO (the envelope accretion rate is low, null in some models), in agreement with KOE08. Fig.~\ref{S201c} shows an extended structure, jet-like, which seems to originate from the YSO (it is only observed at 100~$\mu$m and 160~$\mu$m); CO or OH lines, present in the PACS bands, and observed in regions of jets and shocks 
(van Kempen et al. \cite{van10}) could perhaps explain this feature. This feature, however, is not observed at 4.5~$\mu$m, a {\it Spitzer} band which traces shocks 
(Cyganowski et al. \cite{cyg08}, \cite{cyg09}). 

$\bullet$ S201-e: this source is located in the direction of the massive condensation on the east side, at the waist of Sh~201. Its SED is not constrained at all. The best model 
corresponds to a Stage~I YSO, of intermediate mass.  But numerous models satisfy $\chi ^2-\chi^2$(best) per data point $\leq$3, which span a wide range of parameters.  
 Thus, the nature of S201-e is very uncertain (from a low mass Stage~I YSO to a high mass Stage~II/III YSO).

The three other compact sources (f, g, h) have only {\it Herschel} measurements. Source f has a faint counterpart 
at all {\it Spitzer} wavelengths but it has not been measured by KOE08. Source g, which lacks a 24~$\mu$m counterpart and which is relatively bright at {\it Herschel} wavelengths, is probably a Class~0 source (Sect.~6.3.3).  
It is observed in the direction of the filaments bordering the massive condensation east of Sh~201. This location makes it a very good candidate for 
triggered star formation near the waist of Sh~201. We can estimate the mass of the g and h sources using the {\it Herschel} fluxes (Table~\ref{temperatureYSOs}); 
these masses are respectively $\sim$46~\msol\ and $\sim$4~\msol.  We hypothesize that sources g and h are Class~0 sources. 

The bright {\it Spitzer}-IRAC sources \#1, \#2, \#3, and \#4, located at the waist of Sh~201 and 
identified in Fig.~\ref{S201d}, have been classified as Class~II sources by KOE08. According 
to Ojha et al. (\cite{ojh04}), source \#2 is the main exciting star of Sh~201, whereas sources \#1 and \#3 are early B stars. Source \#4 
is associated with bright extended emission at 24~$\mu$m, as is source \#2; we suggest that it is also an early B star. 
Sources \#5, \#6, and \#7 are Class~I sources according to KOE08. They are bright 24~$\mu$m sources (no fluxes were given by KOE08; we measured these fluxes and  
give them with asterisks in Table~\ref{S201}). None of these sources are detected at 100~$\mu$m. 
Their SEDs are not well-constrained. Most of the models for YSOs \#5 and \#7 indicate Stage~II YSOS: their envelope accretion rates are low,  
even zero in some models, and their disks' masses are high (\#5 and \#7 only). But the mass and age of the central objects are not at all constrained. 
Four other sources have been classified as Class~I in the field of Fig.~\ref{S201d}, which have no 100~$\mu$m counterparts. 
They are \#17667, \#17531, \#17526, \#17620. Except for \#17667, they also have very faint or no 24~$\mu$m counterparts. 
Thus, their classification as Class~I is doubful.

\begin{figure}[tb]
 \includegraphics[width=8cm]{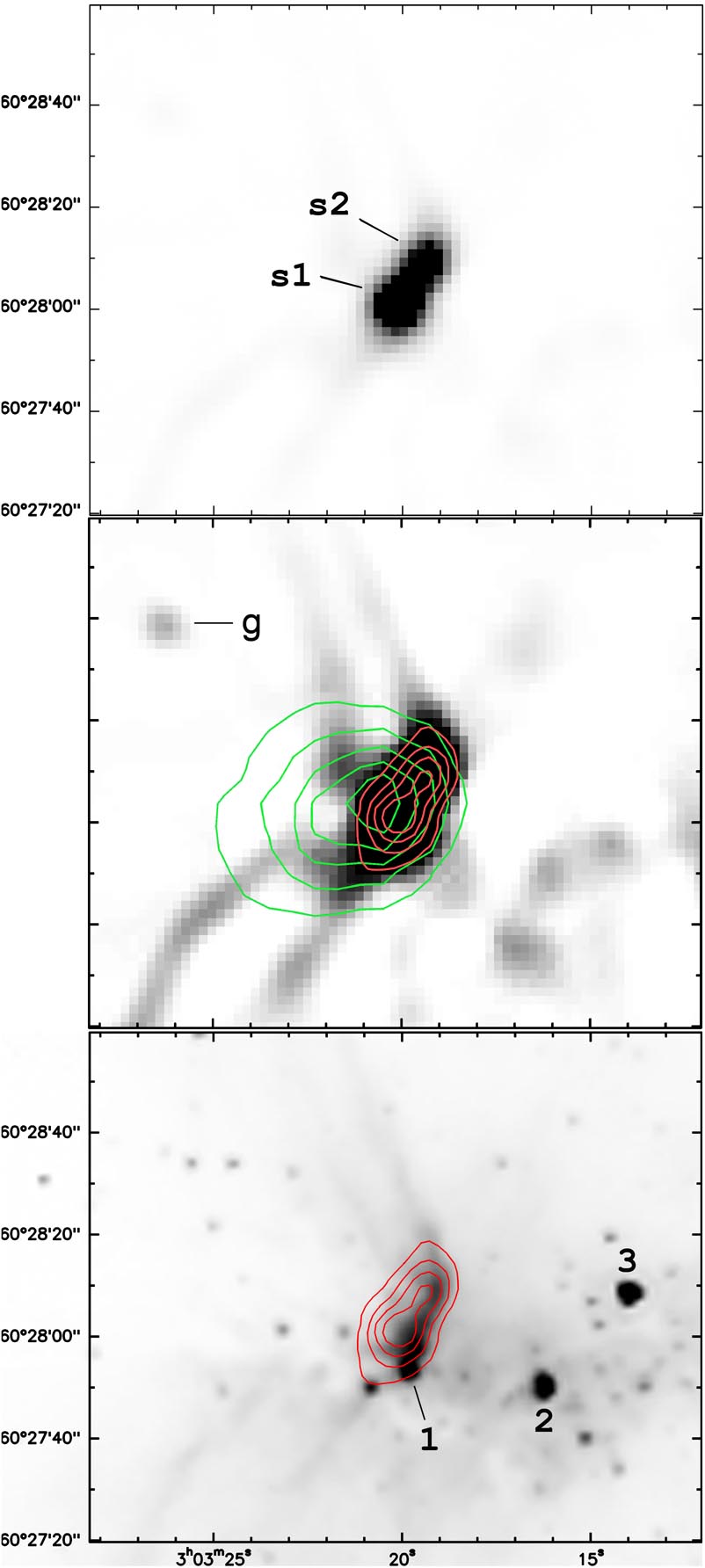}
  \caption{Young sources at the waist of Sh~201. {\it Top:} Unsharp-mask 100~$\mu$m image of the eastern portion of the waist of Sh~201 showing the presence of two point sources. 
  {\it Middle:} Same but deeper 
  image with the column density contours in green showing the dense condensation east of Sh~201.  The column density contour levels are 10$^{22}$, 3$\times$10$^{22}$, 4$\times$10$^{22}$, 5$\times$10$^{22}$, and 6$\times$10$^{22}$~cm$^{-2}$); 
  the red contours are for the 100~$\mu$m emission (levels of 0.5, 1.0, 1.5, and 2.0~Jy/beam). {\it Bottom:} The 100~$\mu$m contours are superimposed on the {\it Spitzer} 4.5~$\mu$m image showing the stars discussed in the text. The coordinates are J2000.} 
  \label{taille}
\end{figure}
 
Two other sources detected by {\it Herschel}, s1 and s2, lie at the eastern part of the waist of Sh~201, between the bright PAH-emitting filaments bordering the central 
ionized region and the massive molecular condensation. They are observed at 100~$\mu$m and 
160~$\mu$m, but are not detected at shorter wavelengths and are not easily discernable at longer wavelengths, probably due to the lower resolution and the presence of the adjacent parental molecular condensation. In Fig.~\ref{taille} 
we show an unsharp-mask image at 100~$\mu$m of the field containing these sources. To create the unsharp-mask, we median filtered the PACS 100~$\mu$m image 
with a square  window of 25\arcsec\ and subtracted the filtered image from the original one. Fig.~\ref{taille} (top) shows the two point sources, of which s1 is the brightest, and  Fig.~\ref{taille} (middle) shows the 100~$\mu$m filaments in the PDR. 
 The 100~$\mu$m flux of s1 and s2 are uncertain by a factor 2 because of their proximity to one another
and because of the bright underlying filaments. s1 lies at   
$\alpha$(2000)$=03^{\rm h}03^{\rm m}20.08^{\rm s}$, $\delta$(2000)$=+60\degr$28$\arcmin$01$\arcsec$
 and has a 100~$\mu$m integrated flux $\sim$80~Jy. s2 lies at  
$\alpha$(2000)$=03^{\rm h}03^{\rm m}19.35^{\rm s}$, $\delta$(2000)$=+60\degr$28$\arcmin$09$\arcsec$ 
and has a 100~$\mu$m integrated flux $\sim$36~Jy. Using the 24~$\mu$m detection limit $\sim$7~mag, we obtain a spectral index in the range 24~$\mu$m--100~$\mu$m 
$\alpha \geq$7.2 for s1 and $\geq$6.65 for s2. These very high indexes indicate that these sources are dominated by their envelopes. These sources are compact 
and their 100~$\mu$m fluxes are large. Thus, they are good candidates for high-mass Class~0 sources. Additionally, their location makes 
them good candidates for triggered star formation.  
An H$_2$O maser has been detected at 22~GHz by Blair et al. (\cite{bla80}) in the direction of star \#1 (Fig.~\ref{taille}); however, due 
to the resolution of these observations ($\sim$1.4\arcmin), it is unclear if it is associated with \#1, s1 or s2.

\section{The region between W5-E and W5-W: W5-NE}

\begin{figure}[tb]
\includegraphics[width=85mm]{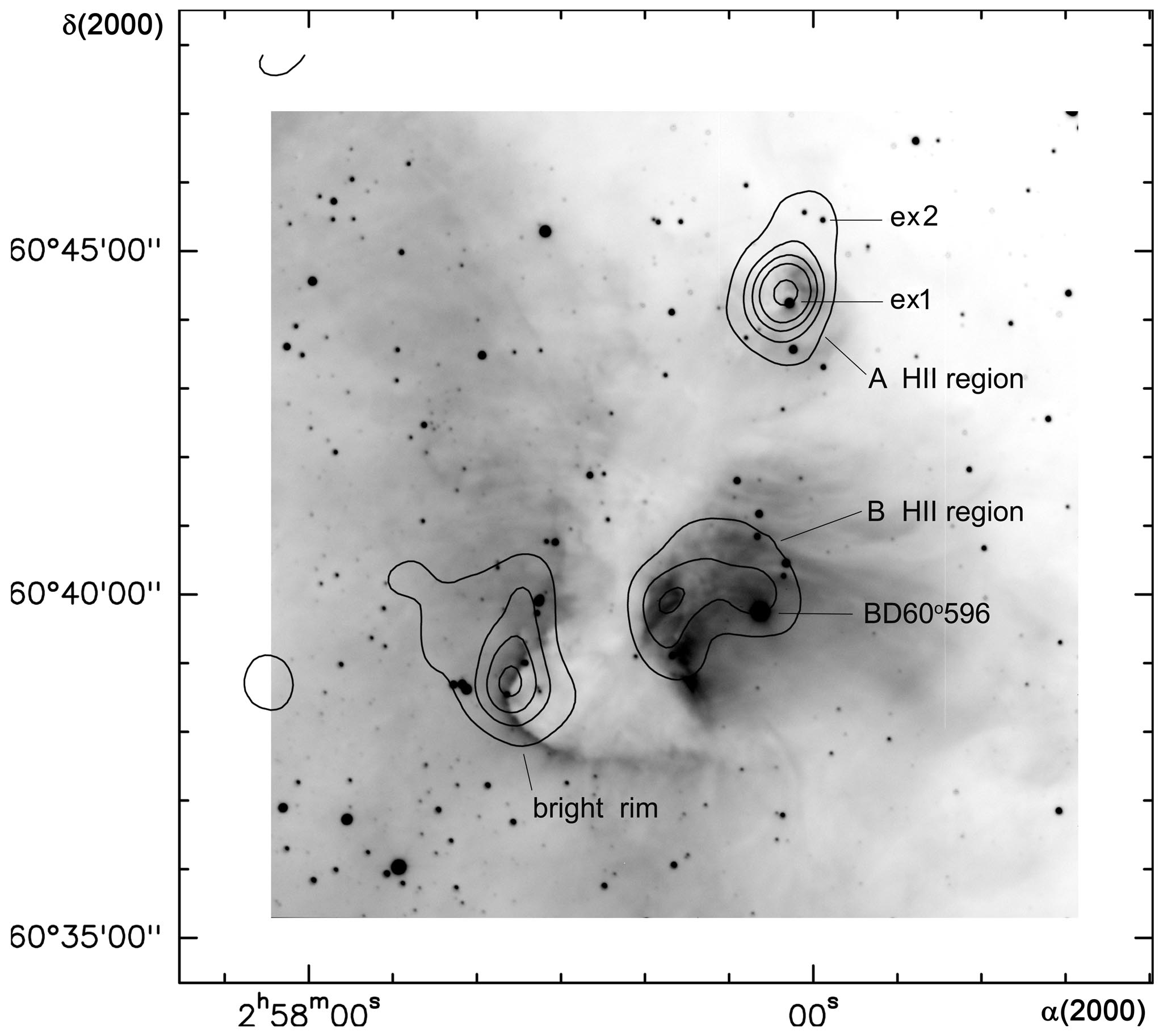}
  \caption{The ionized gas in the region between W5-E and W5-W. The NVSS radio-continuum contours at 20-cm (contour levels 0.002, 0.005, 0.0075, 0.01, and 0.015 Jy/beam) 
   are superimposed on a grey-scale image of the H$\alpha$ emission. The H$\alpha$ image has been obtained at the 1.2-m telescope of the 
   Observatoire de Haute Provence. Two \HII\ regions are present in this field, excited by the stars ex1 and BD+60$\degr$596. 
   Radio continuum emission is also observed at the border of a bright rim.}
  \label{between1}
\end{figure}
 
\begin{figure*}[tb]
\sidecaption
\includegraphics[width=12cm]{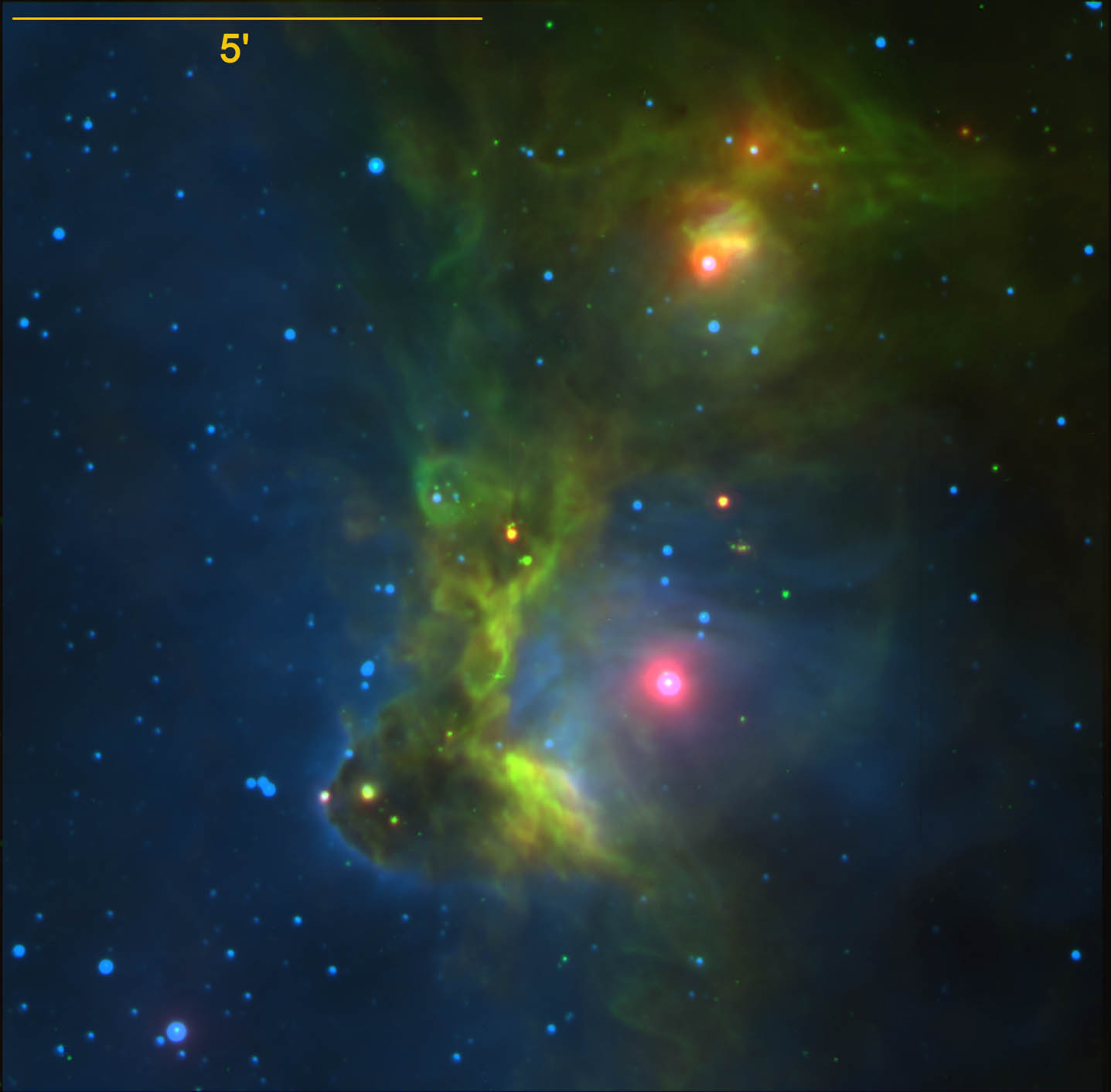}
  \caption{Colour composite image of the region located between W5-E and W5-W. Red is the {\it Spitzer}-MIPS emission from the hot dust at 24~$\mu$m, 
  green is the {\it Spitzer}-IRAC emission at 5.8~$\mu$m from the YSOs and the PDRs (PAH emission), blue is the H$\alpha$ emission of the ionized gas. }
  \label{between2}
\end{figure*}

Molecular material with a rather high column density of $\sim$5$\times$10$^{21}$~cm$^{-2}$ is present between W5-E and W5-W (see Fig.~\ref{column}). 
Two small \HII\ regions are  
located here, surrounded by dense material. A bright rim is also present, turned towards W5-E and its exciting star. Several 
point sources, listed in Table~\ref{between}, are detected at 100~$\mu$m and at longer wavelengths. 

\subsection{The \HII\ regions}

At least two \HII\ regions are present in the region located between W5-E and W5-W. In Fig.~\ref{between1} we show the H$\alpha$ emission of the ionized gas. 
The exciting star of the A \HII\ region is unknown; we identify it with an 
optical star at the center of the optical and IR nebula that we call ex1 (see Fig. \ref{between1} and Fig.~\ref{between2}). Wilking et al. (\cite{wil84}) 
call this IR source IR2, and, using several indicators, propose a spectral type in the range B0--B2 ZAMS for the central star. The B \HII\ region is excited by BD+60$\degr$596, a B1V star (Hiltner~\cite{hil56}) which lies at $\alpha$(2000)$=02^{\rm h} 57^{\rm m} 08^{\rm s}02$, 
$\delta$(2000)$=+60\degr$ 39$\arcmin$ 44$\farcs$1 (a few other spectral types have been proposed for this star: B3III, Garmany \& Stencel \cite{gar92};  
B0.5V-III, Zdanavicius \& Zdanavicius \cite{zda01}).

The NVSS survey at 20-cm 
shows radio-continuum emission from three regions, the A and B \HII\ regions and the border of a bright rim (Fig. \ref{between1}; 
two components are listed in the NVSS catalogue for the B \HII\ region).  
Fig.~\ref{between2} shows in green the {\it Spitzer}-IRAC 5.8~$\mu$m extended emission coming mostly from the PDR of the 
large W5-E and the PDRs surrounding the small \HII\ regions. 
Fig.~\ref{between2} also shows in red the {\it Spitzer}-MIPS 24~$\mu$m extended emission due to the hot dust located inside the ionized regions; it allows 
to locate the exciting stars. A 24~$\mu$m extended emission appears north of the A \HII\ region. It probably points to the presence of another 
early B star (there is an extention of the radio contours in this direction). 
An optical and IR star, that we call ex2, lies at the center of this extended emission.

The three stars (ex1, ex2, and BD+60$\degr$596) are in the catalogue of 
YSOs by KOE08. They are listed in Table~\ref{between}; ex2 and  BD+60$\degr$596 are classified as Class~III 
by KOE08; ex1 is classified as Class~II. This 
classification is uncertain as these stars are at the center of bright 24~$\mu$m extended emission. We can use the $K$ magnitude and $J-K$ colour to estimate 
(a very coarse estimate) the spectral type of these three stars. Using a distance of 2~kpc, the 2MASS photometry of ex1 indicates a B1--B2 spectral type with  
a visual extinction $\sim$3.4~mag. Using the NVSS flux density of 28.8~mJy for the A \HII\ region, we estimate a ionizing photon flux log(N$_{\rm Lyc}$)$=45.9$ 
pointing to a spectral type 
B2V (Smith et al. \cite{smi02}); if we consider the CGPS integrated flux of 0.13~Jy (Taylor et al. \cite{tay03}) we obtain  log(N$_{\rm Lyc}$)$=46.4$ ionizing photons per second, corresponding to a B1V 
spectral type (Smith et al. \cite{smi02}). These results are in agreement with Wilking et al. (\cite{wil84}) spectral type. The 2MASS photometry of ex2 indicates   
a B1--B2 spectral type but with a higher visual extinction $\sim$6.9~mag.  
The 2MASS photometry of BD+60$\degr$596 indicates a spectral type in the range B0V--B1V, with a low visual extinction $\sim$1.4 mag. The integrated CGPS flux density of 0.31~Jy gives 
log(N$_{\rm Lyc}$)$=46.9$ pointing to a B0.5V spectral type (Smith et al. \cite{smi02}); these results are in good agreement with the previous determinations. \\

\subsection{The YSOs}

\begin{figure*}[tb]
\sidecaption
\includegraphics[width=12cm]{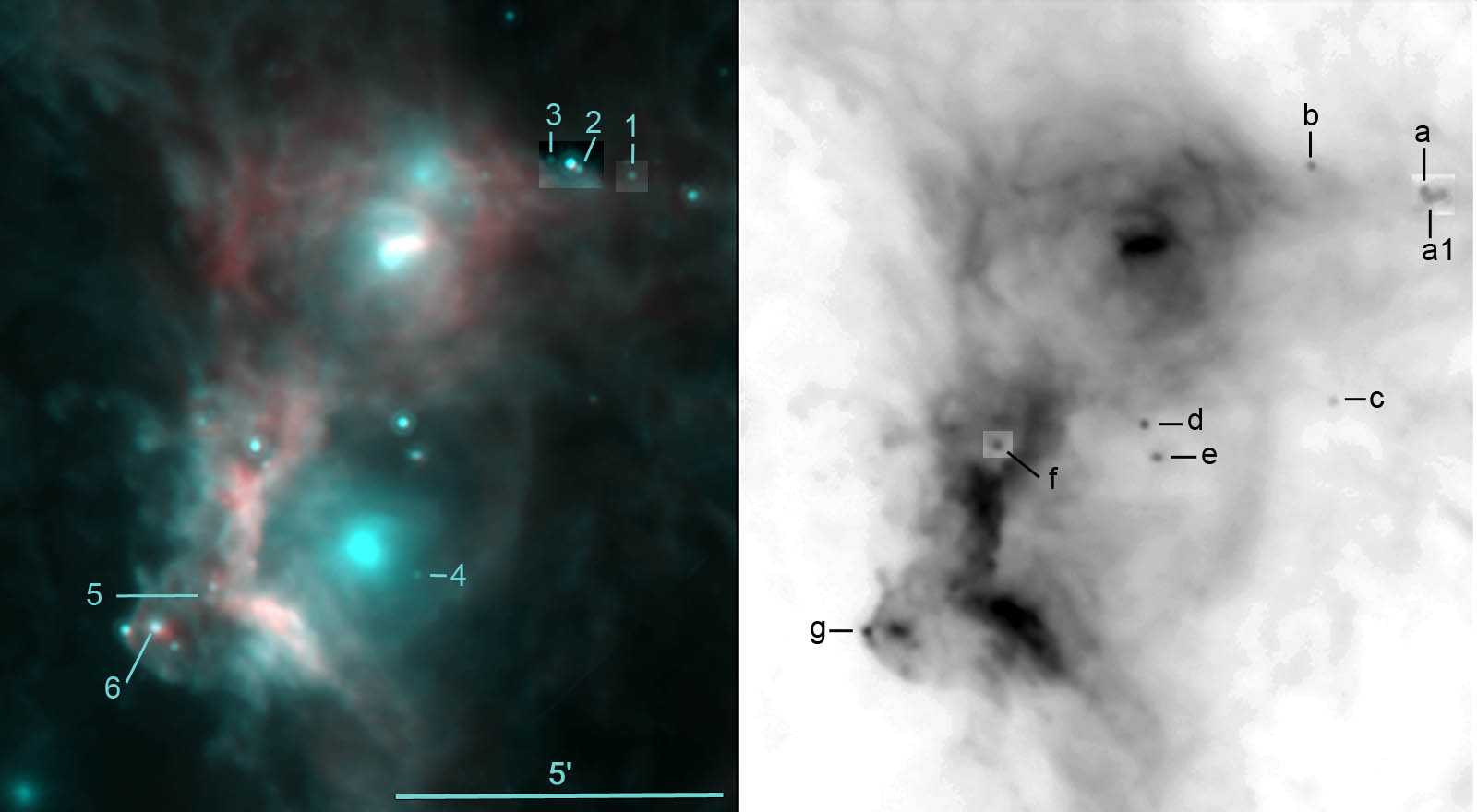}
  \caption{Colour composite image of the region between W5-E and W5-W. {\it Left:} Red is the {\it Herschel}-PACS emission at 100~$\mu$m, and  
  turquoise is the {\it Spitzer}-MIPS emission at 24~$\mu$m. {\it Right:} The point sources detected at 100~$\mu$m and discussed in the text 
  are identified on a grey-scale image of the 100~$\mu$m emission (given in logarithmic units).}
  \label{between3}
\end{figure*}

\begin{figure*}[tb]
\sidecaption
\includegraphics[width=12cm]{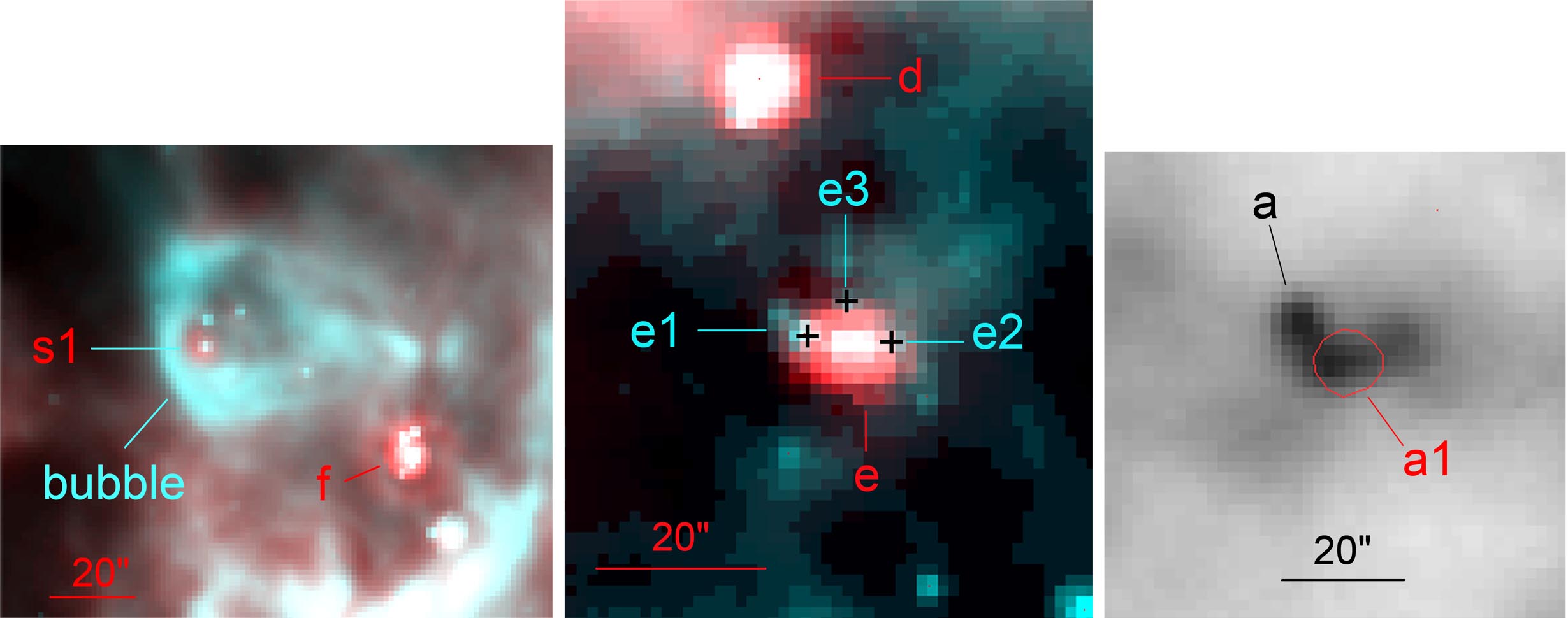}
  \caption{Features discussed in the text. {\it Left:} Colour composite image of a small bubble; red is the 24~$\mu$m image, turquoise is 
  the {\it Spitzer} 8.0~$\mu$m image showing the PAH emission. {\it Middle:} YSO IB-e and its vicinity; red is the 100~$\mu$m emission, turquoise is 8.0~$\mu$m emission. 
  {\it Right:} YSO-a; the underlying grey image is the 100~$\mu$m emission, the red contour shows the position of the 24~$\mu$m Class~I source a1.}
  \label{between4}
\end{figure*}

Figure~\ref{between3} shows the {\it Herschel}-PACS 100~$\mu$m image of the field. The colour image allows us to compare the location of the hot dust 
emitting at 24~$\mu$m with the dust emitting at 100~$\mu$m. The bright extended 24~$\mu$m emission comes from the inside of the \HII\ regions 
and from the PDRs bordering them, whereas 
the extended 100~$\mu$m emission comes only from the PDRs. The A and B \HII\ regions are rather different. 
The B \HII\ region is bordered by a dense PDR on its east side and seems to open on the west side. The low extinction of its exciting star 
indicates that it lies in front of most of the material along this line of sight. The A \HII\ region is more compact, but its influence seems 
to extend over a large region. The extinction of its exciting star is larger than that of BD+60$\degr$596, indicating that it is more embedded. 
A bright feature at 100~$\mu$m, bar-like, lies very close in projection to the ex1 exciting star.

\begin{table*}[h]
\caption{Photometry of point sources: the region between W5-E and W5-W}
\resizebox{18cm}{!}{
\begin{tabular}{llllrrrrrrrrrrrrrr}
\hline\hline \\ [0.5ex]
Name & \# & RA(2000) & Dec(2000) & $J$    & $H$   & $K$   & [3.6] & [4.5]  & [5.8]  & [8.0] & [24]     & Class & S(100)  & S(160)   & S(250)  & S(350) & S(500) \\
     &         & ($\degr$) & ($\degr$) & mag & mag & mag & mag & mag & mag & mag & mag & & Jy & Jy & Jy & Jy & Jy \\          
\hline \\ [0.5ex]
a  &       & 44.11650  & 60.75322  &       &       &       &       &       &      &      &      &    & 1.12  &  2.53: & 3.42:  &  3.03:  &  \\
a1 & 12110 & 44.111756 & 60.751366 & 15.71 & 13.87 & 12.72 & 11.29 & 10.58 & 9.34 & 7.54 & 3.73 &  I &       &        &       &       &        \\
b & 12334 & 44.174880 & 60.759755 &       &       & 14.18 & 12.13 & 10.90 & 9.97 & 8.80 & 4.16 &  I  & 2.35  &  5.03  & 3.59: & 2.89: &  \\  
c & 12297 & 44.164765 & 60.700052 & 15.78 & 13.65 & 12.24 & 10.54 &  9.83 & 9.22 & 8.62 & 6.13 & II & 1.06 & 2.54  & 2.90 & 2.10: & \\
d & 12648 & 44.263389 & 60.694426 & 12.71 & 11.32 & 10.51 &  9.29 &  8.49 & 7.74 & 6.63 & 2.75 &  I & 3.85 & 5.97 & 6.58 & 5.06: & 2.8: \\             
e &       & 44.25698  & 60.68601  &       &       &       &       &       &      &      &      &     & 3.21  & 9.76 & 6.70:  & 3.30: & 2.0:: \\
e1 & 12641 & 44.259853 & 60.686344 &       &       & 13.70 & 11.54 & 10.54 & 9.78 & 9.01 &     &  I  &       &       &       &       & \\
e2 & 12620 & 44.254253 & 60.686155 & 15.17 & 13.60 & 12.72 & 11.49 & 10.90 &      &      &     &  II &        &       &      &       &  \\
e3 & 12632 & 44.257101 & 60.687345 &       &       & 13.47 & 12.02 & 11.33 & 11.18 & 10.48 &   &  II &        &        &      &      & \\        
f & 12922 & 44.339923 & 60.688855 & 12.49 & 11.09 & 10.03 &  8.76 &  8.22 & 7.60 & 6.88 & 2.82 & II & 2.62  &  5.80 & 6.94: &       & \\
g & 13165 & 44.407821 & 60.642469 & 12.00 & 10.99 & 10.38 &  9.56 &  9.19 & 8.34 & 7.24 & 3.11 & II & 5.15: &        &       &       & \\
\hline
ex1 & 12666 & 44.268587 & 60.736758 &  10.29  & 10.01 & 9.84 & 9.76 & 9.59 & 9.13 & 7.76 & 1.91 & II  &       &       &       &       & \\
ex2 &             12609 & 44.251992 & 60.756720 & 11.13 & 10.44 & 10.06 & 9.87 & 9.80 & 9.68 & 9.16 &       & III &       &       &       &       &  \\
BD+60$\degr$596 & 12717 & 44.283529 & 60.662250 & 8.80 & 8.72  &  8.72  & 8.79 & 8.74 & 8.51 & 7.42 &       & III &       &       &       &       &  \\
\hline
1 & 12234 & 44.143079  & 60.756463 &      &       & 14.63 & 12.23 & 11.10 & 10.23 & 9.45  & 5.94 &  I  &    &       &        &       &  \\  
2 & 12313 & 44.170433  & 60.758392 &      &       & 14.89 & 12.49 & 11.28 & 10.48 & 10.05 & 5.98 &  I  &     &      &        &       &  \\
3 & 12381 & 44.185222  & 60.760892 &      &       &       & 14.56 & 13.27 & 11.98 & 11.29 & 7.08 & I   &     &      &        &       &  \\
4 & 12631 & 44.256847  & 60.655770 & 17.40 & 16.31 & 14.47 & 11.84 & 10.80  & 9.95 & 9.01 & 5.64 & I   &     &      &        &       &  \\
5 & 13019 & 44.366178  & 60.650644 &       & 15.68 & 13.95 & 11.90 & 11.16 & 10.18 & 9.12 &      & I   &     &      &        &       &  \\
6 & 13118 & 44.392491  & 60.643065 & 12.57 &       & 11.22 & 10.48 & 10.24 & 7.77  & 5.82 & 2.98 & I   &     &      &        &       &  \\
\hline
s1 & 13025 & 44.367277 & 60.695293 & 12.27 &  11.86 & 11.52 & 10.91& 10.59 & 10.49 & 10.35 & 4.46 & II &     &      &        &       &  \\  [0.5ex]
\hline
\label{between}
\end{tabular}\\
}
\end{table*}

\begin{table*}[h]
\caption{Parameters of the YSOs in the region between W5-E and W5-W, obtained with the SED fitting tool of ROB07}
\resizebox{18cm}{!}{
\begin{tabular}{llllllllll}
\hline\hline \\ [0.5ex]
Name     &  $\chi^2$ & $M_*$ & $T_*$   & $M_{\rm disk}$  & $\dot{M}_{\rm disk}$  & $\dot{M}_{\rm env}$  & $L$     & $i$     & Stage \\
         &           & (\msol )& (K)   &(\msol)          &(\msol~yr$^{-1}$)      & (\msol~yr$^{-1}$)    & (\lsol) & (\degr) &       \\          
\hline \\ [0.5ex]
b       & 116 (--146) & 0.9 (0.7--1.4)  & 3920  & 6.3e-3 (1.0e-3--1.1e-2) & 4.9e-7 (--1.0e-7) & 2.6e-4 (--4.7e-4) & 19 (14--21) & 18 (--41) & I \\ 
c       & 116 (--177)    & 1.9 (--1.5)     & 4300  & 2.1e-4 (--8.6e-4)       & 5.6e-10 (--2.0e-8) & 2.4e-4 (--1.4e-4) & 29 (--20)   & 49 (56) & I \\   
d (1.96 mag) & 273 (--289) & 1.9 (3.3)  & 4150  & 2.6e-1 (--8.4e-4) & 2.5e-6 (--6.4e-9) & 1.2e-4 (--6.2e-4) & 69 (--80) & 18 (41) & I \\
d (10.38 mag)& 199         & 3.2        & 4420  & 8.4e-4            & 6.4e-9            & 6.2e-4            & 80        & 31      & I \\
g       & 16 (--43) & 1.5 (--5.0) & 4224 & 1.7e-2 (3.2e-4--1.3e-1)& 1.5e-6 (4.5e-11--2.7e-6)& 1.5e-4 (1.4e-5--5.9e-4) & 30 (--120)  & 18 (--75) &  I? \\ [0.5ex]
\hline
\label{betweenbis}
\end{tabular}\\
}
\end{table*}

\begin{figure*}[h]
\centering
 \includegraphics[width=150mm]{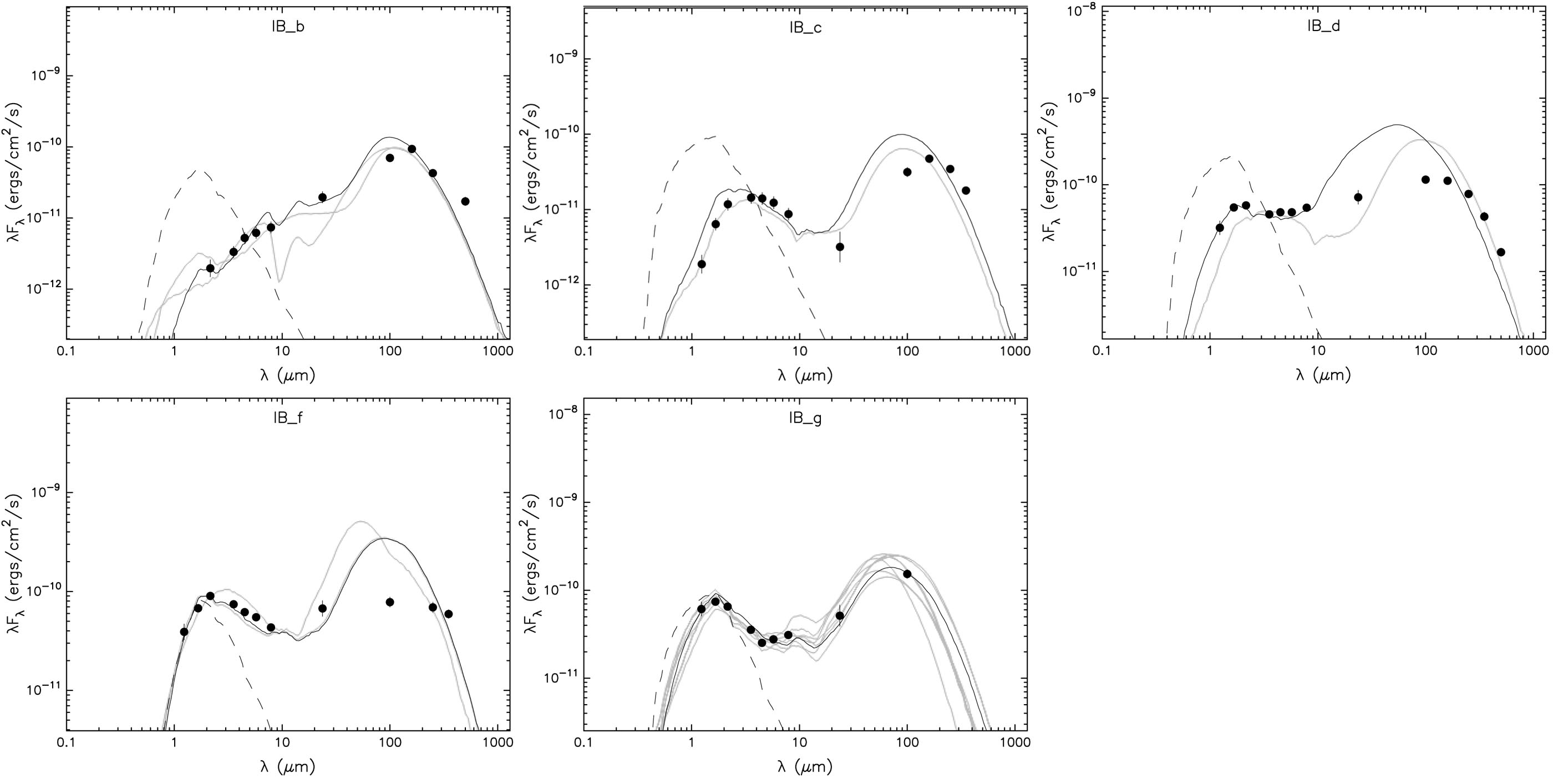}
  \caption{Spectral energy distributions of point sources in the region between W5-E and W5-W. We used the SED fitting tool 
  of ROB07.}
  \label{IBSED}
\end{figure*}

Seven 100~$\mu$m point sources are present in this field, identified in   
Fig.~\ref{between3}, and listed in Table~\ref{between}. We fit some SEDs using the SED fitting-tool of ROB07 (assuming an external visual extinction of 1.96~mag).
The obtained parameters are given in Table~\ref{betweenbis}. We comment on individual point sources below (the name IB refers to ``in between'').

$\bullet$ IB-a: This source is the brightest of a group of possibly three 100~$\mu$m sources. A detailed view of this 
region is given in Fig.~\ref{between4}. It has a very faint 24~$\mu$m counterpart, and thus is a 
candidate Class~0 source. Using only the {\it Herschel} fluxes we obtain for the envelope of this source a dust temperature of 13.7~K and a mass 
of 16.5~\msol\ (Table \ref{temperatureYSOs}). The Class~I YSO a1 in KOE08 catalogue has a faint 100~$\mu$m counterpart, too faint to be measured accurately. 
The western 100~$\mu$m source has no 24~$\mu$m counterpart at all; it is also a candidate Class~0 YSO.

$\bullet$ IB-b: This source is a Class~I YSO according to KOE08. We show its SED in Fig.~\ref{IBSED}. Three models satisfy $\chi ^2-\chi^2$(best) per data point $\leq$3. 
The SED of this source  points to a Stage~I YSO, with a central object of about one solar mass, an envelope of 8.7~\msol\ (Table~\ref{temperatureYSOs}),  
and a rather high accretion rate. 

$\bullet$ IB-c: This source is a Class~II YSO according to KOE08; however a far-IR counterpart appears to be present on all {\it Herschel} images. 
The SED is not well fit as shown by Fig.~\ref{IBSED}; however it points to a Stage~I YSO.  We cannot exclude the existence of two distinct sources: a Class~II YSO adjacent
to a condensation of 8.5~\msol\ (Table \ref{temperatureYSOs}).

$\bullet$ IB-d: This source is a Class~I according to KOE08. We agree with this classification, but again the SED fit is not good.  The parameters of the model 
give a Stage~I source with a massive envelope (20.3~\msol; Table \ref{temperatureYSOs}), and a high envelope accretion rate.  We did try the fits with the extinction as a free parameter;
the fit is better (according to $\chi^2$) with a higher external extinction (10.38~mag), but the fitted model is very similar to the previous ones. 

$\bullet$ IB-e: This 100~$\mu$m source is slightly elongated in the east-west direction. It is a complicated region as a small group of about half a dozen stars 
lies in its direction (see Fig.~\ref{between4}). Only three YSOs (e1, e2, and e3) are in KOE08 catalogue. None of them are associated 
with the 100~$\mu$m source, as shown by Fig.~\ref{between4}. (The 24~$\mu$m source resembles the 100~$\mu$m source; it is elongated east-west).  
Two sources appear at 8.0~$\mu$m in the direction of IB-e (which probably explains its elongation); they are not present in the catalogue of KOE08. 
Source e is probably associated with these two sources, which thus probably are Class~I YSOs.  

$\bullet$ IB-f:  This source has been identified as a Class~II YSO by KOE08. Two other faint stars lie nearby, but f is dominant at all {\it Spitzer} wavelengths. 
The SED is not well fitted by the SED fitting-tool of ROB07, as shown by Fig.~\ref{IBSED}. This source is V$^*$~LW~Cas, a variable star of 
uncertain nature (Kukarkin et al. \cite{kuk71}), not a cataclysmic variable (Downes \& Shara \cite{dow93}), 
not a symbiotic star (Henden \& Munari \cite{hen08}). It is probably evolved, and we do not know if it is  associated  
or not with W5-E.  

$\bullet$ IB-g: This source is a Class~II YSO according to KOE08. It has a 100~$\mu$m counterpart, but nothing is detected at 
longer wavelengths. Numerous models have an acceptable fit to the SED. The model parameters point to an intermediate mass YSO,
probably in Stage~I; but the parameters of the disk are not well contrained.

The field of Fig.~\ref{between3} contains six other Class~I YSOs according to KOE08; they are identified in this figure. YSOs a, a1, b, \#1, \#2, and \#3 
are part of a high column density ridge extending east-west; ex2 is possibly part of this ridge. This structure therefore appears to be an active star forming region, but its origin is unclear. 
YSOs \#1 and \#2 have a very faint counterpart at 100~$\mu$m (too faint to be measured). YSO \#4 is observed in the direction of the B \HII\ region; 
it has a very faint counterpart at 100~$\mu$m. YSO \#5 has no 100~$\mu$m counterpart. It is faint at 24~$\mu$m and therefore its classification by KOE08 as a Class~I is uncertain. 
YSO \#6 is peculiar. It has an extended and bright 100~$\mu$m counterpart, located on the border of a bright condensation. A close examination of the 
{\it Spitzer} images shows that this source is also extended at all {\it Spitzer} wavelengths. This condensation, centered at 
$\alpha$(2000)$=02^{\rm h} 57^{\rm m} 33^{\rm s}02$, $\delta$(2000)$=+60\degr$ 27$\arcmin$ 32$\farcs$ (position of the peak column density, 6.1$\times$10$^{21}$~cm$^{-2}$)
has an integrated flux at 250~$\mu$m $\sim$10.5~Jy in an aperture of radius 20\arcsec. The temperature map indicates a mean temperature $\sim$23~K for 
this extended structure and we estimate its mass to be $\sim$6\msol.

Another peculiar structure is present in the field. It is a small bubble (diameter $\sim$35\arcsec, thus $\sim$0.3~pc), well traced by the PAH emission at 8.0~$\mu$m (see Fig.~\ref{between4}). 
It contains three YSOs, classified as Class~II by KOE08. One of them is especially bright at 24~$\mu$m and possibly extended; it has no 100~$\mu$m counterpart.
We propose that this source, identified as s1 in Table~\ref{between} and Fig.~\ref{between4}, is a B star heating the nearby dust. (Its $K$
 magnitude and $J-K$ colour could correspond to a B5--B7 star with an extinction $A_V\sim$5.6~mag.)

\section{Isolated point sources}

\begin{figure*}[h]
\sidecaption
 \includegraphics[width=12cm]{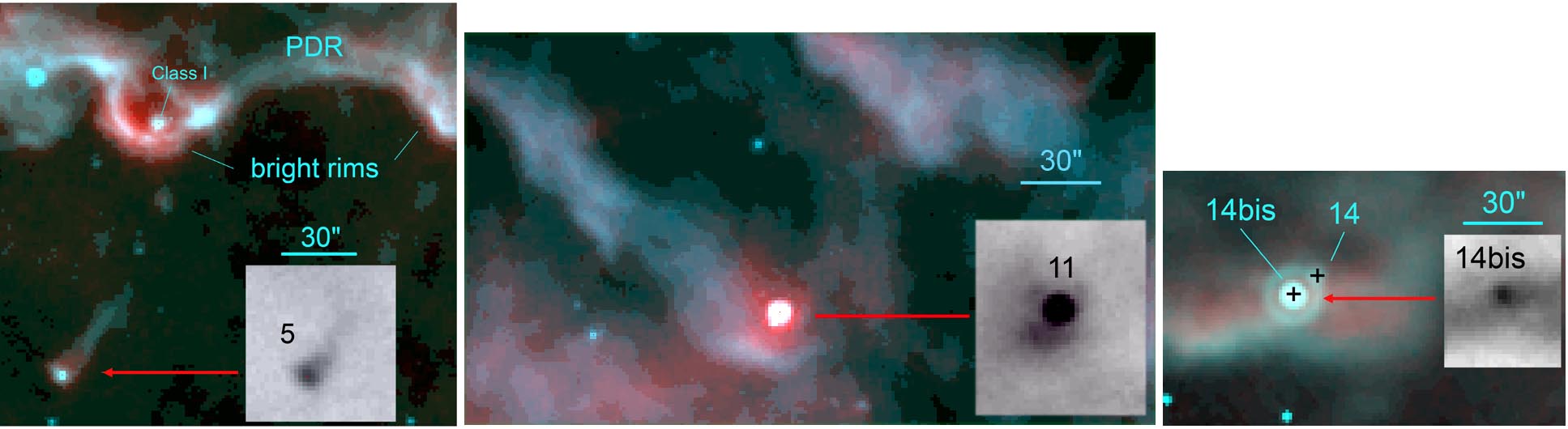}
  \caption{Colour composite images of three isolated sources discussed in the text. Red is the {\it Herschel} 100~$\mu$m emission and turquoise is for {\it Spitzer}
8.0~$\mu$m emission (isolated 5 and 11) or the 24~$\mu$m emission (isolated 14bis). The grey inserts show, at the same scale, the 100~$\mu$m source.}
  \label{isolated5+11}
\end{figure*}

There are some 100~$\mu$m point sources in the W5-E field which lie outside the regions discussed previously. 
We give their measured {\it Herschel} fluxes in Table~\ref{isolated}. Four of these sources lie outside the coverage of the {\it Spitzer} images and 
therefore they have no measured {\it Spitzer} fluxes. 
Four other sources have faint {\it Spitzer} counterparts that have not been measured by KOE08. Another source (\#4) is isolated, bright at 8~$\mu$m and 24~$\mu$m, but curiously, is absent in the KOE08 catalogue. When possible we have meaured the 24~$\mu$m fluxes (these measurements have an asterisk in Table~\ref{isolated}). 
We attempt to classify them, using the spectral index of the SED between 24~$\mu$m and 100~$\mu$m 
(definition in Sect.~5). According to this metric, sources \#1, \#4, \#5 (in agreement with KOE08 classification), \#6, \#7, \#8, and \#9 are Class~I sources.,  
sources \#11 is a flat spectrum source, and source \#12 is a Class~II source (again in agreement with KOE08).

\begin{table*}[tb]
\caption{Photometry of isolated point sources detected at 100~$\mu$m}
\resizebox{18cm}{!}{
\begin{tabular}{llllrrrrrrrrrrrrrr}
\hline\hline \\ [0.5ex]
Name & \# & RA(2000) & Dec(2000) & $J$    & $H$   & $K$   & [3.6] & [4.5]  & [5.8]  & [8.0] & [24]     & Class & S(100)  & S(160)   & S(250)  & S(350) & S(500) \\
     &         & ($\degr$) & ($\degr$) & mag & mag & mag & mag & mag & mag & mag & mag & & Jy & Jy & Jy & Jy & Jy \\          
\hline \\ [0.5ex]
1   &       & 44.96406   & 59.98754   &        &      &      &      &      &       &      &  7.00*  &  & 0.36 & 0.53: & & & \\   
2   &       & 46.04535   & 60.59474   &        &      &      &      &      &       &      &         & outside {\it Spitzer} & 0.64 & 1.04 & 0.39: & & \\
3   &       & 45.86957   & 60.64833   &        &      &      &      &      &       &      &         & outside {\it Spitzer} & 0.27 & & & & \\
4   &       & 45.72951   & 60.65243   &        &      &      &      &      &       &      & 4.23* &  & 1.30 & 1.07 & 0.49: & & \\
5   & 14148 & 44.65653   & 60.69825 &        & 15.85 & 14.03 & 12.08 & 11.13 & 10.19 & 9.18 & 5.44 & I & 0.57 & 1.33 & 0.91 & 0.58: & \\ 
6   & 17344 & 45.62620   & 60.69937 &        & 15.57 & 14.72 & 13.46 & 12.93 & 12.08 & 10.12 & 6.18* & II & 0.32 & & & & \\
7   &       & 45.56116   & 60.70956   &         &     &      &      &      &       &      &  6.41*  &  & 0.44 & 0.53: & & & \\
8   &       & 45.23067   & 60.75611   &        &      &      &      &      &       &      &  5.86*  &   & 0.31 & 0.53: & & & \\   
9   &       & 44.68234   & 60.78960  &        &      &      &      &      &       &      &  6.80*  &    & 0.84 & 0.99 & 0.59: & & \\
10  &       & 46.30267   & 60.81229   &        &      &      &      &      &       &      &         & outside {\it Spitzer} & 0.68 & 1.01 & 0.99: & & \\
11  & 12921 & 44.33975   & 60.88602 & 14.07 & 12.81 & 11.79 & 9.85 & 8.93 & 8.04 & 6.62 &  2.48 &  I  & 2.69 & 5.21 & 4.76 & 3.19 & 1.69: \\
12  & 14630 & 44.77136   & 60.90115 & 10.82 & 9.65 &  8.67 & 7.37 & 6.75 & 6.04 & 4.71  & 1.61  & II & 1.07 & 0.91 &0.45: & & \\
13  &       & 45.81908   & 60.96289   &        &      &      &      &      &       &      &         & outside {\it Spitzer} & 0.78 & 1.88 & 1.48 & 1.45: & \\
14  & 15939 & 45.06133   & 60.99216 &       &      &      & 13.19 & 12.32 & 11.60 & 10.62 & 7.15 & I &  & & & & \\        
14bis & 15965 & 45.06639 & 60.99005 & 14.73 & 12.94 & 11.59 & 9.90 & 9.13 & 8.61 & 7.56 & 4.10 (3.92*) & II & 0.37 & & & & \\  [0.5ex]       
\hline
\label{isolated}
\end{tabular}\\
}
\end{table*}

\begin{table*}[tb]
\caption{Parameters of the isolated YSOs, obtained with the SED fitting tool of ROB07}
\resizebox{18cm}{!}{
\begin{tabular}{llllllllll}
\hline\hline \\ [0.5ex]
Name & $\chi^2$ & $M_*$  & $T_*$ & $M_{\rm disk}$ & $\dot{M}_{\rm disk}$  & $\dot{M}_{\rm env}$  & $L$      & $i$  & Stage \\
     &          & (\msol )& (K)   &(\msol)          &(\msol~yr$^{-1}$)  &(\msol~yr$^{-1}$) & (\lsol)  & (\degr) &       \\          
\hline \\ [0.5ex]
5     & 62 (--95)  & 0.7 (0.4--0.9) & 3900 & 8.6e-4 (5.1e-4--6.6e-2) & 1.9e-9 (--1.7e-7)  & 1.7e-4 (--2.3e-5) & 9 (6--11) & 31 (18--69) & I?  \\
11    & 259  & 4.6  & 11300 & 2.9e-4 & 9.1e-11 & 1.5e-7 & 440 & 87 & II/III? \\
11*   & 21 (--39)  & 0.8 (--1.4) & 3931 & 6.6e-2 (3.4e-3--1.0e-1) & 5.0e-6 (--3.4e-9) & 4.7e-5 (1.5e-5--6.9e-5) & 26 (21--35) & 18 & I \\
12    & 26 (--59)  & 4.9 & 16200 & 7.3e-2  & 9.9e-7  & 0 & 477    & 18  & II \\   
14bis ($A_V=1.96$~mag) & 25 (--52)  & 3.8 (1.9--7.9) & 13940 & 1.3e-2 (2.0e-4--1.4e-1) & 3.8e-8 (1.5e-10--1.1e-7) & 0 (1.7e-5) & 179 (34--2730) & 81 (--69)  & II \\
14bis ($A_V=8.7$~mag) & 2.7 (--29.7) & 3.1 (--2.6) & 12215 (--9400) & 7.3e-2 (--2.7e-3) & 9.0e-8 (7.0e-10--1.2e-7) & 0 & 87 (--43) & 75 (--18) & II \\ [0.5ex]
\hline
\label{isolatedbis}
\end{tabular}\\
}
\end{table*}

\begin{figure*}[tb]
\centering
 \includegraphics[width=150mm]{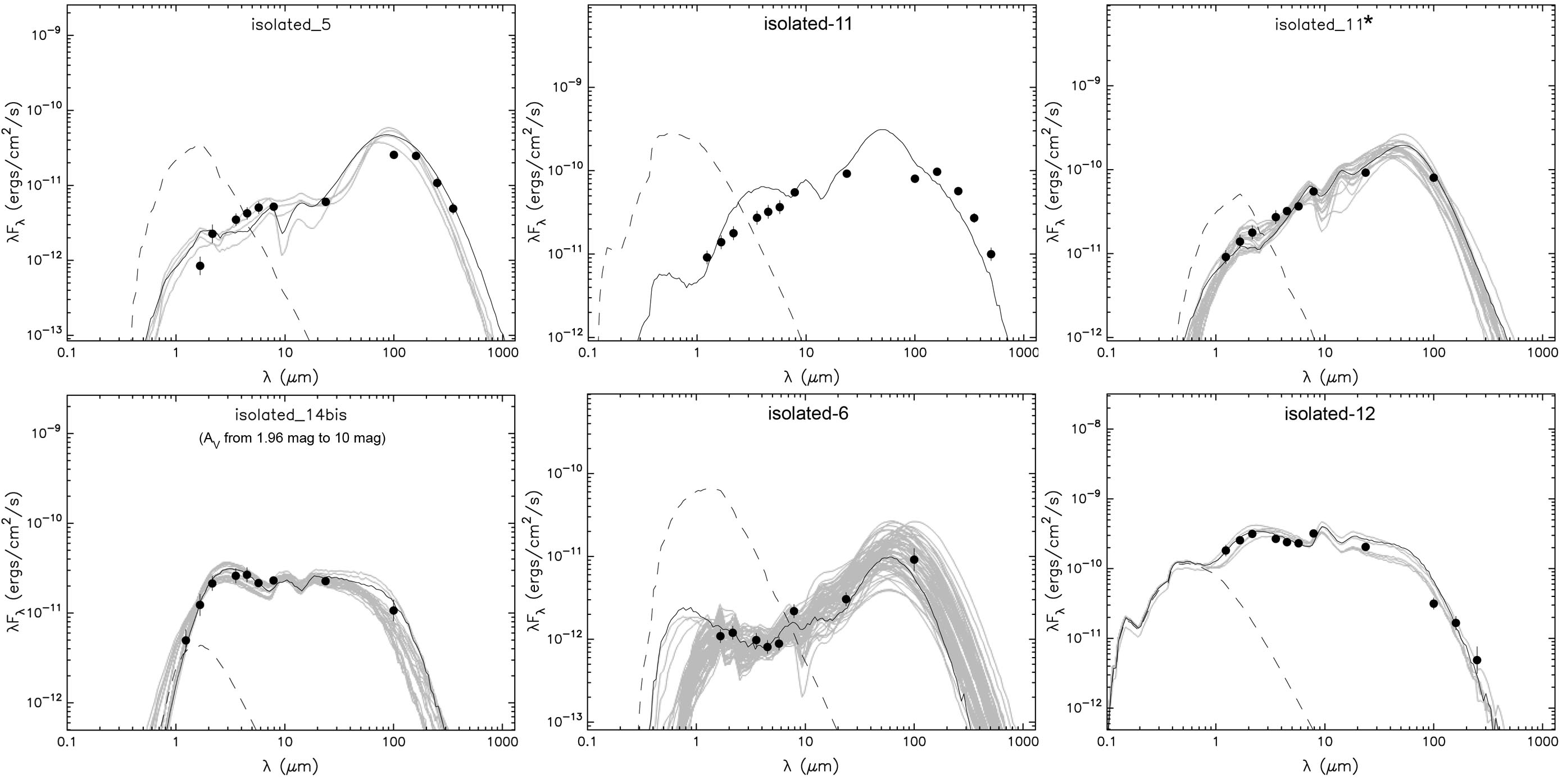}
  \caption{Spectral energy distributions of a few isolated point sources. We used the SED fitting-tool of ROB07. (In isolated-11$^*$ we use only the flux at 100~$\mu$m.) }
  \label{isolatedSEDs}
\end{figure*}

When possible we used the SED fitting tool of ROB07, assuming an interstellar visual extinction of 1.96~mag. The results are the following (see also Table~\ref{isolatedbis}):

$\bullet$ Isolated-5: This source lies at the end of a small globule, not far from an IF and its associated PDR (Fig.~\ref{isolated5+11}). It is possibly the head of a pillar 
separated from the PDR (similar to what is observed in the region of the pillars to the south-east of W5-E; see Sect.~6.2). The emission we detect 
at 100~$\mu$m may also be the counterpart 
to the Class~I YSO identified by KOE08. Five models satisfy $\chi^2-\chi ^2$ (best) per data point $\leq$3, all of which point to a low-mass Stage~I source. 
But we cannot exclude that we are dealing with two distinct sources, a Class~I YSO close to or inside a globule. Using only the Herschel fluxes and the 
modified blackbody model we estimate a mass of 1.7~\msol\ for the globule's head or the YSO's envelope (Table~\ref{temperatureYSOs}).

$\bullet$ Isolated-6: As shown by Fig.~\ref{isolatedSEDs} the SED is not well-constrained. A large range of masses and ages are possible for the central source. 
However, the fact that the envelope accretion rate is never zero, possibly disagrees with KOE08 classification as a Class~II.

$\bullet$ Isolated-11: According to KOE08 this source is a  Class~I YSO.  It is a bright {\it Spitzer} and {\it Herschel} source, 
point-like at 100~$\mu$m (as shown by Fig.~\ref{isolated5+11}). Only one model satisfies $\chi ^2-\chi ^2$(best) per data point $\leq$3
for the whole range of wavelengths.  Its parameters, given in Table~\ref{isolatedbis}, correspond to an intermediate mass Stage~II/III source.
Fig.~\ref{isolatedSEDs}, however, shows that the fit is not satisfactory. When we use only the 100~$\mu$m flux the fit is much better
(SED of isolated-11$^*$).  In this case the SED corresponds to a Stage~I YSO of about 1~\msol. As shown by Fig.~\ref{isolated5+11},
the source lies near an IF, and is surrounded by a small bright rim.  One possible explanation is that, at {\it Herschel}-SPIRE
wavelengths we see the emission of the cold condensation enclosed by the bright rim, whereas at {\it Spitzer} and at 100~$\mu$m
we see the Class~I YSO, possibly a Stage~I YSO recently formed inside this condensation. Better spatial resolution in the FIR
is necessary to confirm this interpretation.

$\bullet$ Isolated-12: Several very similar models satisfy $\chi ^2-\chi^2$(best) per data point $\leq$3 and therefore we give only the best one in Table~\ref{isolatedbis}. All  point to a Stage~II source, in agreement with KOE08 classification. This is an intermediate mass YSO, with no accreting envelope. 

$\bullet$ Isolated-14bis: This source lies in the direction of PDR~2. It is classified as a Class~II YSO by KOE08. It is bright 
at 24~$\mu$m but its 100~$\mu$m counterpart is rather faint (Fig.~\ref{isolated5+11}). A Class~I YSO lies nearby that we call isolated-14  (isolated-14 is very faint at  
24~$\mu$m and has no 100~$\mu$m counterpart; thus, its classification is uncertain). For isolated-14bis several models satisfy $\chi^2-\chi^2$(best) per data point $\leq$3, 
which span a large range of parameters. The classification of isolated-14bis is therefore uncertain. A better fit is obtained if we let free the external extinction (up to 10~mag). 
This source is clearly a Stage~II YSO (of intermediate mass), as all the models give an envelope accretion rate equal to zero and a massive disk.

\end{appendices}
%
\end{document}